\documentclass{jfm}

\usepackage{amsmath,amssymb}
\usepackage{mathrsfs}
\usepackage{xcolor}
\usepackage{subfigure}
\usepackage{bbm}
\usepackage{bm}
\usepackage{pict2e} 
\usepackage{booktabs}
\usepackage{tabularx}
\usepackage{setspace}
\usepackage{mathtools}
\usepackage{tikz}
\usepackage{soul}
\usepackage{graphicx,epstopdf}

\usepackage[colorlinks=true, linkcolor=blue, urlcolor=blue, citecolor=blue]{hyperref}
\usepackage{natbib}

\usepackage{multirow,bigdelim}
\usepackage{makecell}

\begin{document}
\newtheorem{lemma}{Lemma}
\newtheorem{corollary}{Corollary}
%
\title{Efficient simulation of non-classical liquid-vapour phase-transition flows: a method of fundamental solutions}
\shorttitle{Efficient simulation of non-classical liquid-vapour phase-transition flows}
\shortauthor{A. Rana, S. Saini, S. Chakraborty, D. Lockerby and J. Sprittles} 

\author
 {
 {Anirudh S. Rana\aff{1}
\corresp{
 \email{anirudh.rana@pilani.bits-pilani.ac.in}
},
}
Sonu Saini\aff{1},
Suman Chakraborty\aff{2},
Duncan A. Lockerby\aff{3},
\and
 James E. Sprittles\aff{4}
 }

\affiliation
{
\aff{1}
Department of Mathematics, BITS Pilani, Pilani Campus, Rajasthan, India
\aff{2}
Department of Mechanical Engineering, Indian Institute of Technology Kharagpur, India
\aff{3}
School of Engineering, University of Warwick, Coventry CV4 7AL, UK
\aff{4}
Mathematics Institute, University of Warwick, Coventry CV4 7AL, UK
}

\maketitle

\begin{abstract} 
Classical continuum-based liquid vapour phase-change models typically assume continuity of temperature at phase interfaces along with a relation which describes the rate of evaporation at the interface (Hertz-Knudsen-Schrage, for example).
However, for phase transitions processes at small scales, such as the evaporation of nanodroplets, the assumption that the temperature is continuous across the liquid-vapour interface leads to significant inaccuracies \citep{Ward2002, RanaPRL2019}, as may the adoption of classical constitutive relations that lead to the Navier-Stokes-Fourier equations (NSF). In this article, to capture the notable effects of rarefaction at small scales,  we adopt an extended continuum-based approach utilizing the coupled constitutive relations (CCR). In CCR theory, additional terms are invoked in the constitutive relations of NSF equations originating from the arguments of irreversible thermodynamics as well as consistent with kinetic theory of gases. 
The modelling approach allows us to derive new fundamental solutions for the linearised CCR model and to develop a numerical framework based upon the method of
fundamental solutions (MFS) and enables
three-dimensional multiphase micro-flow simulations to be performed  at remarkably low computational cost.
 The new framework is benchmarked against classical results and then explored as an efficient tool for solving three-dimensional phase-change events involving droplets. 

\end{abstract}
\section{Introduction}

When a liquid evaporates it loses energy; that is why sweating cools us down. Evaporation is an effective cooling mechanism widely employed by nature; for example, as a body-temperature control for mammals \citep{Sherwood2005}, and in engineering applications, such as spray drying, spray cooling and microelectronics cooling \citep{Dhavaleswarapu, plawskyreview, lingreview}. The use of pot-in-pot coolers by mankind can be tracked back to Bronze Age civilisation \citep{Evans2000};  a device that relies on the principle of evaporative cooling.

The description of phase-transition processes in micro/nano scales is
intriguing, mainly due to the inability of classical continuum theories to accurately capture the rarefied vapour-flow characteristic. Specifically, as the representative physical length
scale ($\ell$) of the flow becomes comparable to the mean free path ($\lambda $%
) in the vapour, i.e., the Knudsen number $(\mathrm{Kn}=\lambda /\ell)\approx 1$ \citep{Bird1994, Cercignani2000, Sone2002, Struchtrup2005, TorrilhonARFM}. 
The classical continuum description based on the Euler or Navier--Stokes--Fourier (NSF) equations
is valid only in regimes for which $\mathrm{Kn}\lesssim 0.001$---referred to as the hydrodynamic regime. There
are, of course, computational atomistic techniques, such as molecular
dynamics simulations (MD), which are capable of modelling phase transition
at nano-scales. However, due to computational time and memory limitations, it is impractical to use them for multi-scale processes,  which span over a wide range of time and length
scales. 

The behaviour of a dilute vapour phase can readily  be described by the Boltzmann equation, which is an evolution equation for the (velocity) distribution function of vapour molecules  \citep{Bird1994, Cercignani2000}. The Boltzmann equation offers the complete mesoscopic description of the vapour for all range of $\mathrm{Kn}$; however, the exact analytical solution of the Boltzmann equation is intractable in a general case while its numerical solutions are computationally very expensive.   Grad---in his seminal work \citep{Grad1949a, Grad1949b}---proposed an asymptotic solution procedure
for the Boltzmann equation through the method of moment. A variety of other extended fluid dynamics equations, aiming to describe processes under rarefied conditions, have been developed, see e.g.~\cite{Eu1992, MullerandRuggeri1998, Kremer2010, Struchtrup2005, ChakrabortyDurst2007, Dongari2010, Myong2014, TorrilhonARFM}.
These models close the gap between
classical fluid dynamics 
and kinetic theory; that is, they aim at a good description in the transition regime ($0.001\lesssim \mathrm{Kn}\lesssim 1$).

The classical phase transition models, found in the literature, assume that the temperature and the velocity tangential to the interface are continuous across it. However, it is now well established, both experimentally
and theoretically \citep{Ward2002, BondStruchtrup2004, RanaPRL2019}, that such assumptions are invalid at a nano-confined liquid-vapour interface. For such small length scales,  the Knudsen number lies in the transition regime (or beyond) and a 
difference in velocity and temperature jump are observed across the interface---thus it is important to employ models capable of describing strong non-equilibrium effects that  occur at high Knudsen numbers. In  \citet{RanaPRL2019}, the lifetime of an evaporating nanodroplet was studied. The study revealed that,  when the drop size is large ($>$micron),
its diameter-squared decreases in  proportion  to time; known as the $d^2$-law. However, as the diameter approaches sub micron and nano-scales, a crossover to a new behavior is observed, with the diameter now
reducing in  proportion to time (following a `$d$--law'). 
By taking into account the temperature discontinuity at the liquid-vapour interface,the drop's lifetime was correctly predicted. 

Notably, a full theory on boundary conditions is still missing for these nonlinear extended fluid dynamics equations, for which the notion of solvability and stability is more intricate. The approach suggested by \cite{Grad1958}, and recently explored by researchers \citep{Gu&Emerson2007, TorrilhonStruchtrupJCP2008, GuEmerson2009, SBRF2017, RanaLockerbySprittles2018}, violates Onsager symmetry conditions \citep{RanaStruchtrup2016, BRTS2018, SarnaTorrilhon2018}, leading to instabilities which influence  the convergence of numerical schemes, and prevent convergence of moments \citep{TorrilhonSarna2017, ZinnerOttinger2019}. 
    
The second law of thermodynamics (entropy inequality) plays an important role in finding constitutive relations in the bulk \citep{deGrootMazurBook1962, Gyarmati1970, Harten1983, Lebonetal2008}, in designing numerical schemes \citep{Tadmor2006, KumarMishra2012, chandrashekar_2013}, as well as, in developing physically admissible boundary conditions \citep{BondStruchtrup2004, STPRL2007, KjelstrupBedeauxBook2008, Schweizer2016NonequilibriumTO, RanaStruchtrup2016, RGS2018, SarnaTorrilhon2018, BRTS2018}. For the latter, one determines the entropy generation at the boundary and finds the boundary conditions as phenomenological laws that guarantee positivity of the entropy generation at the boundary. For example, in \citet{STPRL2007}, \citet{RanaStruchtrup2016}, \citet{SarnaTorrilhon2018} and \citet{BRTS2018} the linearized second law (taking a quadratic entropy) was employed to determine boundary conditions for the linearized version of the moment equations. The resulting phenomenological boundary conditions were thermodynamically consistent for all processes and complied with the Onsager reciprocity relations \citep{Onsager38}. 

Over the last few years, an impressive body of research work has been devoted to the development of  constitutive theories whose closed conservation laws guarantee the second law \citep{OttingerPRE2010, MTPRLreply2010, MT2012, LIU201347, PaolucciandPaolucci2018}.  The Burnett equations, which are obtained perturbatively from the kinetic theory, are unstable \citep{Bobylev2006} and due to lack of any coherent second law these equations lead to physically inadmissible solutions \citep{StruchtrupNedler2020}. The Burnett equation are unstable due to lack of a proper entropy inequality. On the other hand, the moment equations (Grad 13-moment equations, regularized 13-moment equations, etc.) are accompanied by proper entropy inequalities, and are stable, but only in the linear cases \citep{STPRL2007, RanaStruchtrup2016, TorrilhonSarna2017}. In \cite{RGS2018}, the coupled constitutive relations (CCR) were obtained---as an enhancement to the NSF equations---by considering a correction term to the non-convective entropy flux (in addition to the classic entropy flux: heat-flux/thermodynamic temperature). As a result, the CCR adds several second-order terms to the NSF equations, in the bulk and in the boundary conditions, capturing important rarefaction effects in moderately rarefied conditions---such as the Knudsen minimum, and non-Fourier heat transfer---which cannot be described by classical hydrodynamics \citep{RGS2018}.

Through this article, we are going to utilize the CCR theory, thereby some comments about these equations are in order. The CCR theory originates from the arguments of irreversible thermodynamics. In CCR theory, additional terms appear in the constitutive relations of NSF equations, where the coupling between constitutive relations for the heat-flux vector and stress tensor is introduced via a coupling coefficient ($\alpha_0$). For $\alpha_0 = 0$, the coupling vanishes and one obtains the classical NSF relations. Besides, for the flow conditions  considered in the article (steady state and linearized equations), the CCR system (conservation laws plus CCR) reduces to the linearized Grad 13-moment equations for $\alpha_0 = 2/5$; a value obtained for  Maxwell molecule (MM) gases \citep{Struchtrup2005}. Thus, under suitable assumptions, this article provides a unified framework for NSF, Grad-13 and the CCR system. More precisely, we explore the method of fundamental solutions (MFS) for the CCR system (hence, NSF, Grad-13) that can be useful for constructing solutions over a wide range of free-stream profiles and complex geometries.  

The MFS \citep{KUPRADZE1964} is a mesh-free numerical technique for solving partial differential equations based on using the fundamental solutions (Green's functions). The main advantage that the MFS has over the more classical numerical methods, such as finite difference method, finite element method, is that its solution procedure does not require discretisation of the interior of the computational domain, thus a significant amount of computational effort and time is saved. Furthermore, the MFS does not require the (very computationally expensive) 3D mesh generation, as a result, the MFS is ideally suited for the problems involving moving interfaces and phase-change processes.
The first steps towards utilizing MFS  for rarefied gas flows were taken by \cite{LC2016}, who derived Green's functions for the Grad 13 system and utilized them to solve some canonical problems. The present study further develops a MFS-based framework (MFS--CCR) for phase-transition boundary-value problems, which, as we shall show later, will involve derivation of new Green's functions generated by a singular mass source/sink term.
%
In this article, a set of temperature-jump and velocity-slip boundary conditions for a liquid-vapour interface will be derived for the linearised CCR system and implemented within the framework of MFS. However, it should be noted that the developed MFS methodology in this article can also be applied to other type of liquid-vapour interface boundary conditions, such as the classical Schrage law \citep{Schrage1953}, statistical rate theory \citep{Ward1999} and phenomenological approach based on Hertz-Knudsen-Schrage relation \citep{LIANG2017105}. We do not address these boundary conditions in this article, instead leaving this for a future investigation.

The remainder of this paper is organized as follows. In Section \ref{sec:The inearized NSF and CCR systems}, we introduce the linearized and dimensionless CCR model. In Section
\ref{sec: Green's functions}, the derivation of Green's functions for the CCR model is presented, followed by the derivation of thermodynamically consistent boundary conditions at the phase interface in Section \ref{Sec: Liquid-vapour interface boundary conditions}. A brief summary of the method of fundamental solutions
and implementation is given in Section \ref{sec: Method of fundamental solutions}. In Section \ref{subsec: Validation and verification of CCR-MSF} the numerical scheme is applied to evaporation from a spherical droplet and low-speed
 gas flow around a sphere (for which the analytical solutions exist) in order to validate our numerical scheme. 
To illustrate the utility of fundamental solutions for complex geometries, in Sections \ref{sec: Motion of two spherical non-evaporating droplets}, \ref{sec: Interaction of two evaporating droplets} and \ref{sec: evaporation of a deformed droplet} we solve the motion of two spherical non-evaporating droplets with different orientations, evaporation of two interacting droplets, and evaporation of a deformed droplet, respectively.  Concluding remarks are made in Section \ref{sec: Conclusion and future directions}.

\section{The linearized NSF, Grad-13 and CCR systems}
\label{sec:The inearized NSF and CCR systems}
In this article  we shall only consider flow conditions where the deviations
from a constant equilibrium state---given by a constant reference density $%
\hat{\rho}_{0}$, a constant reference temperature $\hat{T}_{0}$, and all
other fields zero---are small. Thereby, the governing equations can be
linearized with respect to the reference state. The governining equations are 
 put into dimensionless form by introducing%
\begin{equation}
\mathbf{r}=\frac{\mathbf{\hat{r}}}{\hat{\ell} }\text{, \quad }\rho =\frac{\hat{\rho}-%
\hat{\rho}_{0}}{\hat{\rho}_{0}}\text{,\quad and }T=\frac{\hat{T}-\hat{T}_{0}}{\hat{T}%
_{0}}\text{,}  \label{dimensionless 1}
\end{equation}%
where $\mathbf{r}$ is the position vector in Cartesian coordinates, $\hat{\ell}$ is the reference length scale, and $\rho $ and $T$ are the
dimensionless perturbations in the density and temperature from their
reference values, respectively; hats denote dimensional quantities. The
dimensionless velocity vector $\mathbf{v}$, the stress tensor $\bm{\Pi }$
and the heat-flux vector $\mathbf{q}$ are  
\begin{equation}
\mathbf{v}=\frac{\mathbf{\hat{v}}}{\sqrt{\mathrm{\hat{R}}\hat{T}_{0}}}\text{, }\quad%
\bm{\Pi }=\frac{\bm{\hat{\Pi}}}{\hat{\rho}_{0}\mathrm{\hat{R}}\hat{T}_{0}}%
\text{,\quad and }\mathbf{q}=\frac{\mathbf{\hat{q}}}{\hat{\rho}_{0}\left( \mathrm{%
\hat{R}}\hat{T}_{0}\right) ^{3/2}}\text{,}  \label{dimensionless 2}
\end{equation}%
where $\mathrm{\hat{R}}$ is the gas constant. The perturbation in pressure $p$ is
given by a linearized equation of state 
\begin{equation}
p  = \frac{\hat{\rho}\hat{R}\hat{T}}{\hat{\rho}_{0}\hat{R}\hat{T}_{0}}-1 = 
\left( 1+\rho \right) \left( 1+\theta \right)-1 \approx \rho +\theta 
\end{equation}%
where the last equation assumes small perturbations $\rho $ and $\theta $, hence
 $\rho \theta $ is assumed to be negligible in comparison to $\rho +\theta$.

Accordingly,
dimensionless and steady-state (linearized) conservation laws for mass,
momentum, and energy read%
\begin{equation}
\nabla \cdot \mathbf{v}=0\text{,\quad }\nabla p+\nabla \cdot \bm{\Pi }=0\text{%
,\quad and }\nabla \cdot \mathbf{q}=0\text{.}
\label{conservation laws vector form}
\end{equation}%
It should be noted that conservation laws (\ref{conservation laws vector
form}) contain the stress tensor and heat flux vector as unknowns, hence
constitutive equations are required for these quantities. If the CCR closure
is adopted, the (linearized) constitutive relations for $\bm{\Pi }$ and $%
\mathbf{q}$ read \citep{RGS2018}
\begin{equation}
\bm{\Pi }=-2\mathrm{Kn}\langle \nabla \mathbf{v}\rangle -2\alpha _{0}%
\mathrm{Kn}\langle \nabla \mathbf{q}\rangle \text{,\quad and }\mathbf{q}=-\frac{%
c_{p}\mathrm{Kn}}{\Pr }\left( \nabla T+\alpha _{0}\nabla \cdot \bm{\Pi }%
\right)  \label{CCR relations vector form}
\end{equation}%
where, $c_{p}=5/2  (=\hat{c}_{p}/\hat{R})$ is the specific heat for mono-atomic gases, $\Pr $ is the
Prandtl number and the Knudsen number $\mathrm{Kn}=\hat{\mu}_{0}/\left( \hat{%
\rho}_{0}\sqrt{R\hat{T}_{0}}\hat{\ell} \right) $, where $\hat{\mu}_{0}$ is the
viscosity of the gas at the reference state. Furthermore, the angular
brackets in (\ref{CCR relations vector form}) denote the traceless and
symmetrical component of the tensor, for instance 
\begin{equation}
\langle \nabla \mathbf{v}\rangle =\frac{1}{2}\left( \nabla \mathbf{v+}\nabla 
\mathbf{v}^{T}\right) -\frac{1}{3}\left( \nabla \cdot \mathbf{v}\right) 
\mathbf{I}\text{,}  \label{traceless tensor}
\end{equation}%
where $\mathbf{I}$ is the identity matrix.

Since the constitutive relations (\ref{CCR relations vector form}) for the
fluxes $\bm{\Pi}$ and $\mathbf{q}$ are coupled (in contrast to NSF), these equations are
referred as the coupled constitutive relations (CCR). The benefit of the CCR model
compared to extended moment equations (Grad-13 equations, for instance) is
that it retains full thermodynamic structure without any restriction \citep{RGS2018}.

In the above equations the phenomenological coefficient $\alpha _{0}$
appears, which is chosen such that the CCR model agrees with the Burnett
coefficients (in the sense of the Chapman--Enskog expansion of the Boltzmann equation); the NSF
equations are obtained when $\alpha _{0}=0$. The value of $\alpha _{0}$
for the Maxwell molecule gases is 2/5; for
other power potentials it can be computed from the relation $\alpha _{0}=%
\frac{\Pr \varpi _{3}}{5}$, where $\varpi _{3}$ is the Burnett coefficient \citep{Struchtrup2005, RGS2018}.

It is important to note here, in steady state, the linearized CCR model
reduces to the linearized Grad's-13 moment equations for $\alpha _{0}=2/5$
(i.e., for Maxwell molecule gases). Hence, using the naming convention of \cite{LC2016, CSRSL2017}, the solution of the CCR equations (\ref{conservation laws vector form}--\ref{CCR relations vector form}) will still be referred as the Generalised Gradlet.

\section{Green's functions: the Gradlet, thermal Gradlet, and sourcelet}
\label{sec: Green's functions}
In this section, we shall write the fundamental solutions (Green's
functions) for the CCR system (\ref{conservation laws vector form}--\ref{CCR relations vector form}), that will be used further in constructing the new
solutions for complex geometries and phase-change problems.

\subsection{The Gradlet and Thermal Gradlet}

The fundamental solutions sought here correspond to the steady-state
response of the vapour with regard to a point body force $\mathbf{f}$ and a heat source of
strength $g$, acting at the singularity point at the origin ($%
\mathbf{r}^{s}=0$), i.e., the Green's functions associated with the following
equations 
\begin{equation}
\label{conservation laws perturebed}
\nabla \cdot \mathbf{v} =0\text{,\quad }  
\nabla p+\nabla \cdot \bm{\Pi } =\mathrm{Kn}\,\,\mathbf{f}\delta \left( 
\mathbf{r}\right)\text{,\quad and}\quad  
\nabla \cdot \mathbf{q} =\mathrm{Kn}\,\,g\delta \left( \mathbf{r}\right)\text{,}
\end{equation}
where $\delta\left(\mathbf{r}\right) $ is  the three-dimensional Dirac
delta-function.
The solutions to (\ref{conservation laws perturebed}) along with  (\ref{CCR
relations vector form}) are obtained via a Fourier transformation.   The velocity $\mathbf{v}^{(Gr)}$,
pressure $p^{(Gr)}$, and stress tensor $\bm{\Pi }^{(Gr)}$, at any point $%
\mathbf{r}$, are found as 
\begin{subequations}
\label{Gradlet solutions}
\begin{eqnarray}
\mathbf{v}^{(Gr)}{\left (\mathbf{r}\right)} &=&\frac{1}{8\pi }\mathbf{J}{\left (\mathbf{r}\right)} \cdot \mathbf{f}+\frac{3\alpha _{0}^{2}c_{p}\mathrm{%
Kn}^{2}}{4\pi \Pr}\mathbf{K}{\left (\mathbf{r}\right)} \cdot \mathbf{f}\text{,}  \label{Gradlet velocity} \\
p^{(Gr)}{\left (\mathbf{r}\right)} &=&\frac{\mathrm{Kn}}{4\pi }\frac{\mathbf{f\cdot r}}{|\mathbf{r}|^{3}}\text{,%
}  \label{Gradlet pressure} \\
\bm{\Pi }^{(Gr)}{\left (\mathbf{r}\right)} &=&\frac{3\mathrm{Kn}}{4\pi }\left( \mathbf{f\cdot r}%
+2\alpha _{0}\mathrm{Kn}g\right) \mathbf{K}{\left (\mathbf{r}\right)}\text{.}  \label{Gradlet stress}
\end{eqnarray}%
\end{subequations}
Here, we have introduced the abbreviations 
\begin{equation}
\mathbf{J}{\left (\mathbf{r}\right)}=\frac{\mathbf{r}\mathbf{r}}{|\mathbf{r}|^{3}}+\frac{\mathbf{I}}{|%
\mathbf{r}|}\text{,\quad and }\quad \mathbf{K}{\left (\mathbf{r}\right)}=\frac{\mathbf{%
r}\mathbf{r}}{|\mathbf{r}|^{5}}-\frac{1}{3}\frac{\mathbf{I}}{|\mathbf{r}|^{3}}\text{,}  \label{Oseen-Burgers tensor}
\end{equation}%
to denote the Oseen-Burger tensor and the third decaying harmonic tensor \citep{Lamb1945},
respectively.
The temperature $T^{(Gr)}$ and heat flux $\mathbf{q}^{(Gr)}$ are obtained as 

\begin{subequations}
\label{Gradlet  solutions2}
\begin{eqnarray}
T^{(Gr)}{\left (\mathbf{r}\right)} &=&\frac{\Pr }{4\pi c_{p}}\frac{g}{|\mathbf{r}|}\text{,}
\label{Gradlet temperature} \\
\mathbf{q}^{(Gr)}{\left (\mathbf{r}\right)} &=&\frac{\mathrm{Kn}}{4\pi }\frac{g}{|\mathbf{r}|^{3}}\mathbf{r}-%
\frac{3\alpha _{0}c_{p}\mathrm{Kn}^{2}}{4\pi \Pr } \mathbf{K}{\left (\mathbf{r}\right)} \cdot \mathbf{f}\text{.%
}  \label{Gradlet  heat flux}
\end{eqnarray}%
\end{subequations}
Here, it is assumed that all the field variables are measured relative to
their equilibrium values at infinity, hence all field variables vanish as $%
|\mathbf{r}|\rightarrow \infty$.

One can easily verify that, for $\alpha _{0}=0$, the fundamental solutions (%
\ref{Gradlet solutions}--\ref{Gradlet solutions2}) reduce to the
solution of Stokes (Stokeslet) and the heat equation (Thermal Stokeslet),
which is available in the works of Oseen and Burger \citep{Lamb1945} and the recent work
of \cite{LC2016}. Moreover, for $\alpha _{0}=2/5$, (\ref{Gradlet
solutions}--\ref{Gradlet solutions2}) reduce to the Grad-13 moment
equations---the Gradlet and the thermal Gradlet---obtained by \cite{LC2016}. Thus, one can also think of (\ref{Gradlet solutions}--\ref%
{Gradlet solutions2}) as solution of 13 moment equations (steady-state and
linearized) with general power potentials in the collision operator for the Boltzmann equation.
\subsection{The Sourcelet}

The linearity of the equations (\ref{conservation laws vector form}-\ref{CCR
relations vector form}) allows one to obtain a class of singularity
solutions that are readily applicable to various type of boundary-value
problems. 
Note that the fundamental solutions (Gradlet and thermal Gradlet) obtained in the previous section create no mass flux. Due to Gauss' theorem, one may draw any boundary enclosing the Stokeslet/Gradlet, and find that there can  be no mass flux across it  ($\nabla \cdot \mathbf{v} =0$ everywhere).  Therefore, it follows that, to capture phase-change events (and here we are particularly interested in the liquid-vapour interface) an additional fundamental solution are required---one that produces a source of mass. Hence, we consider the solution of
the following equations 
\begin{equation}
\nabla \cdot \mathbf{v}=h\delta (\mathbf{r})\text{, }\nabla p+\nabla \cdot 
\bm{\Pi }=0\text{, and }\nabla \cdot \mathbf{q}=0\text{,}
\label{conservation laws perturebed source-sink}
\end{equation}%
Where $h$ is the strength of the mass source.    Again, the fundamental solutions of (\ref{conservation laws perturebed source-sink} and \ref{CCR relations vector form}) are derived via Fourier
transform (See appendix \ref{sec:appendixA0} for detailed derivation), yielding 
\begin{subequations}
\label{solutions source}
\begin{eqnarray}
\mathbf{v}^{(S)}{\left (\mathbf{r}\right)} &=&\frac{h}{4\pi }\frac{\mathbf{r}}{|r|^{3}}\text{,}
\label{velocity source} \\
\bm{\Pi }^{(S)}{\left (\mathbf{r}\right)} &=&\frac{3\mathrm{Kn} h}{2\pi }\mathbf{K}{\left (\mathbf{r}\right)}\text{,}
\label{stress source} \\
p^{(S)}{\left (\mathbf{r}\right)} &=&T^{(S)}{\left (\mathbf{r}\right)}=0\text{, and }\mathbf{q}^{(S)}{\left (\mathbf{r}\right)}=0\text{.}
\label{other source}
\end{eqnarray}%
As such, the pressure, temperature and heat flux vanish for this case. 

\subsection{Equations collected}

Combining the Gradlet, thermal Gradlet (\ref{Gradlet solutions}-\ref{Gradlet
solutions2}) and sourcelet (\ref{solutions source}), the flow
fields induced by Generalised Gradlet (a point mass source, point force, and point heat
source) are  
\end{subequations}
\begin{subequations}
\label{solution all 1}
\begin{eqnarray}
\mathbf{v}{\left (\mathbf{r}\right)} &=&\frac{1}{8\pi }\mathbf{J}{\left (\mathbf{r}\right)}\cdot \mathbf{f}+\frac{3\alpha _{0}^{2}c_{p}\mathrm{%
Kn}^{2}}{4\pi \Pr}\mathbf{K}{\left (\mathbf{r}\right)}\cdot \mathbf{f}+\frac{h}{4\pi }\frac{\mathbf{r}}{|r|^{3}}\text{,}
\label{velocity solution all} \\
p{\left (\mathbf{r}\right)} &=&\frac{\mathrm{Kn}}{4\pi }\frac{\mathbf{f\cdot r}}{|r|^{3}}\text{,}
\label{presure solution all} \\
\bm{\Pi }{\left (\mathbf{r}\right)} &=&\frac{3\mathrm{Kn}}{4\pi }\left( \mathbf{f\cdot r}+2\alpha
_{0}\mathrm{Kn}g+2h\right)\mathbf{K}{\left (\mathbf{r}\right)} \text{,} \label{stress solution all}
\end{eqnarray}%
and 
\end{subequations}
\begin{subequations}
\label{solution all 2}
\begin{eqnarray}
T{\left (\mathbf{r}\right)} &=&\frac{\Pr }{4\pi c_{p}}\frac{g}{|r|}\text{,}  \label{temperature solution all}
\\
\mathbf{q}{\left (\mathbf{r}\right)} &=&\frac{\mathrm{Kn}}{4\pi }\frac{g}{|r|^{3}}\mathbf{r}-\frac{%
3\alpha _{0}c_{p}\mathrm{Kn}^{2}}{4\pi \Pr }\mathbf{K}{\left (\mathbf{r}\right)}\cdot \mathbf{f}\text{.}
\label{heat solution all}
\end{eqnarray}
\end{subequations}

These solutions need to be supplemented with appropriate boundary conditions, which are formulated in the  next section.
\section{Liquid-vapour interface boundary conditions}
\label{Sec: Liquid-vapour interface boundary conditions}
In this section, we shall derive the phase interface boundary conditions for the CCR system within the framework of irreversible thermodynamics \citep{KjelstrupBedeauxBook2008}.
In order to establish (thermodynamically consistent) boundary conditions at the phase interface, one determines the entropy generation at the interface, and finds the boundary conditions as phenomenological laws that guarantee positivity
of the entropy generation rate. For the linearized CCR equations, the entropy generation rate at the interface is  \citep{BRTS2018, RGS2018}
\begin{align}
\label{entropyProdwall}
\Sigma _{\text{surface}}&=-\left(\mathbf{v}-\mathbf{v}^{I}\right)\cdot \mathbf{n}\left[
p-p_{sat}+\left( \mathbf{n}\cdot \bm{\Pi}\cdot \mathbf{n}\right) \right]
-\mathbf{q}\cdot \mathbf{n}\left[T-T^{l}+\alpha _{0}\left( 
\mathbf{n}\cdot \bm{\Pi}\cdot \mathbf{n}\right) \right]\nonumber\\
&\quad
-\sum_{i=1}^{2}\left( \mathbf{t}^{(i)}\cdot \bm{\Pi }\cdot \mathbf{n}%
\right) \left[ \mathbf{t}^{(i)}\cdot \left( \mathbf{v-v}^{l}\right) +\alpha
_{0}\mathbf{t}^{(i)}\cdot \mathbf{q}\right] \geq 0\text{,}
\end{align}%
where $\mathbf{n}$ is a unit normal pointing from a boundary point into the gas, and $\left\{ \mathbf{t}^{(i)}\right\} _{i=1,2}$%
 are orthonormal tangent vectors to the boundary. Here, $\mathbf{v}^{I}$
denotes the velocity of the interface, $T^{I}$ denotes the temperature of
the liquid at the interface and $p_{sat}{\left( T^{I}\right) }$ is the
saturation pressure. 

The positivity of entropy generation rate $\Sigma
_{surface}$ is ensured by adopting the following boundary equations%

\begin{subequations}
\label{BCs evaporation}
\begin{eqnarray}
\left(\mathbf{v}-\mathbf{v}^I\right)\cdot \mathbf{n} &=& -\eta _{11}\left(p-p_{sat}+\mathbf{n}\cdot \bm{\Pi }\cdot \mathbf{n} \right)+\eta _{12}\left(T-T^{I}+\alpha _{0}\mathbf{n}\cdot \bm{\Pi }\cdot \mathbf{n}\right)\\
\mathbf{q}\cdot \mathbf{n} &=&\eta _{12}\left(p-p_{sat}+\mathbf{n}\cdot \bm{\Pi }\cdot \mathbf{n} \right)-(\eta _{22}+2\tau_0)\left(T-T^{I}+\alpha _{0}\mathbf{n}\cdot \bm{\Pi }\cdot \mathbf{n}\right)
\end{eqnarray}
\end{subequations}
%
and%
\begin{subequations}
 \label{BC slip}
\begin{eqnarray}
\mathbf{t}^{(1)}\cdot \bm{\Pi }\cdot \mathbf{n} &=&-\varsigma\left( 
 \mathbf{v-v}^{I} +\alpha _{0} \mathbf{q}\right)\cdot\mathbf{t}^{(1)} \text{,} \label{BC slip 1} \\
\mathbf{t}^{(2)}\cdot \bm{\Pi }\cdot \mathbf{n} &=&-\varsigma \left(\mathbf{v-v}^{I} +\alpha _{0}\mathbf{q}\right)\cdot\mathbf{t}^{(2)}\text{.}  \label{BC slip 2}
\end{eqnarray}
\end{subequations}
Through boundary conditions (\ref{BCs evaporation}), the evaporative mass and heat flux
are governed by the difference between pressure and saturation
pressure, and temperature difference across the interface. Furthermore, the boundary conditions  (\ref{BC slip 1}-\ref{BC slip 2}) relates the
shear stress to the tangential velocity slip and  thermal
transpiration---a flow induced by a tangential heat flux---which is a second-order effect \citep{Sone2002, Hadjiconstantinou2003, Cercignani2010}.
The boundary conditions (\ref{BCs evaporation}-\ref{BC slip}) are an extension to the
classical Hertz-Knudsen-Schrage law for evaporation \citep{Schrage1953}, taking into account the temperature jump, velocity slip and some second-order effects.  

In boundary conditions (\ref{BCs evaporation}), $\eta _{ij}$'s are Onsager resistivitity coefficients, which are obtained from the asymptotic ($\mathrm{Kn}\rightarrow 0$) kinetic theory \citep{Sone2002}, as (see appendix \ref{sec:appendixPBC} for details)
\begin{equation}
\label{PBC coeffcients}
\eta_{11}=0.9134\sqrt{\frac{2}{\pi}}\frac{\vartheta}{2-\vartheta}\text{, }\eta_{12}=0.3915\sqrt{\frac{2}{\pi}}\frac{\vartheta}{2-\vartheta}\text{, and }\eta_{22}=0.1678\sqrt{\frac{2}{\pi}}\frac{\vartheta}{2-\vartheta}\text{.} 
\end{equation}
These coefficients were obtained  assuming evaporation/condensation coefficient
$\vartheta$ is independent of the impact energy of molecules and that all vapour
molecules that are not condensing are thermalized, i.e., the accommodation coefficient, $\chi=1$.
The temperature-jump coefficient  $\tau_0 = 0.8503\sqrt{2/\pi}$ and velocity-slip coefficient $\varsigma = 0.8798\sqrt{2/\pi}$..  Throughout this article, we shall take $\vartheta=0$ for the canonical boundaries and $\vartheta=1$ (a value largely accepted in literature) for the phase-change boundaries.

Although in the development of our MFS approach, we use equation (\ref{BCs evaporation}-\ref{BC slip}) as the boundary conditions, this methodology is in principle capable of accommodating other type of phase-interface boundary conditions, such as statistical rate theory \citep{Ward1999} and phenomenological approach based on Hertz-Knudsen-Schrage relation \citep{LIANG2017105}. We do not address these boundary conditions in this article but should be worthy of future investigation.

In next sections, we will introduce, and use, the method of fundamental (MFS)
solutions for CCR (CCR-MFS), which is computationally economical, and show that it
provides reliable solutions of the CCR and NSF equations in good agreement
to the known closed-form analytical results.

\section{Method of fundamental solutions for the CCR equations (CCR-MFS)}
\label{sec: Method of fundamental solutions}
\begin{figure}
\centering
\includegraphics[scale=0.6]{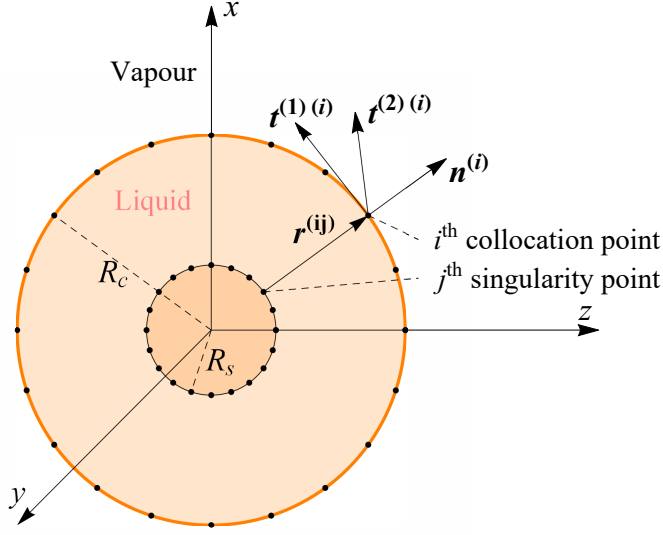}
\caption{\label{fig:msflayout}
Schematic representation of the collocation and singularity points layout; $\mathbf{n}^{(i)}$, $\mathbf{t}^{(1)(i)}$, and $\mathbf{t}^{(2)(i)}$ represent the unit normal and both the tangential vectors, respectively at the $i$th collocation point.}
\end{figure}

Let us consider $N_{c}$ collocation points on the boundary
 at $\left\{ \mathbf{r}^{c(j)}\right\} _{j=1}^{N_{c}}$,
while $N_{s}$ singularity points are located outside the computational domain (i.e. inside the solid/liquid body)
with position vectors $\left\{ \mathbf{r}^{s(j)}\right\} _{j=1}^{N_{s}}$; see Figure \ref{fig:msflayout}.
The flow-field variables are given by a superposition of the Gradlets,
thermal Gradlets and sourcelets, formulated in (\ref{solution all 1}-\ref{solution
all 2}), as 
\begin{subequations}
\label{solution 1 superimposed}
\begin{eqnarray}
\mathbf{v}^{(i)(Gr)} &=&\sum_{j=1}^{N^{s}}\frac{1}{8\pi }\mathbf{J}{\left (\mathbf{r}^{(ij)}\right)}\cdot \mathbf{f}^{(j)}+\frac{3\alpha _{0}^{2}c_{p}\mathrm{Kn}^{2}}{%
4\pi \Pr}\mathbf{K}{\left (\mathbf{r}^{(ij)}\right)}\cdot \mathbf{f}^{(j)}  \label{v1 superimposed} \\
\mathbf{v}^{(i)(S)} &=&\sum_{j=1}^{N^{s}}\frac{h^{(j)}}{4\pi }\frac{%
\mathbf{r}^{\left( ij\right) }}{|\mathbf{r}^{\left( ij\right) }|^{3}}\text{,}
\label{v2 superimposed} \\
\mathbf{v}^{(i)} &=& \mathbf{v}^{(i)(Gr)}+\mathbf{v}^{(i)(S)}\text{,}\label{v total} \\
p^{(i)} &=&\sum_{j=1}^{N^{s}}\frac{\mathrm{Kn}}{4\pi }\frac{\mathbf{f}^{(j)}%
\mathbf{\cdot r}^{\left( ij\right) }}{|\mathbf{r}^{\left( ij\right) }|^{3}}\text{,}
\label{pressure superimposed} \\
\bm{\Pi}^{(i)} &=&\sum_{j=1}^{N^{s}}\frac{3\mathrm{Kn}}{4\pi }\left( \mathbf{f%
}^{(j)}\mathbf{\cdot r}^{\left( ij\right) }+2\alpha _{0}\mathrm{Kn}%
g^{(j)}+2h^{(j)}\right)\mathbf{K}{\left (\mathbf{r}^{(ij)}\right)}\text{,}  \label{stress superimposed}
\end{eqnarray}%
and 
\end{subequations}
\begin{subequations}
\label{solution 2 superimposed}
\begin{eqnarray}
T^{(i)} &=&\sum_{j=1}^{N^{s}}\frac{\Pr }{4\pi c_{p}}\frac{g^{(j)}}{|\mathbf{r}%
^{\left( ij\right) }|}\text{,}  \label{temperature superimposed} \\
\mathbf{q}^{(i)} &=&\sum_{j=1}^{N^{s}}\frac{\mathrm{Kn}}{4\pi }\frac{g^{(j)}%
\mathbf{r}^{\left( ij\right) }}{|\mathbf{r}^{\left( ij\right) }|^{3}}-\frac{%
3\alpha _{0}c_{p}\mathrm{Kn}^{2}}{4\pi \Pr }\mathbf{K}{\left (\mathbf{r}^{(ij)}\right)}\cdot \mathbf{f}%
^{(j)}\text{.}  \label{heat-flux superimposed}
\end{eqnarray}%
\end{subequations}
are fundamental solutions for the CCR equations. Here, $\mathbf{r}^{\left(
ij\right) }=\mathbf{r}^{c\left( i\right) }-\mathbf{r}^{s\left( j\right) }$
are the displacement vectors from the $j$th singularity site to the $i$th
collocation point.

There are total of $5\times N^{s}$ unknowns in (\ref%
{solution 1 superimposed}-\ref{solution 2 superimposed})---$5$ at each
singularity site---which need to be specified via suitable boundary
conditions. The $4$ boundary conditions in (\ref{BCs evaporation}--\ref{BC slip}) need to be satisfied at every collocation point. Hence, it
turns out, one additional boundary condition is required (at each collocation point), we use
the following boundary conditions: 
\begin{subequations}
\label{mass flux BC split}
\begin{align}
\mathbf{v}^{(i)(S)}\cdot \mathbf{n}^{(i)} &= -\eta _{11}\left(p^{(i)}-p_{sat}^{(i)}+\mathbf{n}^{(i)}\cdot \bm{\Pi }^{(i)}\cdot \mathbf{n} ^{(i)}\right)\nonumber\\
&\quad+\eta _{12}\left(T^{(i)}-T^{(i)I}+\alpha _{0}\mathbf{n^{(i)}}\cdot \bm{\Pi}^{(i)}\cdot \mathbf{n}^{(i)}\right)\text{,}\label{mass flux BC1 split}\\
\left(\mathbf{v}^{(i)(Gr)}-\mathbf{v}^{\left( i\right)I}\right) \cdot 
\mathbf{n}^{\left( i\right) } &= 0\label{mass flux BC2 split}\text{,}
\end{align}
\end{subequations}
Notably, we have split the boundary condition (\ref{BCs evaporation}a) into two components (\ref{mass flux BC1 split}) and (\ref{mass flux BC2 split}). Clearly, these two boundary conditions are sufficient for condition (\ref{BCs evaporation}a) to hold.  The boundary conditions for the normal heat-flux (\ref{BCs evaporation}b) and shear-stress (\ref{BC slip}) read
\begin{subequations}
\label{BC othrs}
\begin{align}
\mathbf{q}^{(i)}\cdot \mathbf{n}^{(i)} &= \eta _{12}\left(p^{(i)}-p_{sat}^{(i)}+\mathbf{n}^{(i)}\cdot \bm{\Pi }^{(i)}\cdot \mathbf{n} ^{(i)}\right)\nonumber\\
&\quad-\left(\eta _{22}+2\tau_0\right)\left(T^{(i)}-T^{(i)I}+\alpha _{0}\mathbf{n^{(i)}}\cdot \bm{\Pi}^{(i)}\cdot \mathbf{n}^{(i)}\right)\text{,}\\
\mathbf{t}^{(1)\left( i\right) }\cdot \bm{\Pi }^{\left( i\right) }\cdot 
\mathbf{n}^{\left( i\right) } &=-\varsigma\left(\mathbf{v}^{\left( i\right) }\mathbf{-v}^{\left(
i\right) I}+\alpha _{0}\mathbf{%
q}^{\left( i\right) }\right)\cdot \mathbf{t}^{(1)\left(
i\right) }\text{,}  \label{BC 3 matrix} \\
\mathbf{t}^{(2)\left( i\right) }\cdot \bm{\Pi }^{\left( i\right) }\cdot 
\mathbf{n}^{\left( i\right) } &=-\varsigma\left(\mathbf{v}^{\left( i\right) }\mathbf{-v}^{\left(
i\right) I}+\alpha _{0} \mathbf{%
q}^{\left( i\right) }\right)\cdot \mathbf{t}^{(2)\left(
i\right) }\text{.} \label{BC 4 matrix}
\end{align}
\end{subequations}
For convenience, we write the set of equations (\ref{mass flux BC split}-\ref{BC othrs}) in
 matrix form :  

\begin{equation}
\sum_{j=1}^{N^{s}}\mathbf{\mathscr{L} }^{ij}\mathbf{u}^{\left( j\right) }=%
\mathbf{b}^{\left( i\right) }\text{,}  \label{Big matrix}
\end{equation}%
where $\mathbf{u}^{\left( j\right) }$ is the vector containing the $5$ 
degrees of freedom at each singularity point, i.e., 
\begin{equation*}
\mathbf{u}^{\left( j\right) }=\left\{ g^{\left( j\right) }\text{, }%
f_{1}^{\left( j\right) }\text{, }f_{2}^{\left( j\right) }\text{, }%
f_{3}^{\left( j\right) }\text{, }h^{\left( j\right) }\right\}\text{,}
\end{equation*}%
Where $\mathscr{L} ^{ij}$ is a $5\times 5$ coefficient matrix, and $
\mathbf{b}^{\left( i\right) }$ is a $5\times 1$
vector, which can be readily extracted using a symbolic software; we used
Mathematica{\textregistered} for this. The inhomogeneous part $\left\{ \mathbf{b}^{\left(
i\right) }\right\} _{i=1}^{N^{c}}$ contains the properties of the interface, namely, $T^{(i)I}$, $p^{(i)}_{sat}$ and $\mathbf{v}^{\left( i\right)I}$. The $%
5N^{c}\times 5N^{s}$ linear system (\ref{Big matrix}) is solved for $\left\{ 
\mathbf{u}^{\left( i\right) }\right\} _{i=1}^{N^{s}}$ using  an iterative
quasi-minimal residual method \citep{QMR1991}. Moreover, the normal and the tangent
vectors~$\left\{ \mathbf{n}^{(j)}\text{, }\mathbf{t}^{(1)\left( j\right) }%
\text{, }\mathbf{t}^{(1)\left( j\right) }\right\} _{j=1}^{N^{c}}$ are
obtained using Householder reflection formula for vector orthogonalization
\citep{LOPES2013}.

\section{Results and discussion}
\label{sec: Results and discussion}
The CCR-MFS described above involves some numerical parameters specifically, how many collocation ($N^c$) and singularity ($N^s$) points to take, and the location of singularity points outside of  the computational 
domain. These parameters govern the overall efficiency of the numerical scheme, by striking a balance between numerical error and computational time. Throughout this study, we consider an equal number of collocation and singularity points, i.e., $N^c=N^s$ and take $\gamma \in \left(0,1\right)$ as a geometry-dependent parameter, which governs how far  the singularity points are  from the boundary.  For instance, for the spherical geometry shown in Figure \ref{fig:msflayout}, we choose $\gamma=R_s/R_c$.

In order to validate our code and to establish the numerical accuracy of the CCR-MFS scheme, we
validate our results for a case of an  evaporating spherical droplet, the analytic solutions to this problems are
readily available in \citep{RanaLockerbySprittles2018} for the case of Grad-13 equations. Additionally, in order to establish the accuracy of our models, we consider a slow flow around a sphere (evaporative and non-evaporative) for various values of Knudsen number and compare our solutions with results from kinetic theory. 

\subsection{Validation and verification of CCR-MSF}
\label{subsec: Validation and verification of CCR-MSF}
\subsubsection{Validation case: evaporation from a spherical droplet}

\begin{figure}
\centering
\includegraphics[scale=0.5]{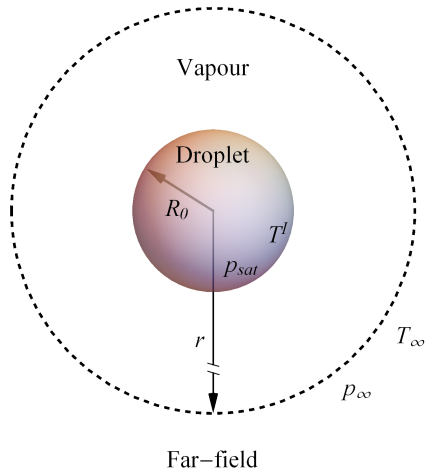}
\caption{\label{fig:evaporationdroplet}
Schematic representation of a liquid droplet surrounded by its own vapour. The equations are made dimensionless with respect to the  far-field conditions ($T_{\infty}$, $p_{\infty}$). The temperature and the pressure deviations vanish as $r\rightarrow\infty$.}
\end{figure}

First, we consider a liquid droplet of fixed radius with a given interface temperature $%
T^{I}$ and the corresponding saturation pressure $p_{sat}$, immersed in
its own vapour; see Figure \ref{fig:evaporationdroplet}. The far-field conditions are given by $T_{\infty }=0$, $%
p_{\infty }=0$; that is, we consider the far-field conditions and the
droplet radius ($R_0$) for non-dimensionalisation. Throughout this article we are not considering dynamics within the droplets, surface tension, etc. see \citet{RanaPRL2019}.

Assuming the spherical symmetry of this problem, the analytic solutions for
the radial velocity $v_{r}$, heat-flux $q_{r}$, normal stress $\Pi _{rr}$,
and the temperature $T$ are obtained from (\ref{conservation laws vector
form}-\ref{CCR relations vector form}) as \citep{RanaLockerbySprittles2018}
\begin{equation}
v_{r}=\frac{c_{1}}{r^{2}}\text{, }q_{r}=\frac{c_{2}}{r^{2}}\text{, }\Pi
_{rr}=\frac{4\mathrm{Kn}\left( c_{1}+\alpha _{0}c_{2}\right) }{r^{3}}\text{, 
}T=\frac{\Pr }{c_{p}\mathrm{Kn}}\frac{c_{2}}{r}\text{, and }p=0\text{.}
\label{excat solution CCR}
\end{equation}%
Here, $r$ is the radial direction; $c_{1}$ and $c_{2}$ are constants of
integration. Solutions (\ref{excat solution CCR}) can also be derived from the fundamental solutions (\ref{velocity solution all}--\ref{stress solution all}), which are given in Appendix \ref{sec:appendixA}.

The flow is driven by (i) a unit pressure difference while the temperature
of the liquid is equal to the far-field temperature (i.e., $p_{sat}=1$ and $%
T^{I}=T_{\infty }=0$) or (ii) a unit temperature difference while the
saturation pressure in the liquid is same as the far-field pressure (i.e., $%
p_{sat}=p_{\infty }=0$ and $T^{I}=1$). In both cases, the constants of
integration, namely, the  mass-flux ($c_{1}$) and heat-flux ($c_{2}$)---per unit
area---are obtained from boundary conditions (\ref{BCs evaporation}) at $r=1$. For Grad-13 (i.e., $\alpha_0=2/5$) these are obtained as    
\begin{subequations}
\label{exact solutions CCR droplet}
\begin{eqnarray}
c_{1}^{p} &=&0.220562+\frac{0.0613424+0.0960246\mathrm{Kn}}{0.138548+\mathrm{%
Kn}(0.716085+\mathrm{Kn})}\text{,}  \label{c1 p exact} \\
c_{2}^{p} &=&c_{1}^{\tau}=-\frac{\mathrm{Kn}(0.108868+0.551404\mathrm{Kn})}{%
0.138548+\mathrm{Kn}(0.716085+\mathrm{Kn})}\text{,}  \label{c2 p exact} \\
c_{2}^{\tau} &=&\frac{0.152568\mathrm{Kn}+\mathrm{Kn}^{2}(0.924355+1.37851%
\mathrm{Kn}))}{0.0406847+\mathrm{Kn}(0.348827+\mathrm{Kn}(1.00974+\mathrm{Kn}%
))}\text{.}\label{c2 T exact}
\end{eqnarray}%
\end{subequations}
Here, superscript ``$p$\textquotedblright\ and ``$\tau$\textquotedblright\
denote the pressure-driven case ($p_{sat}=1$ and $T^{l}=T_{\infty }=0$) and
temperature-driven case ($p_{sat}=p_{\infty }=0$ and $T^{I}=1$),
respectively. The results of these two problems can be combined to evaluate
total evaporative mass and heat-flux from the droplet \citep{RanaPRL2019}. Due to the
microscopic reversibility of the evaporation and condensation processes the
Onsager reciprocity relations hold, which give $\left(
c_{2}^{p}=c_{1}^{\tau}\right) $ \citep{CHERNYAK1989, RanaLockerbySprittles2018}.

The mass and the heat-flux predicted by the CCR--MFS are obtained by
integrating $\mathbf{v}^{(Gr)}$ (\ref{v1 superimposed}, \ref{v2 superimposed}%
) and $\mathbf{q}^{(Gr)}$ (\ref{heat-flux superimposed}) over the droplet
surface, i.e.,
\begin{eqnarray}
c_{1}^{(MFS)} &=&\frac{1}{4\pi }\int_{0}^{\pi }\int_{0}^{2\pi }\mathbf{v}%
^{(Gr)}\cdot \mathbf{n}d\varphi d\theta =\frac{1}{4\pi }%
\sum_{i=1}^{N^{s}}h^{(i)}\text{,} \\
c_{2}^{(MFS)} &=&\frac{1}{4\pi }\int_{0}^{\pi }\int_{0}^{2\pi }\mathbf{q}%
^{(Gr)}\cdot \mathbf{n}d\varphi d\theta =\frac{\mathrm{Kn}}{4\pi }%
\sum_{i=1}^{N^{s}}g^{(i)}\text{.}
\end{eqnarray}

\begin{figure}
\includegraphics[scale=0.55]{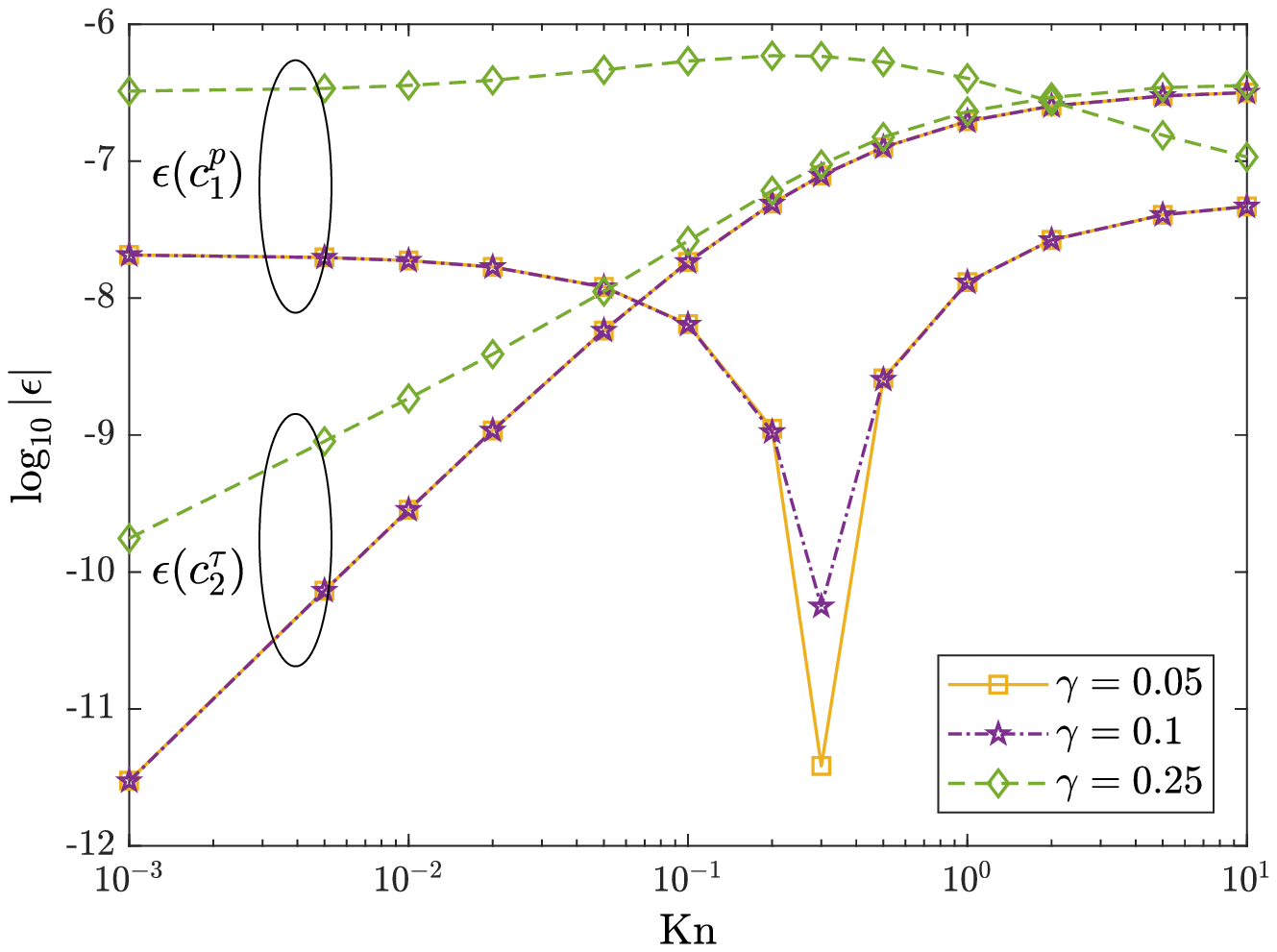}\hfill
\includegraphics[scale=0.55]{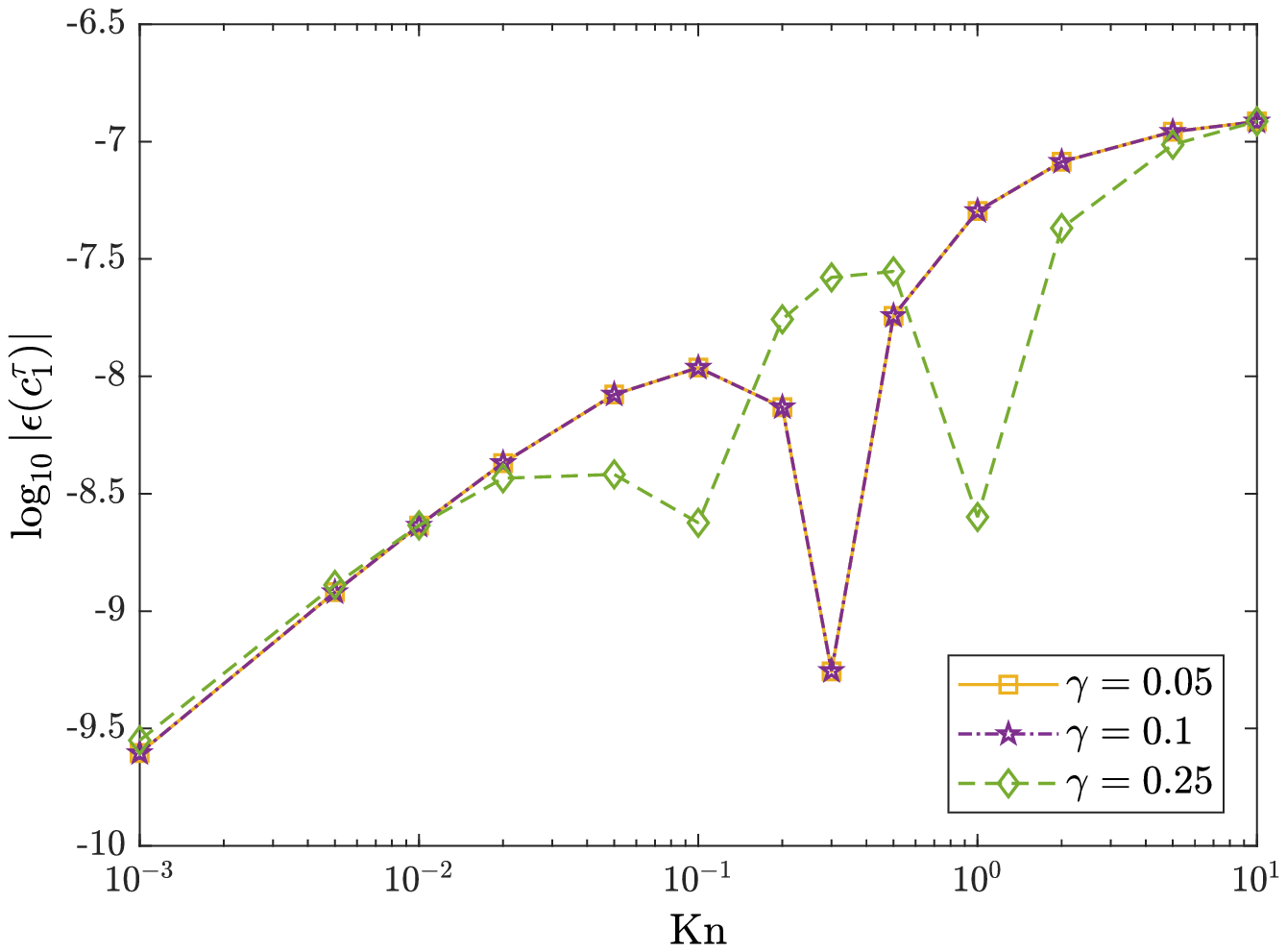}
\caption{\label{fig:errorplots}Log-log plot of errors against Knudsen number with three different singularity locations $\gamma=R_s/R_c=$0.05, 0.1, 0.25 and with $N^c=N^s=112$: (left) error in $c_1^{p}$ and $c_2^{\tau}$  and (right) error in $c_1^{\tau}$.}
\end{figure}

Let us first briefly consider the convergence characteristics of the CCR–MFS. The configuration for the collocation and singularity points is shown in Figure \ref{fig:msflayout}. We define errors $\epsilon$ between the numerical
prediction of $c_1^{p}$, $c_1^{T}$, and $c_2^{T}$ to the analytical values (\ref{exact solutions CCR droplet}) as
\begin{eqnarray}
\epsilon\left(c_1^{p}\right) = c_1^{p\left(MFS\right)}-c_1^{p}\text{, }\epsilon\left(c_2^{\tau}\right) = c_2^{\tau\left(MFS\right)}-c_2^{\tau} \text{, and } \epsilon\left(c_1^{\tau}\right) = c_1^{\tau\left(MFS\right)}-c_1^{\tau}\text{.}
\end{eqnarray}
Figure \ref{fig:errorplots} shows the numerical errors of MFS (with $\alpha_0=2/5$) in  $c_1^{p}$, $c_2^{\tau}$ (left) and $c_1^{\tau}$ (right) for three different singularity site locations ($\gamma=R_s/R_c$=0.05, 0.1, 0.25), and various values of the Knudsen number. The
result indicates very high accuracy with a moderate number of points ($N^c=112$). The best results are obtained when the singularity sites are further away from the collocation nodes, i.e.,  $\gamma$ is small. However, it leads to poorer conditioning of the equation system (\ref{Big matrix}), which in turn results in larger least-square errors or computational time. 

As \citet{LC2016}, numerical tests were also conducted by increasing the number of collocation points, as well as,  by shifting the singularities sites (i.e., breaking the symmetry of the collocation and singularity points); the results were found to be satisfactory. Henceforth, the numerical simulations are performed with $\gamma=0.1$, unless otherwise stated. 

\begin{figure}
\includegraphics[scale=0.55]{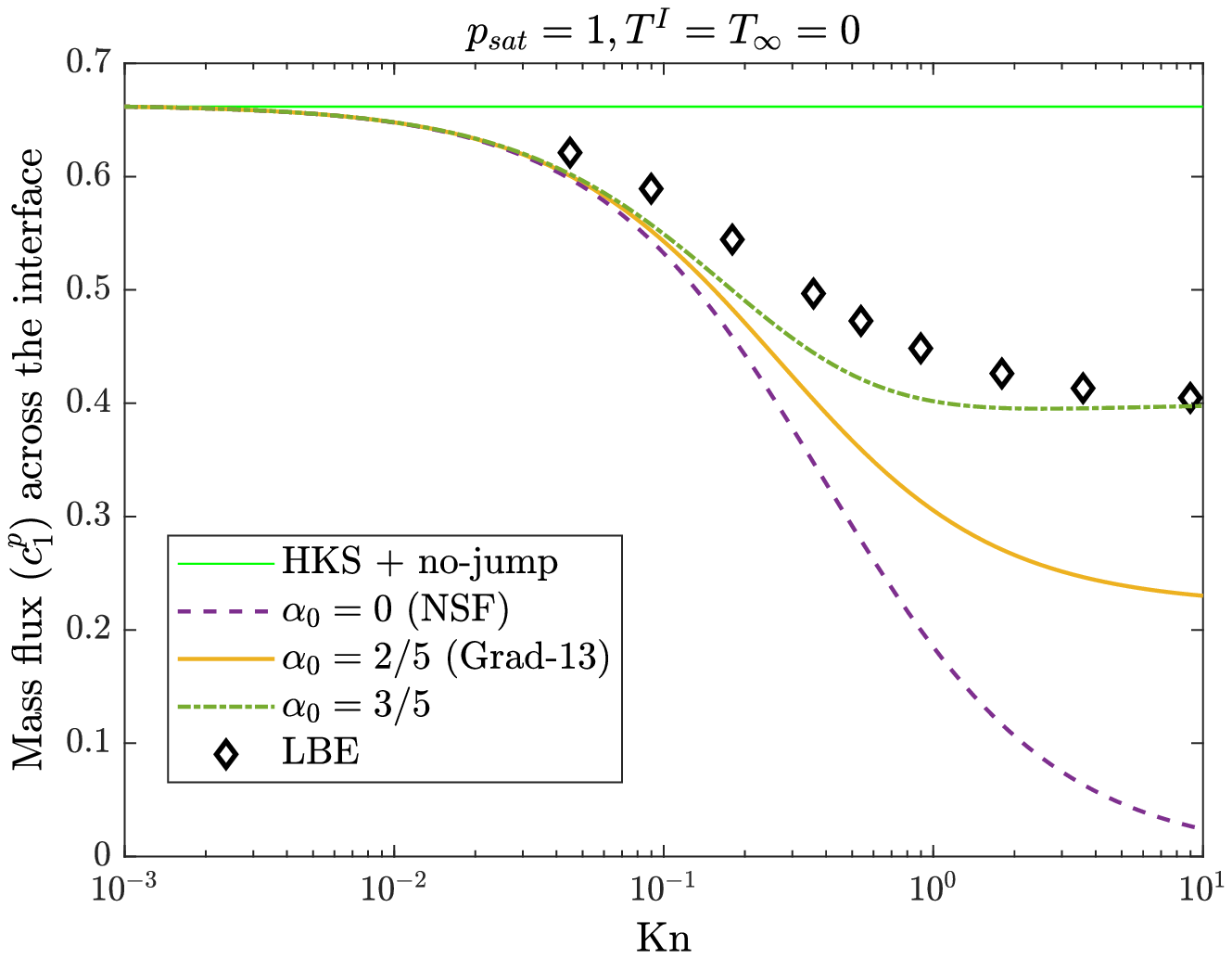}\hfill
\includegraphics[scale=0.55]{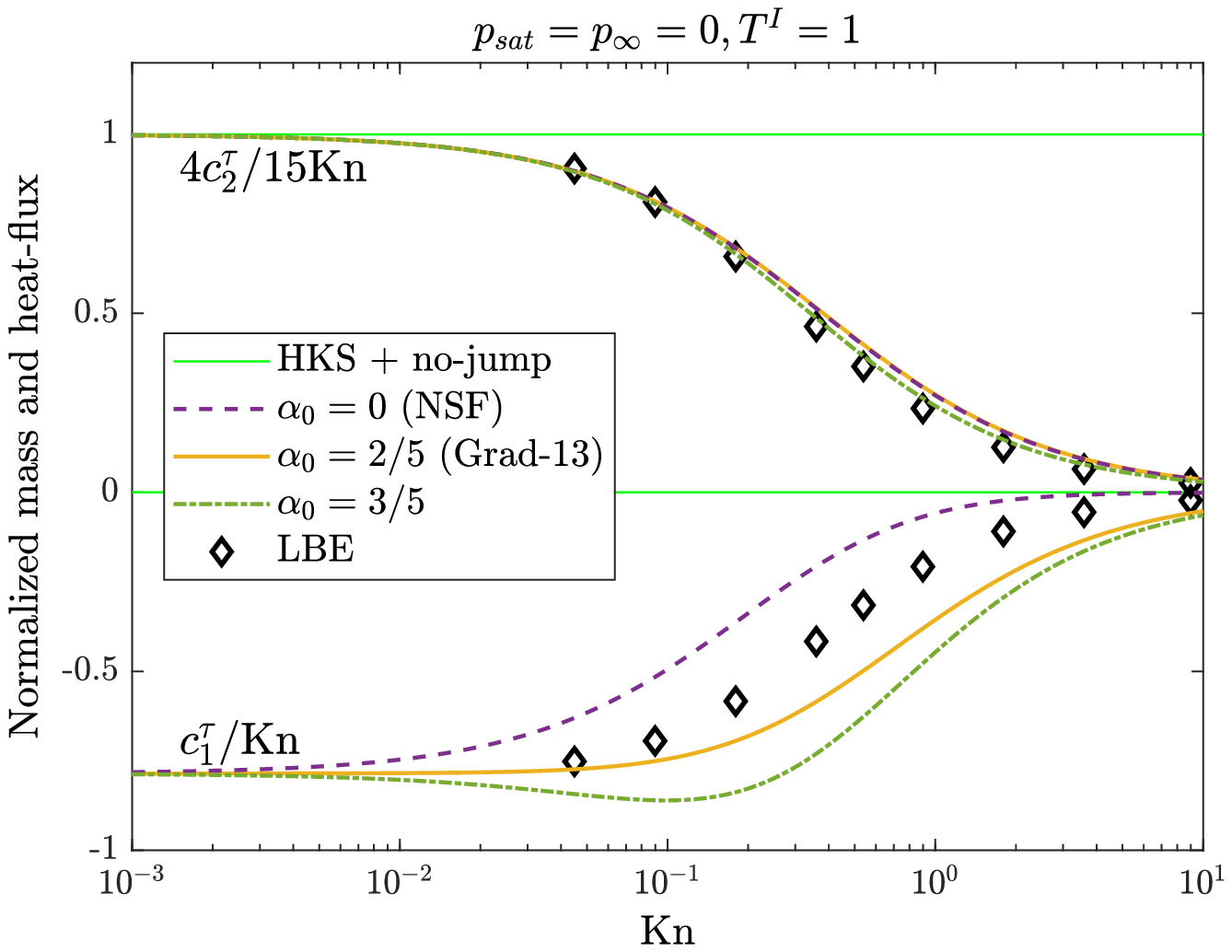}
\caption{\label{fig:myplotc1andc2all}
Mass and heat-flux coefficients as a function of the Knudsen number: (a) mass-flux ($c_1^{p}$) in pressure-driven case and (b) normalized mass ($c_1^{\tau}/\mathrm{Kn}$) and heat-flux ($4c_2^{\tau}/15\mathrm{Kn}$) coefficients  in the temperature-driven case. The results obtained from MFS with different theories are compared. The symbols denoting the results from \cite{SoneEvaporation1998} and (green)  thin line representing the analytic results from classical boundary conditions.  The number of collocation nodes used in each
case are $N^c=N^s$=112 and $\gamma$=0.1.}
\end{figure}

Figure \ref{fig:myplotc1andc2all} shows the variations in the mass and heat-flux coefficients ($c_1$ and $c_2$) with a range of Knudsen numbers; the numerical parameters are included in the caption. In order to put the CCR models into perspective, we compare our predictions from different theories, namely, NSF ($\alpha_0=0$),  Grad-13 ($\alpha_0=2/5$) and ($\alpha_0=3/5$) against the results (symbols) from the linearized Boltzmann equations (LBE) by \cite{SoneEvaporation1998}. The solutions of the NSF equations with classical boundary conditions,
i.e., assuming the (linearized) Hertz-Knudsen-Schrage (HKS) law for evaporation \citep{Schrage1953} and the continuity of temperature
across the interface, are also included  (green-thin lines) in Fig.~\ref{fig:myplotc1andc2all} for the comparison.  These boundary conditions can be obtained from (\ref{BCs evaporation}a) by taking $\eta_{12}=0$ and $T=T^{I}$ instead of (\ref{BCs evaporation}a), which are only valid for $\mathrm{Kn}\rightarrow 0$ \citep{Sone2002, RanaLockerbySprittles2018}.

For the pressure-driven case (cf., Fig.~\ref{fig:myplotc1andc2all}a), mass flux
goes from liquid to vapour (i.e., $c_1^{p}\geq 0$) and  heat flows from vapour to liquid (i.e., $c_2^{p} = c_1^{\tau}\leq 0$) because the enthalpy of phase change must be supplied to the droplet to keep its
temperature constant. All models agree in the limit of $\mathrm{Kn}\rightarrow 0$ for the mass flux; the NSF with temperature jump boundary conditions give better results over classical theories as Knudsen number grows. Interestingly,  the NSF with jump boundary conditions predicts a zero mass flux in the free-molecular regime, results contradicting the kinetic theory predictions. On the other hand, the Grad-13 theory gives a non-zero mass flux, but still predicts almost half of the mass flux of  the LBE result. By performing an asymptotic  expansion, and matching the mass flux as $\mathrm{Kn}\rightarrow \infty$, we find $\alpha_0=3/5$, a value extensively used in this study, with this model consequently providing the best agreement with the LBE.

For the temperature-driven case (cf., Fig.~\ref{fig:myplotc1andc2all}b), the Fourier's law with no-jump boundary condition gives heat-flux $c_2^{\tau} = 15\mathrm{Kn}/4$; while the cross effects, i.e., the heat-flux due to pressure difference or the mass-flux due to temperature difference vanish. We have used this value to normalize the heat-flux computation in Figure \ref{fig:myplotc1andc2all}b. All models, except the no-jump boundary conditions (green-thin lines), match reasonably well with the LBE data, with $\alpha_0=3/5$ giving a slight improvement over other models. However, the cross effects  ($c_1^{\tau}$)  are not well captured by CCR theory---which requires resolution of the Knudsen layer---these are beyond the capabilities of the models considered in this paper; see \citet{RanaLockerbySprittles2018} for a discussion on .

\subsubsection{Slow flow around a rigid sphere}
Let us now consider the case of low-speed rarefied gas flow around a rigid ($%
\vartheta =0$) spherical stationary  particle of radius $R_{0}$. The particle is
assumed to be isothermal---that is the solid-to-gas conductivity ratio is
large---with $T^{I}=0$, i.e., the far-field temperature and the temperature
of the particle are same. The net force exerted on the sphere by gas is
defined as 
\begin{equation}
\mathbf{F}^{(Gr)}=\int_{0}^{\pi }\int_{0}^{2\pi }\left( p^{(Gr)}\mathbf{I}+%
\mathbf{\Pi }^{(Gr)}\right) \cdot \mathbf{n}d\varphi d\theta =\mathrm{Kn}%
\sum_{i=1}^{N^{s}}\mathbf{f}^{(i)}\text{.}
\label{dragforceMFS}
\end{equation}
The drag force is the projection of the net force (\ref{dragforceMFS}) onto the stream-wise direction.
The Stokes formula for drag force (in the direction
of flow) exerted by a spherical particle on the fluid flow has the form \cite{Lamb1945} :%
\begin{equation}
\label{Stokes formula}
F^{S}=-6\pi \mathrm{Kn}u_{\infty }\text{.}
\end{equation}
where $u_\infty$ is the far-field velocity.

\begin{figure}
\centering
\includegraphics[scale=0.55]{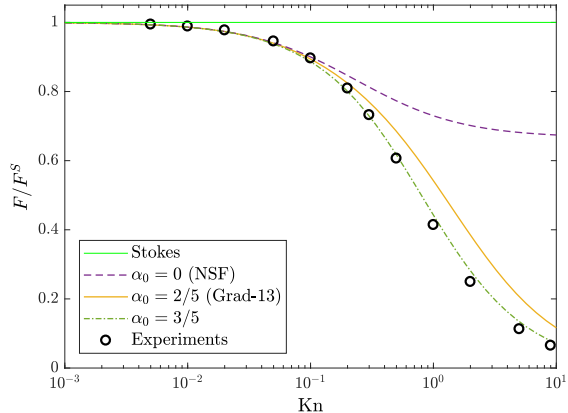}
\caption{\label{fig:dragforcesphere}
The normalized drag force computed from the CCR-MFS with $\alpha_0 = 0$ (NSF), $\alpha_0 = 2/5$ (Grad-13), and $\alpha_0 = 3/5$  against Knudsen number. The experimental data of Millikan (1923) (fitted by Allen \& Raabe (1982)) is plotted (circles) for comparison; the results are normalized with  Stokes drag (\ref{Stokes formula}).}
\end{figure}

The Stokes formula is only valid for $\mathrm{Kn}\rightarrow 0$, and requires corrections at finite Knudsen number. Figure \ref{fig:dragforcesphere} shows the normalized (with Stokes drag (\ref{Stokes formula})) drag coefficient versus the Knudsen number. As expected, all theories agree in the small Knudsen limit. As the value of Knudsen number increases, the normalized drag decreases due to increasing slip (which is not embedded in Stokes formula).  Notably, the normalized drag force reaches a finite value from NSF with velocity-slip boundary conditions as $\mathrm{Kn}\rightarrow \infty$, while the extended theories (CCR) predict a vanishing drag in this limit--an observation tantamount to experimental results. Nevertheless, between Grad-13 ($\alpha_0=2/5$) and CCR with phenomenological coefficient  $\alpha_0=3/5$, the latter gives the best quantitative agreement for values of
the Knudsen number beyond unity (i.e. well beyond its apparent limits of applicability).

\subsubsection{Slow flow around a spherical liquid droplet}

In this section, we consider uniform flow (say, along the z-direction) of a saturated vapour over a spherical liquid droplet with a uniform temperature. The far-field temperature of the surrounding vapour is same as that of the droplet, i.e., $T^{I}=T_{\infty}=0$, $p_{sat}=p_{\infty}=0$, and $\mathbf{v}_{\infty}=-\mathbf{v}^{I} = \left\{0,0,u_{\infty}\right\}$.  We further assume that the droplet size remains fixed (quasi-steady state assumption),  and deformation of the droplet is negligible, that is, the capillary number is very small.
%
%
\begin{figure}
\centering
\includegraphics[scale=0.55]{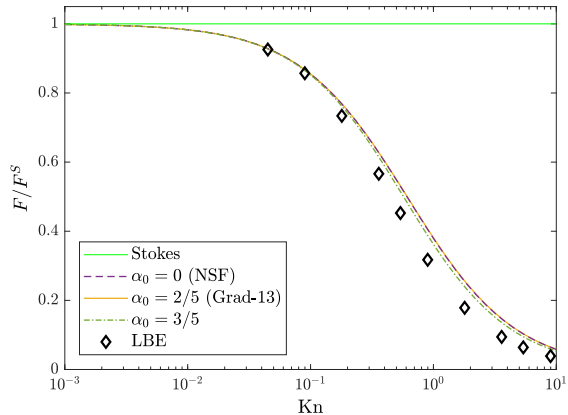}
\caption{\label{fig:dragforceevaporationb}
The normalized drag force on a liquid droplet vs Knudsen number. The results computed from the CCR-MSF with $\alpha_0 = 0$ (NSF), $\alpha_0 = 2/5$ (Grad-13), and $\alpha_0 = 3/5$ are compared with LBE data (symbols) of \cite{SoneEvaporationdrag1994}.}
\end{figure}

Figure \ref{fig:dragforceevaporationb} shows the normalized drag force on a liquid droplet vs  Knudsen number, while allowing  phase-change at the interface. This problem has been considered by \cite{SoneEvaporationdrag1994} from LBE, and is clearly important from an  engineering point of view. Due to the motion of the vapour phase, evaporation and condensation take place on the surface of the droplet thereby slightly reducing the drag on the sphere; cf. Figure \ref{fig:dragforcesphere} and \ref{fig:dragforceevaporationb}. Curiously, all theories, except Stokes, give reasonably accurate results.  Note that, in this case, the net mass and heat-flux from the droplet are zero with evaporation on the front side of the sphere and condensation on the back. 

The problems considered in previous sections typically allow for analytic solutions, which are often not available for various practical problem of interest. In the following sections, we will develop cases of increasing complexity,
starting first with the motion over two particles at a finite distance
from each other.

\subsection{Motion over two solid spheres}
\label{sec: Motion of two spherical non-evaporating droplets}
\begin{figure}
\centering
\includegraphics[scale=0.5]{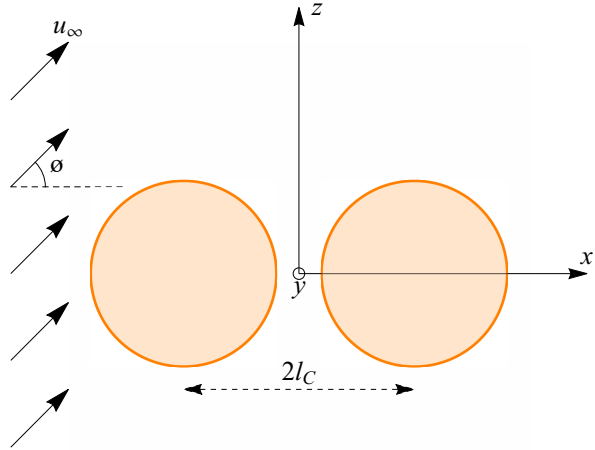}
\caption{\label{fig:dubletlayout}
Schematic for the flow around two solid spheres. $2l_C$ is the center-to-center distance and $\o$ is the angle between flow direction and the line joining the centers. }
\end{figure}

Let us consider a flow over two solid spheres with equal radius with  both fixed in shape  and size . The flow is characterized by the  following parameters: radii of the spheres (i.e., the Knudsen number), center-to-center distance $2l_C$, and the angle $\o$  between the flow direction and the center-line; see Figure \ref{fig:dubletlayout}.

For an axisymmetric case ($\o=0$), the theoretical results for the drag force (which is equal for both the spheres) calculated by \citet{StimsonandJeffery1926} read 
\begin{equation}
\label{StimsonandJefferysolution}
F^{S}=-6\pi \mathrm{Kn}u_{\infty}\beta
\end{equation}
where $\beta$ is a correction coefficient which may be written in the form
\begin{equation}
\label{StimsonandJefferysolutioncorrection}
\beta =\frac{4}{3}\sinh{\alpha} \sum_{k=1}^{\infty }\frac{n\left( n+1\right) 
}{\left( 2n-1\right) \left( 2n+3\right) }\left( 1-\frac{4\sinh ^{2}{\left( n+%
\frac{1}{2}\right) \alpha} -\left( 2n+1\right) ^{2}\sinh ^{2}\alpha }{2\sinh
\left( 2n+1\right) \alpha +\left( 2n+1\right) \sinh 2\alpha }\right).
\end{equation}
with  $\alpha = \cosh{l_C}$. 

\begin{figure}
\centering
\includegraphics[scale=0.5]{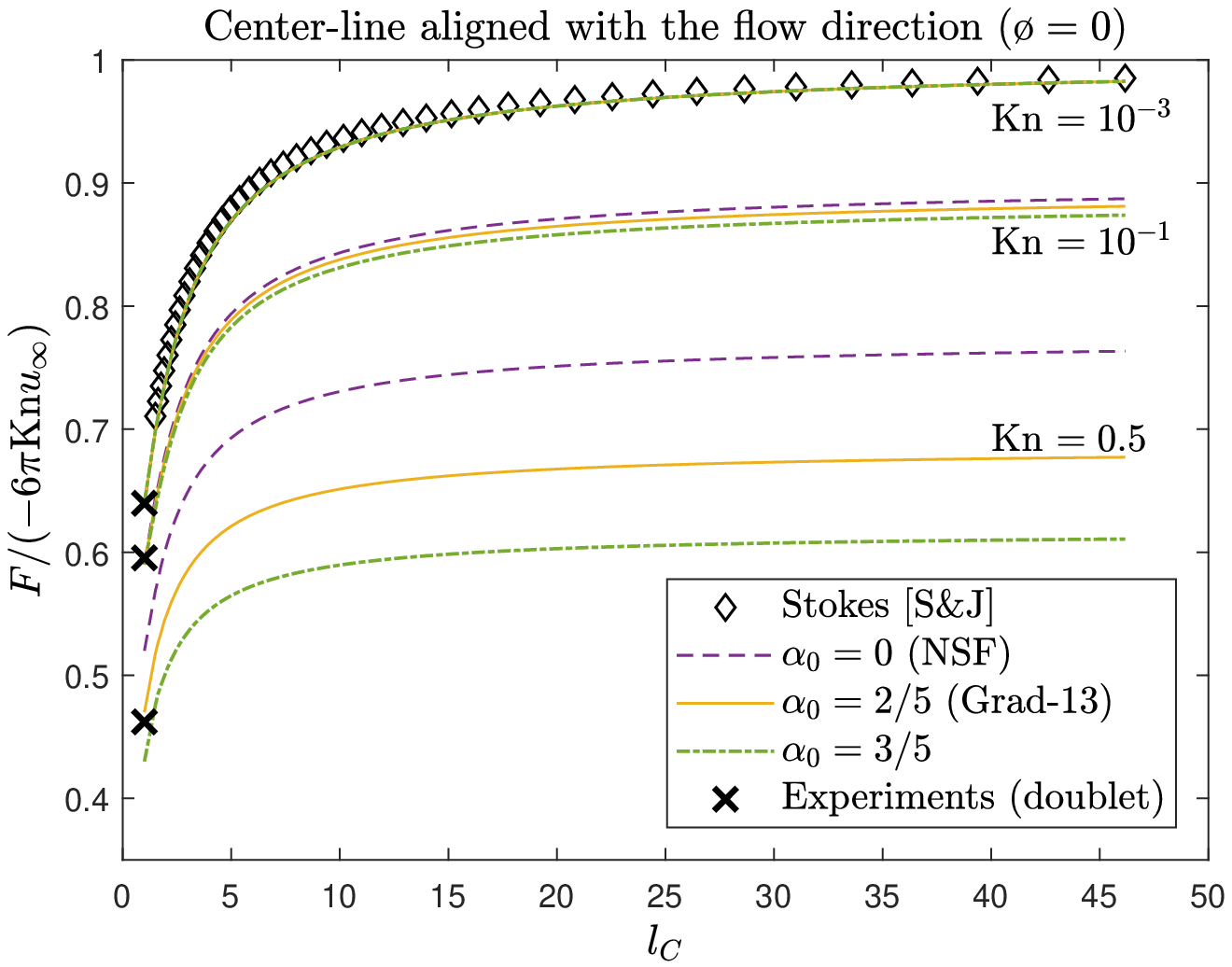}\hfill
\includegraphics[scale=0.5]{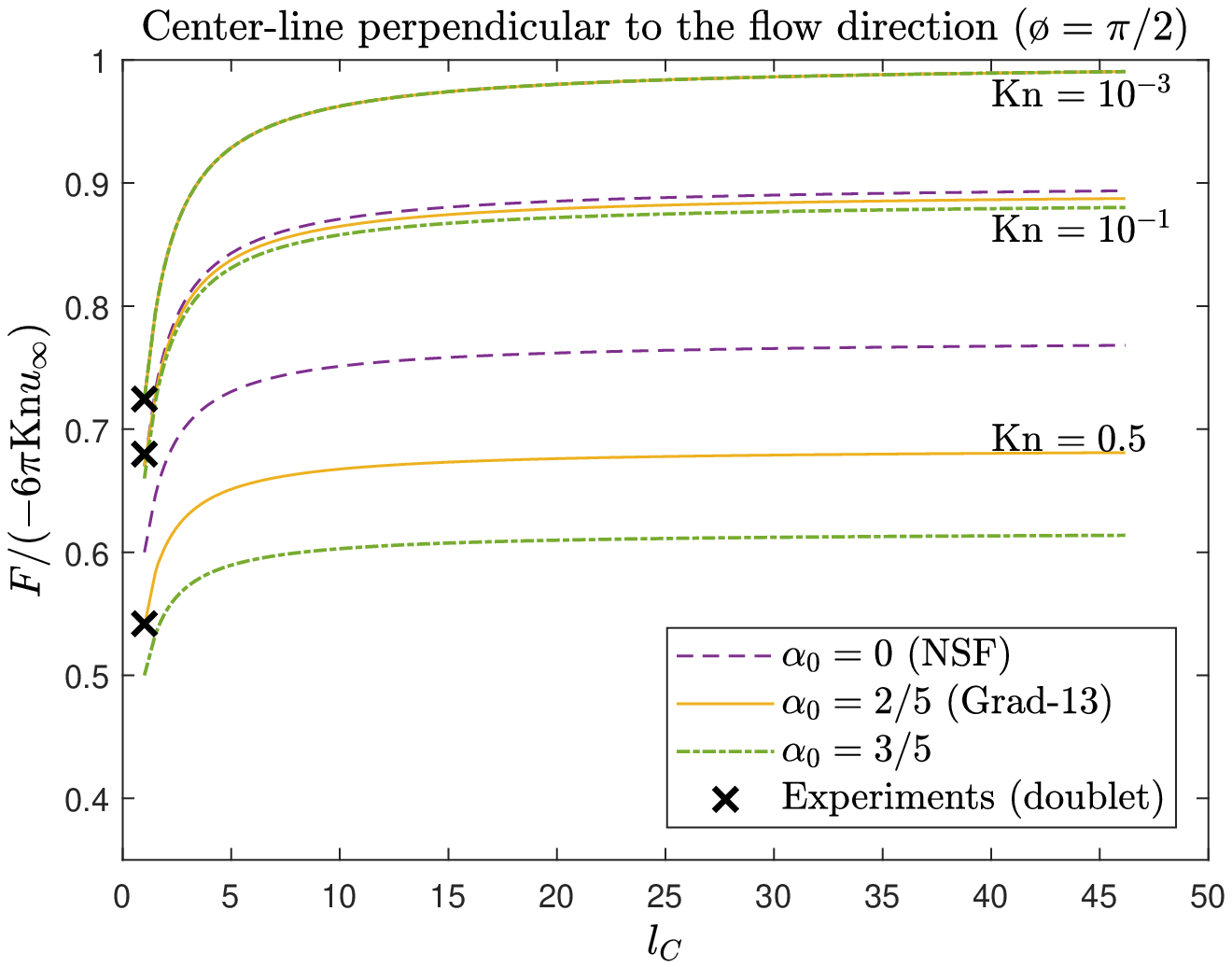}
\caption{\label{fig:twosphereSandJ0plots}
The drag force on two equal spheres vs center-to-center distance: (a) flow direction parallel to their line of centers and (b) flow direction perpendicular to their line of centers. The results computed from the CCR-MSF with $\alpha_0 = 0$ (NSF), $\alpha_0 = 2/5$ (Grad-13), and $\alpha_0 = 3/5$ are compared with Stokes' solution ($\diamond$) derived by \cite{StimsonandJeffery1926}. The experimental results (denoted by $\mathbf{\times}$) for a doublet ($\l_C = 1/2$) obtained by \citet{Chengdoublet1988} are included for comparison.}
\end{figure}

In Figure \ref{fig:twosphereSandJ0plots}a, the drag force (normalized with the Stokes' drag (\ref{Stokes formula}) for the single sphere) for an axisymmetric case ($\o=0$) are plotted as a function of center-to-center distance for different values of the Knudsen number.  The results from \cite{StimsonandJeffery1926}, which are only valid in the limiting case $\mathrm{Kn}\rightarrow 0$, are included for comparison ($\diamond$);as expected for  small value of Knudsen number ($\mathrm{Kn} = 10^{-3}$) our results match with \cite{StimsonandJeffery1926}. The experimental results for $\l_C = 1/2$ obtained by \citet{Chengdoublet1988} (and given by the empirical formula (\ref{empiricalformula})) for $\mathrm{Kn} = 10^{-3}$,  $10^{-1}$ and $0.5$ are also included ($\mathbf{\times}$) for comparison.
For small Knudsen numbers ($\mathrm{Kn} \lesssim10^{-1}$), all models (NSF, Grad-13, CCR) agree, while at large Knudsen numbers these are markedly different, with the NSF model overpredicting the drag compared to Grad-13 and CCR with $\alpha_0=3/5$. The effect of proximity is clearly visible on drag force, which decreases as $l_C$ is reduced (i.e., spheres gets closer). Furthermore, in the limit  $l_C\rightarrow \infty$ the drag approach to the single sphere case (cf., Fig.~\ref{fig:dragforcesphere}). 
Also, as with the case of a single sphere (cf., Fig.~\ref{fig:dragforcesphere} and \ref{fig:twosphereSandJ0plots}a), the force monotonically decreases with increases in Knudsen number, for all models.

For the nonaxisymmetric case ($\o=\pi/2$), the normalized drag force is plotted in Figure \ref{fig:twosphereSandJ0plots}b as a function of $l_C$,  with $\mathrm{Kn} = 10^{-3}$,  $10^{-1}$ and $0.5$. The experimental results of \citet{Chengdoublet1988} for a doublet ($\l_C = 1/2$) are denoted by symbols ($\mathbf{\times}$). The overall drag force in the nonaxisymmetric case is higher compared to the axisymmetric case (cf., Figs.~\ref{fig:twosphereSandJ0plots}a and b) and as before, the NSF theory over predicts the drag.

\subsubsection{Drag on a doublet}
An interesting case arises when one considers $\l_C\rightarrow 1/2$, i.e., when both spheres are in contact with one-another---known as a doublet. \citet{Chengdoublet1988} carried out experiments to study
the effects of orientation (modulated by an electric electric field) on the measured drag force. 

\begin{figure}
\centering
\includegraphics[scale=0.5]{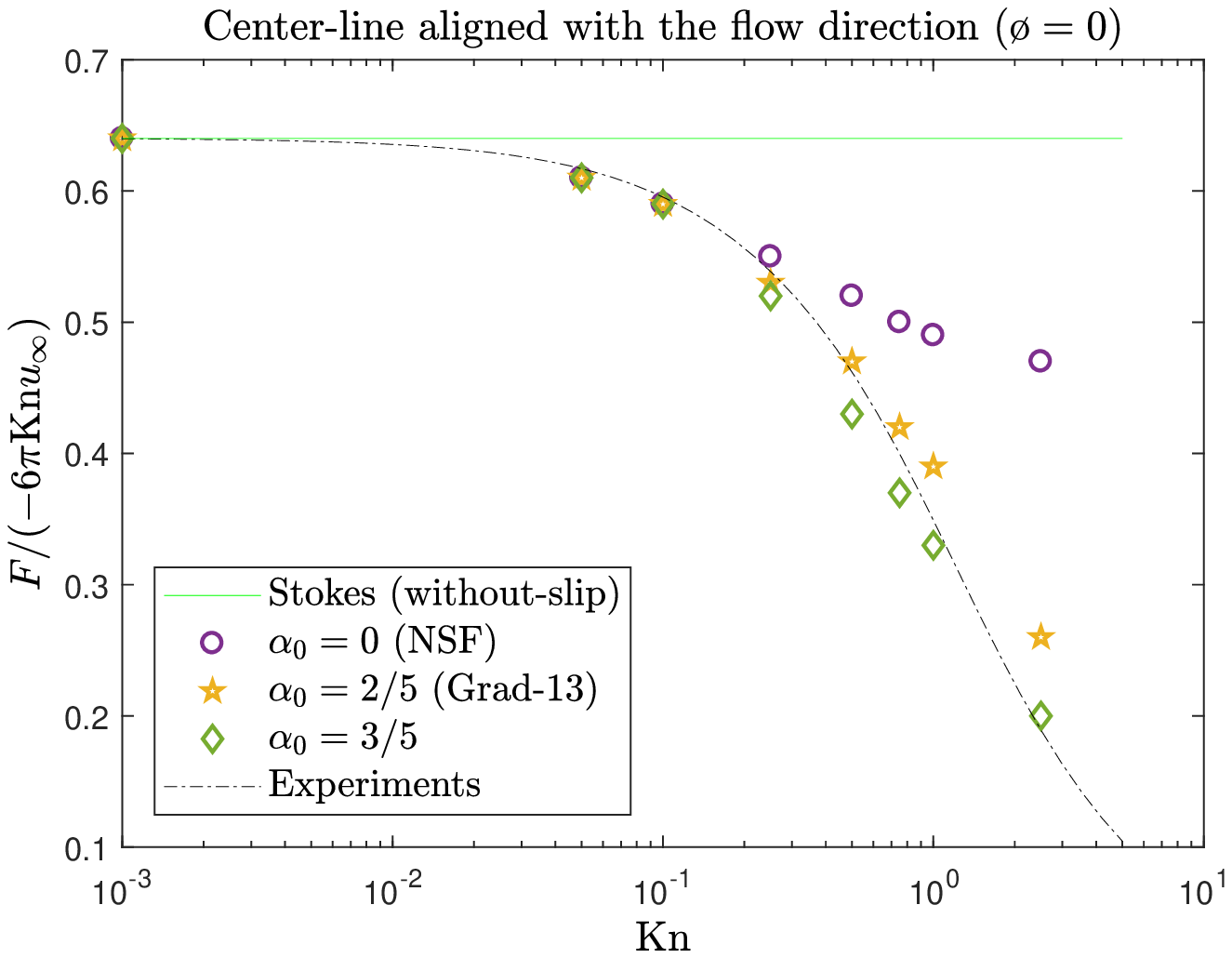}\hfill
\includegraphics[scale=0.5]{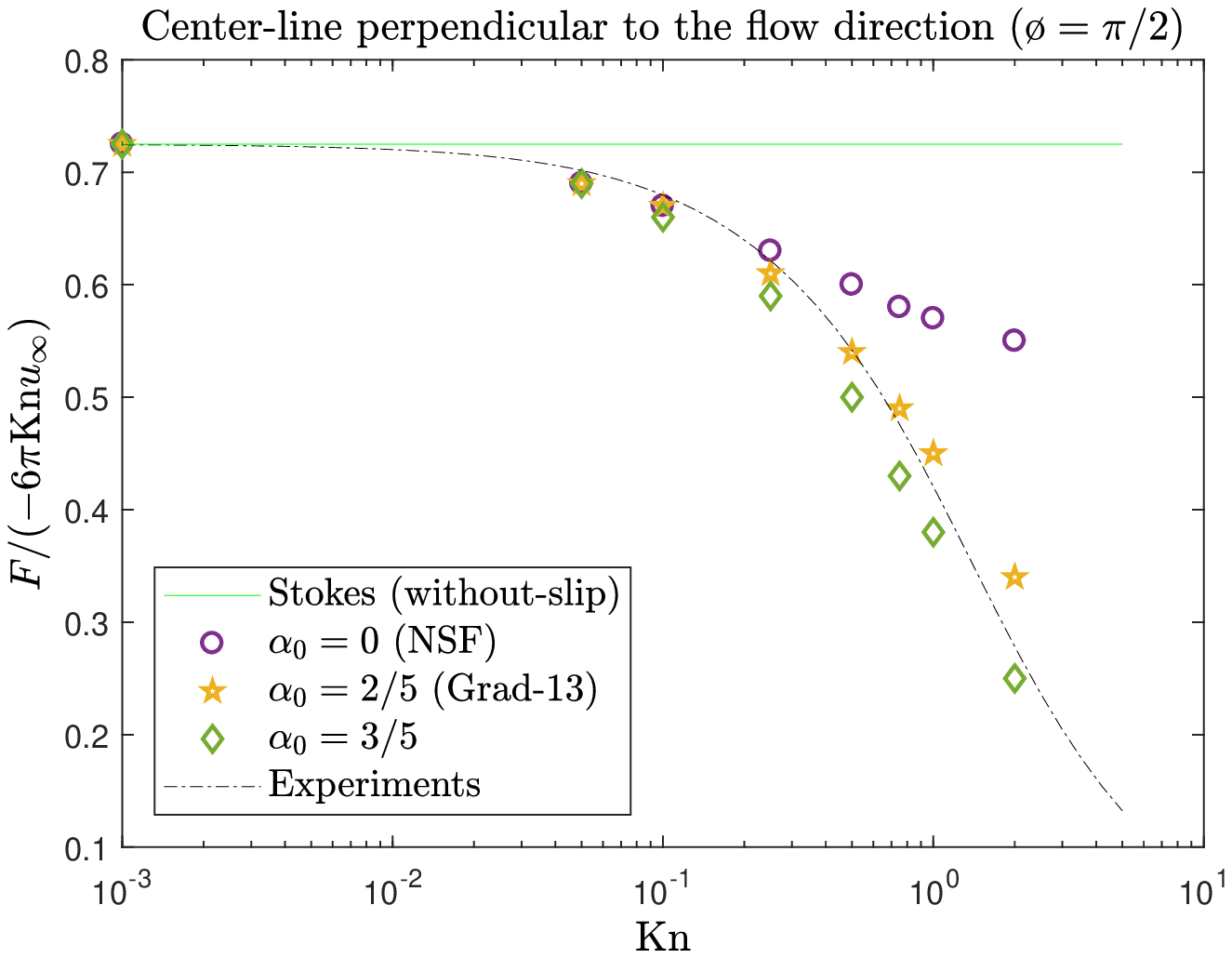}
\caption{\label{fig:doubletspheredrag}
The drag force on a doublet: (a) flow parallel to its line of centers (axisymmetric case) and  (b) flow perpendicular to its line of centers (nonaxisymmetric case). The experimental
data is taken from the empirical formula (\ref{empiricalformula}) obtained in \citet{Chengdoublet1988}.}
\end{figure}

Figure \ref{fig:doubletspheredrag} shows a comparison of the different models, computed using CCR-MFS, over a wide range of the Knudsen number.  Here,  the experimental
data for the drag force acting on a doublet is taken from the empirical formula  from \cite{Chengdoublet1988}, as
\begin{equation}
\label{empiricalformula}
F = -\frac{4\pi\mathrm{Kn}u_{\infty}}{2^{1/3}}\frac{\phi_0}{1+\frac{\mathrm{Kn}}{\phi_0}\left[ 1.142+0.558\exp\left(-0.999\frac{\phi_0}{\mathrm{Kn}}\right) \right]}\text{,}
\end{equation}
where $\phi_0 = 1.21$ or $\phi_0 = 1.37$ for flow moving in parallel ($\o = 0$) or in perpendicular ($\o = \pi/2$) direction to the center line, respectively. 

As shown in Figures \ref{fig:doubletspheredrag}, reassuringly for $\mathrm{Kn}\rightarrow 0$, the results converges to the without-slip solution of the Stokes equation. For larger values of the Knudsen number, the force decreases due to slip, and as $\mathrm{Kn}\rightarrow \infty$, it  reaches a finite value for the NSF theory; while from (\ref{empiricalformula}) it vanishes. Remarkably, from the results of Grad-13 and CCR models, we find that the force indeed vanishes for large Knudsen
numbers. The results obtained from the Grad-13 and CCR models are both in good agreement with the experimental results; a sightly better match at larger Knudsen number is obtained by CCR with $\alpha_0=3/5$.  

\begin{figure}
\centering
\includegraphics[scale=0.4]{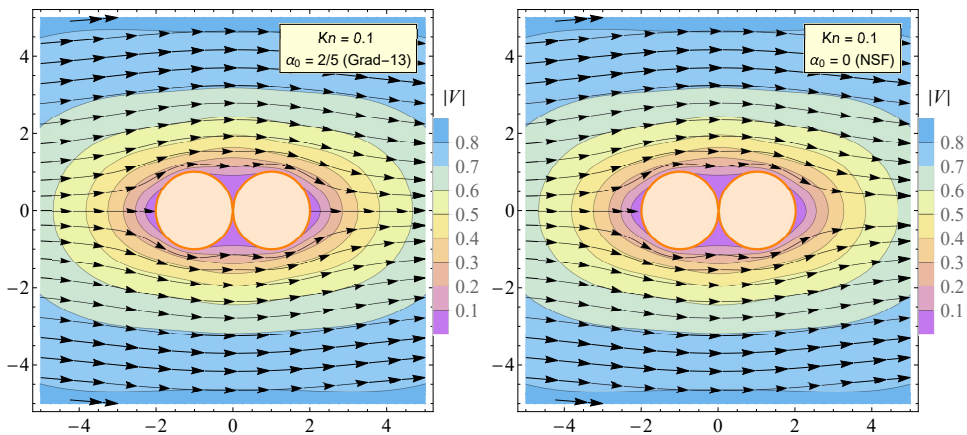}
\caption{\label{fig:twospheresP1Kn0p1}
Stream-lines and speed contours from the (left) Grad-13 and (right) NSF equations for Knudsen number 0.1: flow over a  doublet with flow direction along the center-line ($\o=0$).  Vertical and horizontal axis represents $x$- and $y$-directions. ($N_c = N_s = 650$ and $\gamma = R_s/R_c = 0.1$) }
\end{figure}

\begin{figure}
\centering
\includegraphics[scale=0.4]{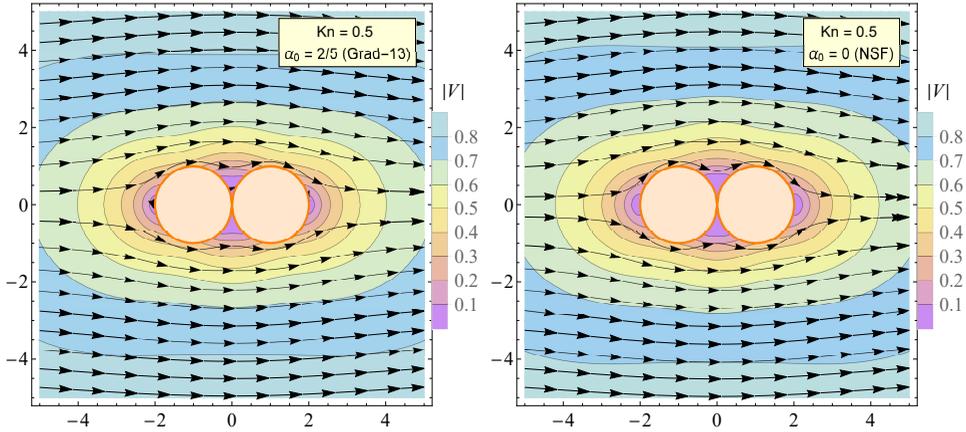}
\caption{\label{fig:twospheresP1Kn0p5}
Stream-lines and speed contours from the (left) Grad-13 and (right) NSF equations for Knudsen number 0.5: flow over a  doublet with flow direction along the center-line ($\o=0$).  Vertical and horizontal axis represents $x$- and $y$-directions. ($N_c = N_s = 650$ and $\gamma = R_s/R_c = 0.1$) }
\end{figure}

Figures \ref{fig:twospheresP1Kn0p1} and \ref{fig:twospheresP1Kn0p5} show the stream-lines and the
speed contour  for
the case $\mathrm{Kn} = 0.1$ and $\mathrm{Kn} = 0.5$, respectively.  The horizontal and vertical axis represent $x$- and $z$-axis, respectively, with flow in $x$-direction. The left hand side of these figures shows the Grad-13 ($\alpha_0=2/5$) results for the case $\o=0$, and the right hand side show the results computed from the NSF ($\alpha_0=0$) equations.  Due to linearity of the equations, the speed contours are symmetric about $x=0$ plane, resulting in equal drag forces on both spheres.  

These contours show that, as to be
expected, the slip-velocity at the bottom and top surfaces increases with Knudsen number,
resulting in reduced drag forces. For $\mathrm{Kn} = 0.1$ (Fig.~\ref{fig:twospheresP1Kn0p1}), the flow line pattern predicted by the Grad-13 (left) and NSF (right) are very similar. The slip-velocity is maximum near the top and bottom surface, which is virtually the same for NSF and Grad-13 theories, thus giving similar predictions for the drag reduction (cf., Fig.~\ref{fig:twosphereSandJ0plots}). However, for $\mathrm{Kn} = 0.5$ (Fig.~\ref{fig:twospheresP1Kn0p5}), the Grad-13 theory predicts a larger slip and thus greater drag reduction.
The results from CCR with $\alpha_0=3/5$ are similar to those obtained from the Grad-13 theory, hence those are not shown here.

\begin{figure}
\centering
\includegraphics[scale=0.4]{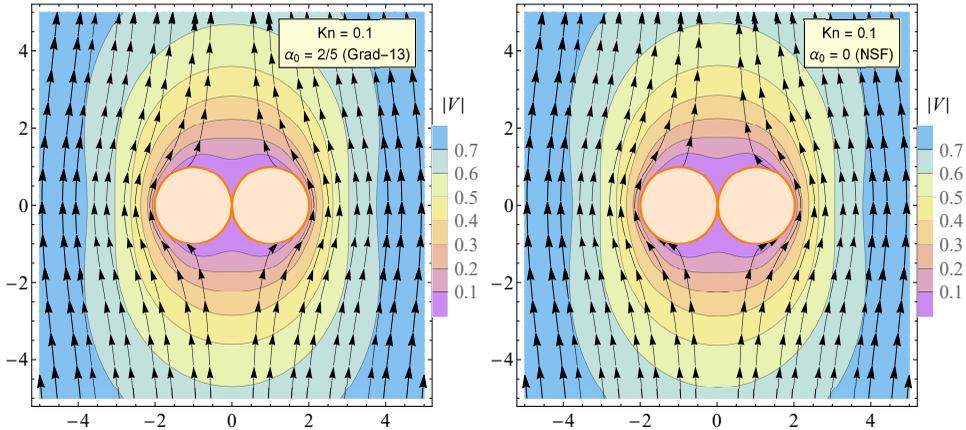}
\caption{\label{fig:twospheresP2Kn0p1}
Stream-lines and speed  contours from the (left) Grad-13 and (right) NSF equations for Knudsen number 0.1: flow over a doublet with flow direction perpendicular to the center-line ($\o=\pi/2$).  Vertical and horizontal axis represents $x$- and $y$-directions. ($N_c = N_s = 650$ and $\gamma = R_s/R_c = 0.1$) }
\end{figure}

\begin{figure}
\centering
\includegraphics[scale=0.4]{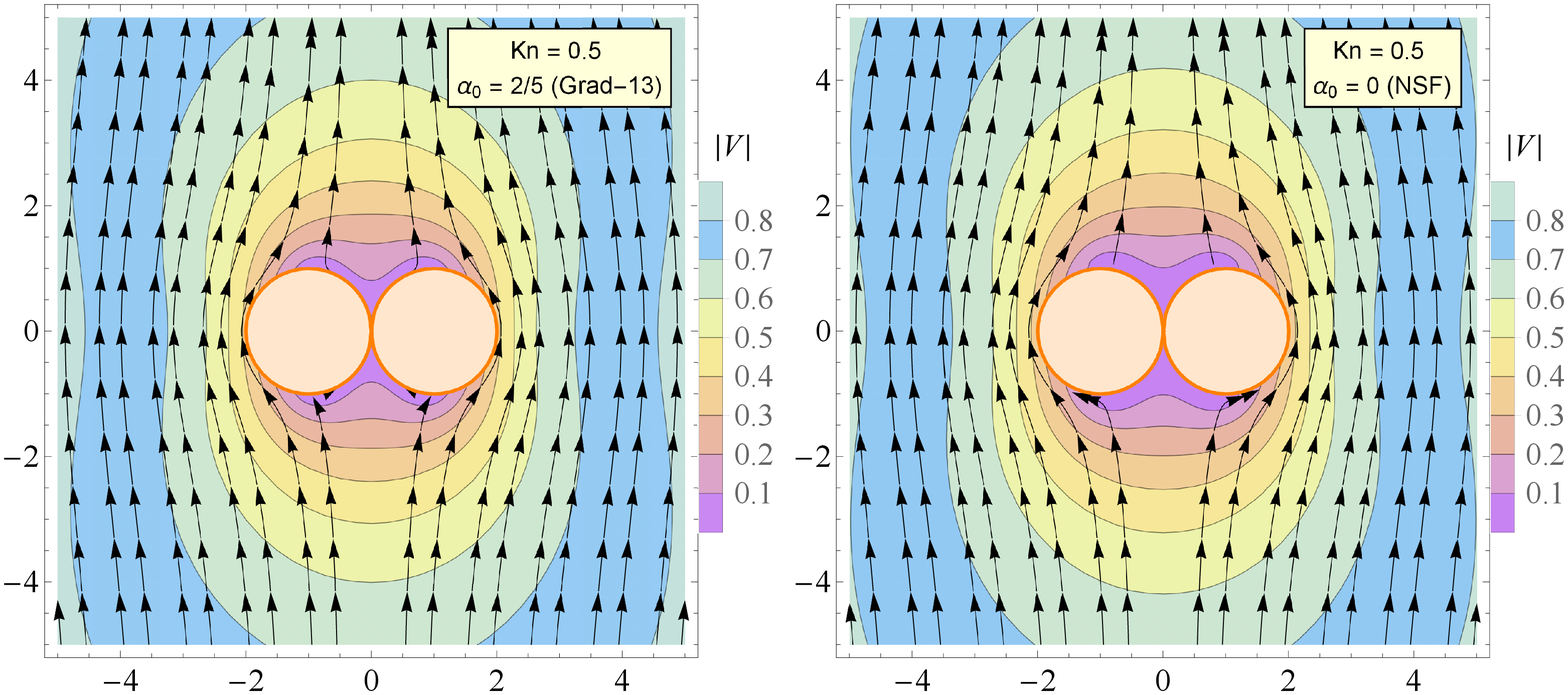}
\caption{\label{fig:twospheresP2Kn0p5}
Stream-lines and Mach contours from the (left) Grad-13 and (right) NSF equations for Knudsen number 0.5: flow over a  non-evaporative ($\vartheta = 0$) doublet with flow direction perpendicular to the center-line ($\o=\pi/2$).  Vertical and horizontal axis represents $x$- and $y$-directions. ($N_c = N_s = 650$ and $\gamma = R_s/R_c = 0.1$) }
\end{figure}

In Figures \ref{fig:twospheresP2Kn0p1} and \ref{fig:twospheresP2Kn0p5},  we again show the stream-lines and speed contours predicted by both theories at $\mathrm{Kn}=0.1$ and $\mathrm{Kn}=0.5$, respectively, for the case when flow (in $z$-direction) is in perpendicular direction to the center-line, i.e., $\o = \pi/2$. Unlike the previous case, this is a truly 3D setup, where no symmetry exist in the azimuthal direction---with the MFS it is remarkably simple to solve this fully 3D flow.  The slip velocity increases with increasing the Knudsen number thus reducing the drag coefficient. 

\subsection{Interaction of two evaporating droplets}
\label{sec: Interaction of two evaporating droplets}
The evaporation dynamics of two interacting droplets suspended in vapour will be investigated here; a problem relevant for dense spray applications. We consider two droplets placed is a saturated vapour with center-to-center distance $2l_C$. Again, we consider the shape, size and the surface temperature of the droplet fixed. 

As before, we consider two cases, i.e.,  (i) evaporation is driven by temperature difference (i.e., $%
p_{sat}=p_{\infty }=0$ and $T^{I}=1$) and (ii) evaporation due to pressure difference ($p_{sat}=1$ and $%
T^{I}=T_{\infty }=0$).  The distance between two droplets and the Knudsen number are varied in order to investigate the proximity and  rarefaction effects. The results from three models are compared NSF, Grad-13 and CCR with $\alpha_0=3/5$.

\begin{figure}
\centering
\includegraphics[height=50mm]{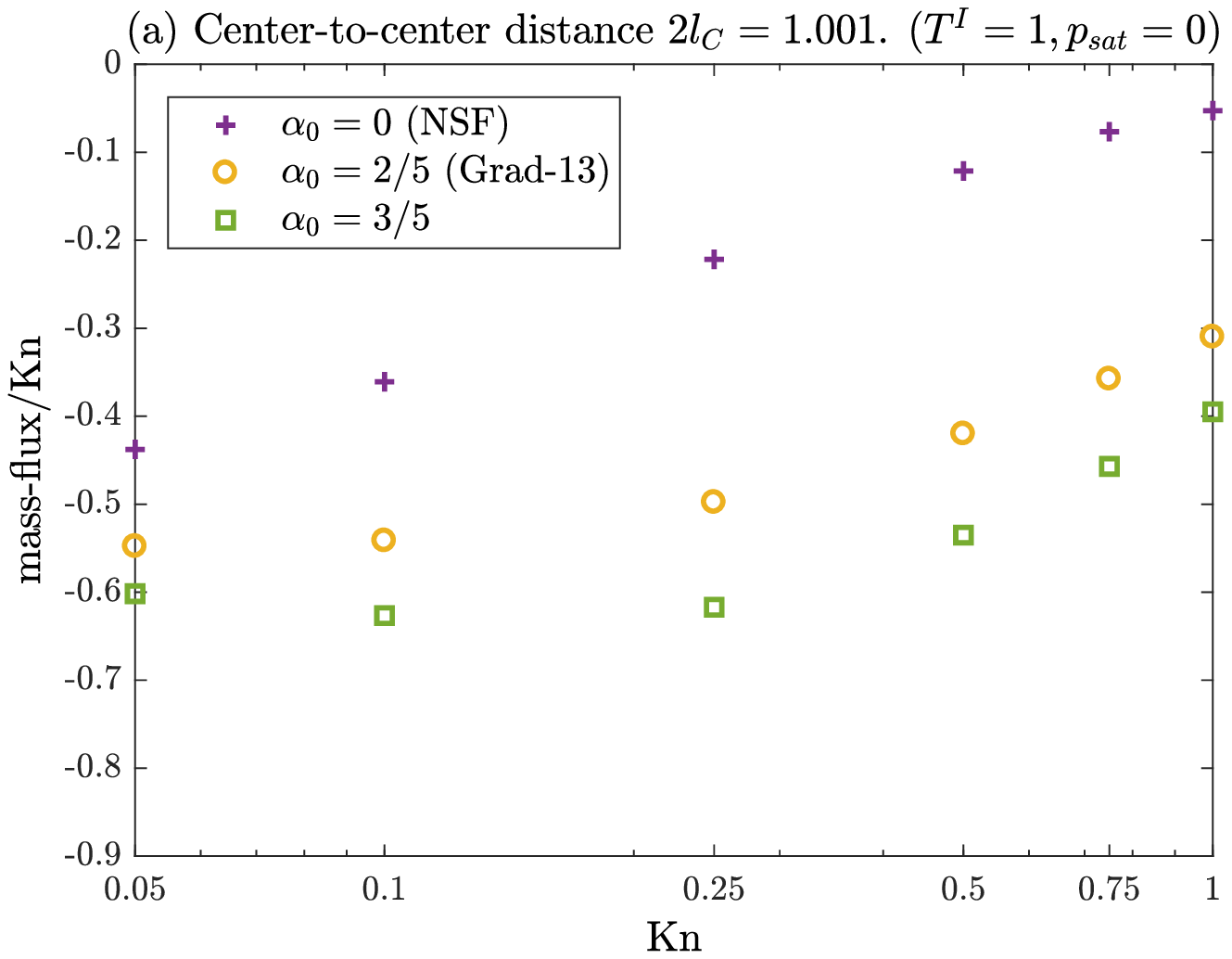}
\hfill
\includegraphics[height=50mm]{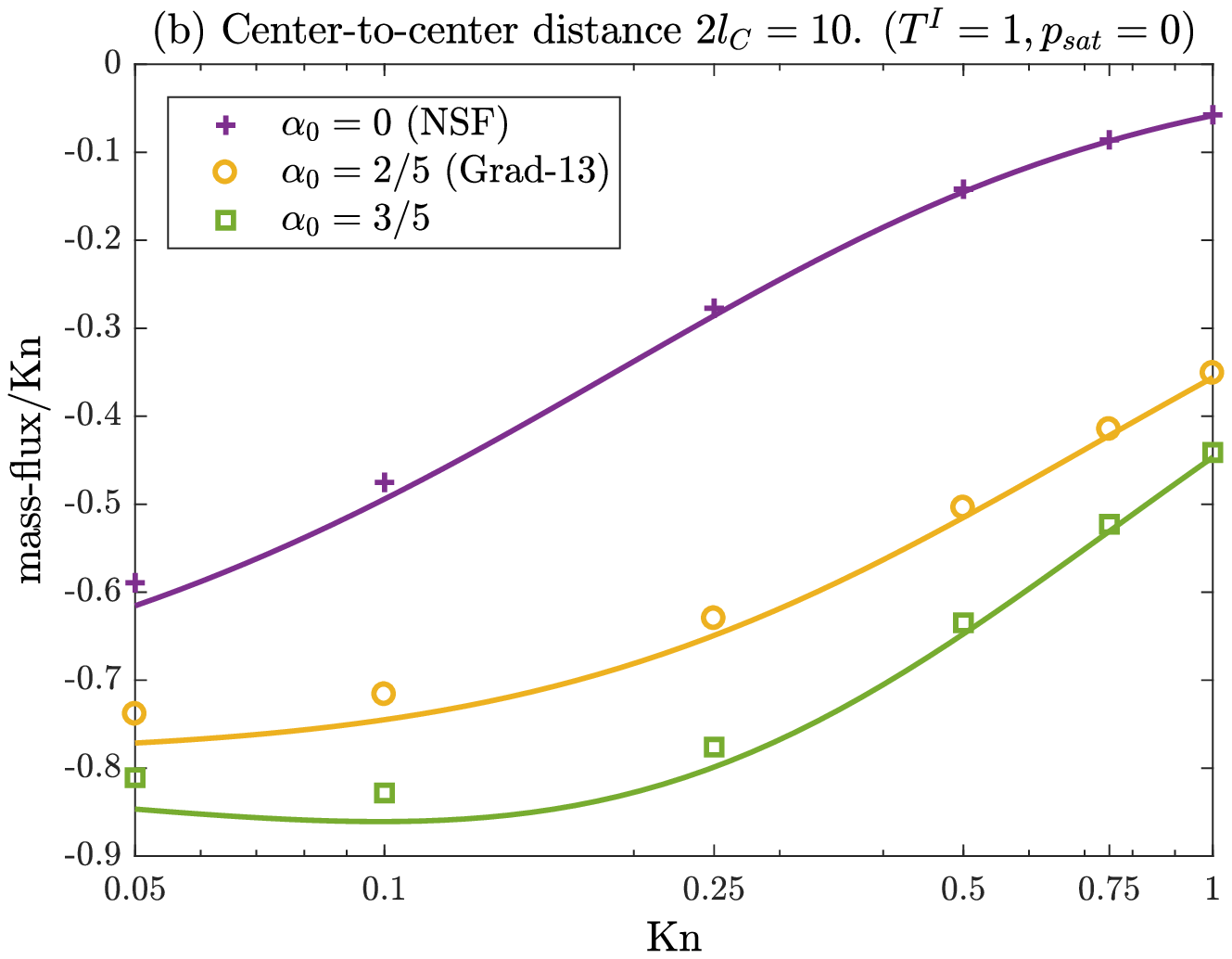}
\\
\includegraphics[height=50mm]{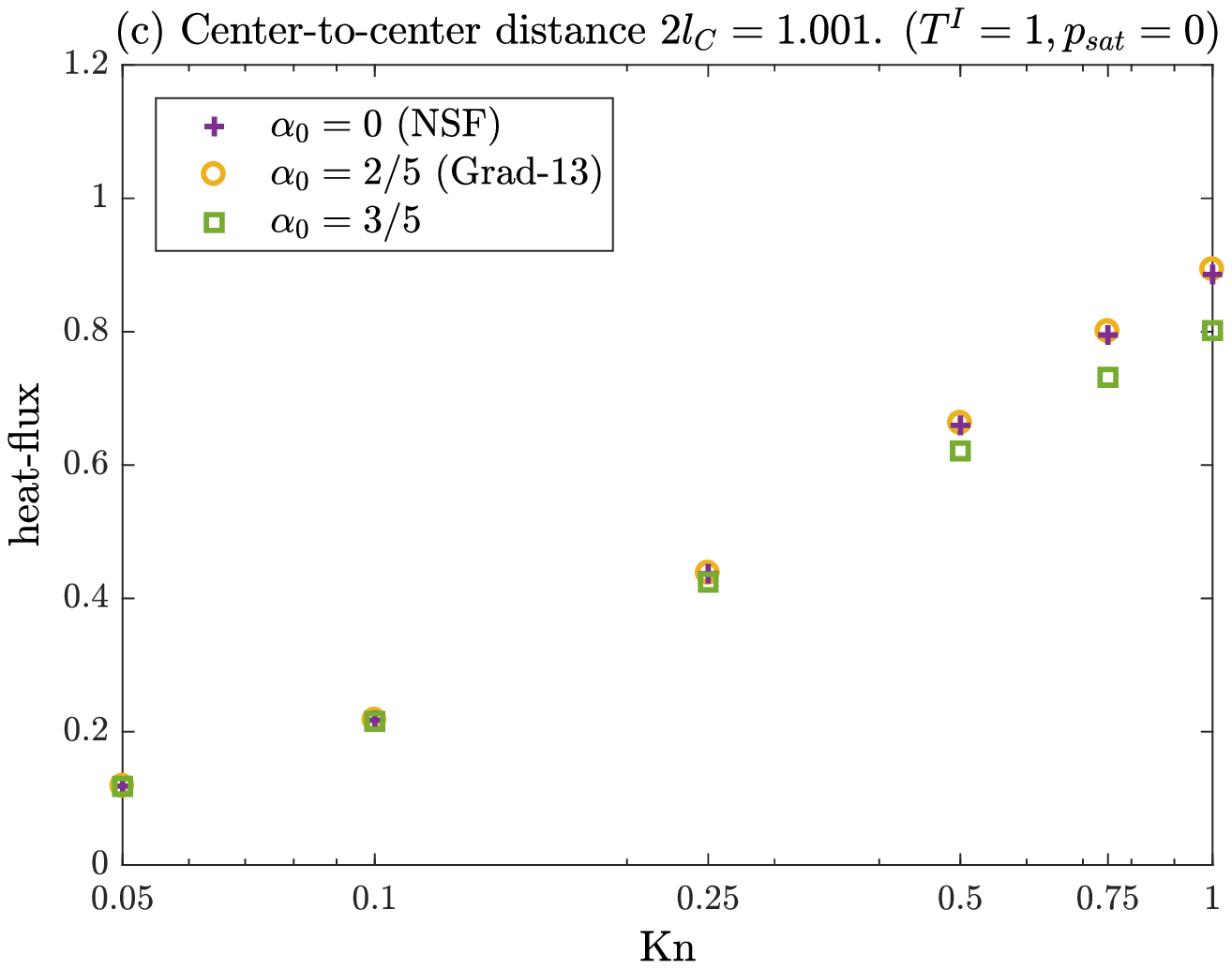}
\hfill
\includegraphics[height=50mm]{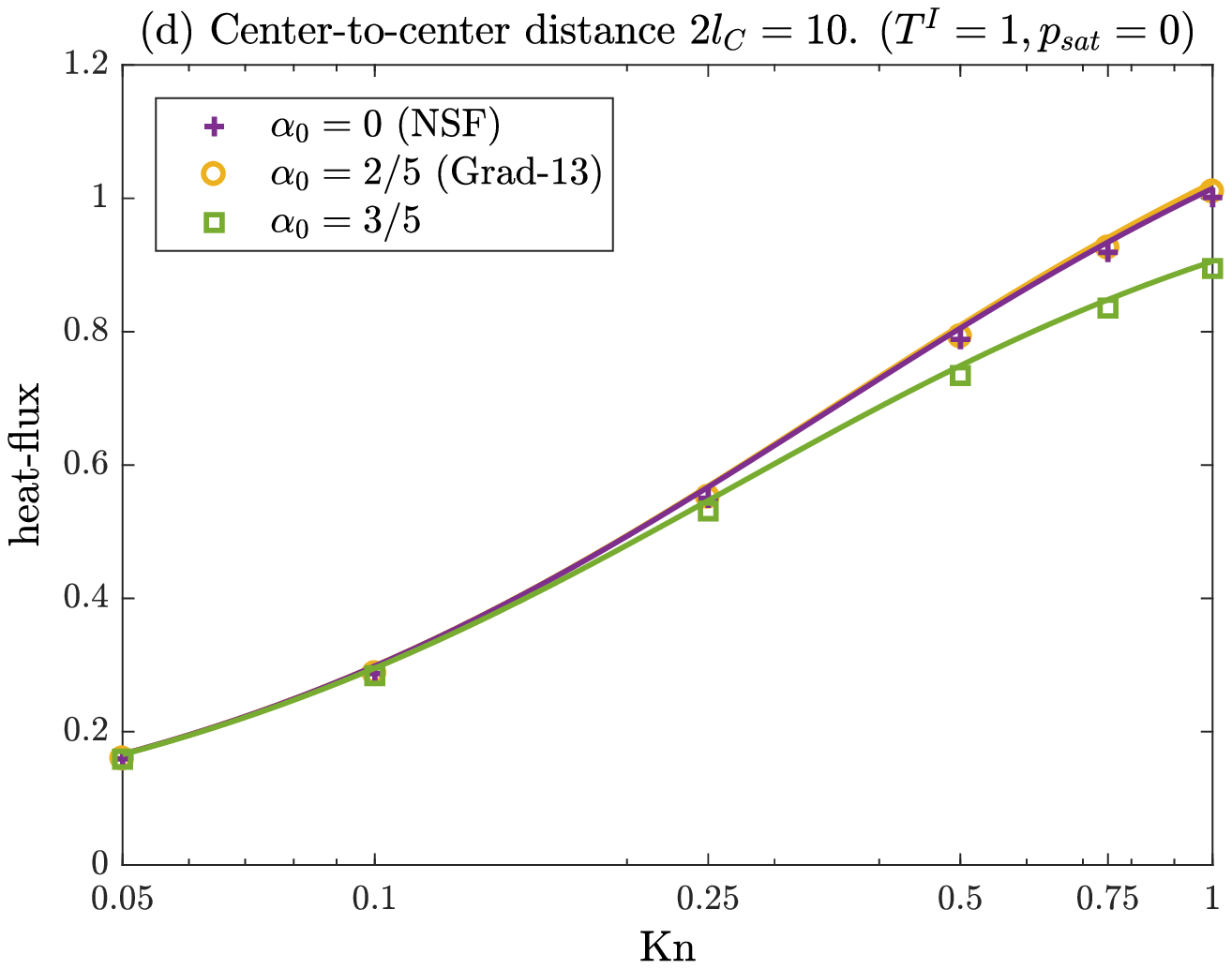}
\caption{\label{fig:twodropletstemperature}Two interacting droplets for the temperature driven case ($T^{I}=1, p_{sat}=0$): Profiles of the (a) mass-flux with center-to-center distance $2l_C=0.001$, (b) mass-flux with $2l_C=10$ (c) heat-flux with $2l_C=0.001$ and (d) heat-flux with  $2l_C=10$ as functions of the Knudsen number for NSF, Grad-13 and CCR with $\alpha_0=3/5$.  Symbols denote MFS solutions and curves represent analytic solution from a single droplet.}
\end{figure}

\underline{Temperature driven case}: In Figures  \ref{fig:twodropletstemperature}, we show the mass-flux and the heat-flux  per unit area as a function of Knudsen number for the temperature-driven case. The mass-flux reduces monotonically with the Knudsen number while the NSF theory predicts a higher mass-flux than CCR. Figures  \ref{fig:twodropletstemperature}a and \ref{fig:twodropletstemperature}b illustrate the mass-flux with center-to-center distance $2l_C=0.001$ (i.e., when the droplets are next to each other) and  $2l_C= 10$ (i.e., at a distance where proximity effects are negligible), respectively. The corresponding heat-flux is shown in Figures  \ref{fig:twodropletstemperature}c and \ref{fig:twodropletstemperature}d. 

When droplets are next to each other a shielding effect is observed, and, as a result, the mass-flux reduces (cf. Figures  \ref{fig:twodropletstemperature}a and b).  At $\mathrm{Kn}=0.05$, all three models predict about $29\%$ reduction in the mass-flux as compared to a single droplet (i.e., $l_C\rightarrow\infty$). The reduction in the mass-flux decreases with increase in Knudsen number. For $\mathrm{Kn}=1$, NSF, Grad-13 and CCR with $\alpha_0=3/5$ predicts about $9\%$ , $13\%$ and $12\%$ reduction in mass-flux, respectively.  

The mass-fluxes computed from the MFS (symbols) for center-to-center distance $10$ are shown in Figure \ref{fig:twodropletstemperature}b; the analytic results from a single droplet are also plotted (continuous lines) for comparison. As expected, as the droplets are moved further apart, the results converges to the single droplet cases; the mass-flux is within $5\%$ (for $\mathrm{Kn}=0.05$) and $1\%$ (for $\mathrm{Kn}=1$) for the single droplet case. 

Similarly, the proximity effects on the heat-flux are visible in Figures \ref{fig:twodropletstemperature}c and \ref{fig:twodropletstemperature}d, for $2l_C=0.001$ and $2l_C=10$, respectively. Again, the heat flux reduces as $l_C\rightarrow 1/2$, and as the value for the Knudsen number increases the proximity effects diminish. The percentage difference in the heat-flux from the single droplet cases varies from about $29\%$ to $12\%$ for $2l_C= 1.001$ and from  about $4\%$ to $1\%$ for $2l_C= 10$, as the value of Knudsen number varies from $0.05$ to 1.

\begin{figure}
\centering
\includegraphics[scale=0.45]{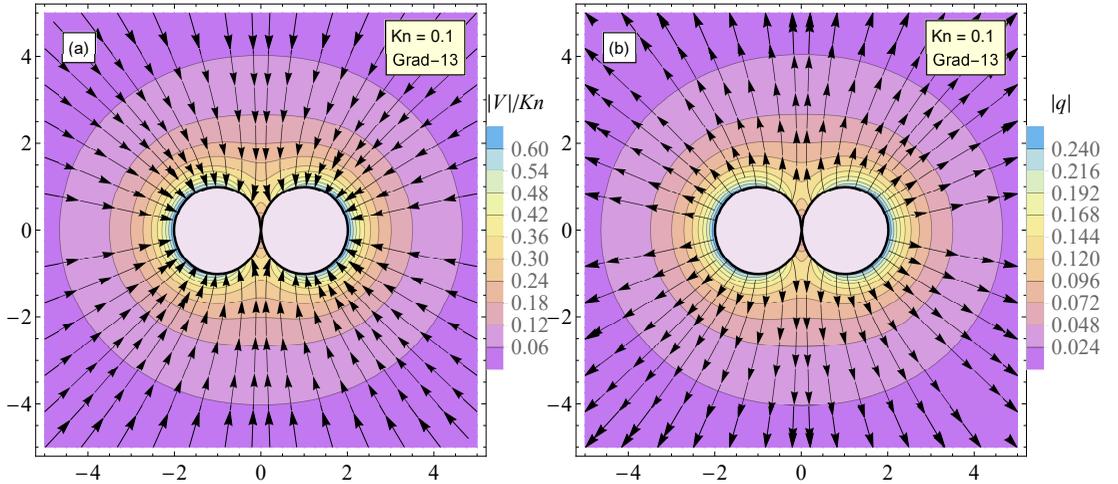}
\caption{\label{fig:G13evporatingdrops2p001} Two droplets  placed adjacent to each other ($2l_c=1.001$) for the temperature driven case ($T^{I}=1, p_{sat}=0$): (a) streamlines and speed contours (b) heat-flux lines and magnitude, obtained from the Grad-13 MFS solution. Vertical and horizontal axis represents $x$- and $y$-directions. ($N_c = N_s = 650$ and $\gamma = R_s/R_c = 0.1$).}
\end{figure}

The vaporization dynamics of a pair of droplets placed adjacent to each other is
markedly different from the single droplet counterpart. In Figure \ref{fig:G13evporatingdrops2p001}, (a) the stream lines superimposed on the velocity magnitude contours, and (b) the heat-flux lines and the magnitude contours are shown. The heat goes out of the hot droplets due to negative temperature gradients while the condensation occurs in order to compensate for the heat loss. Figure \ref{fig:G13evporatingdrops2p001} shows that the mass-flux and the heat-flux at the confined region between droplets are
sufficiently lower than outer regions. In this region, the vapour temperature is relatively high, thus a smaller temperature jump which leads to a reduced mass-flux and heat-flux.

\begin{figure}
\centering
\includegraphics[height=50mm]{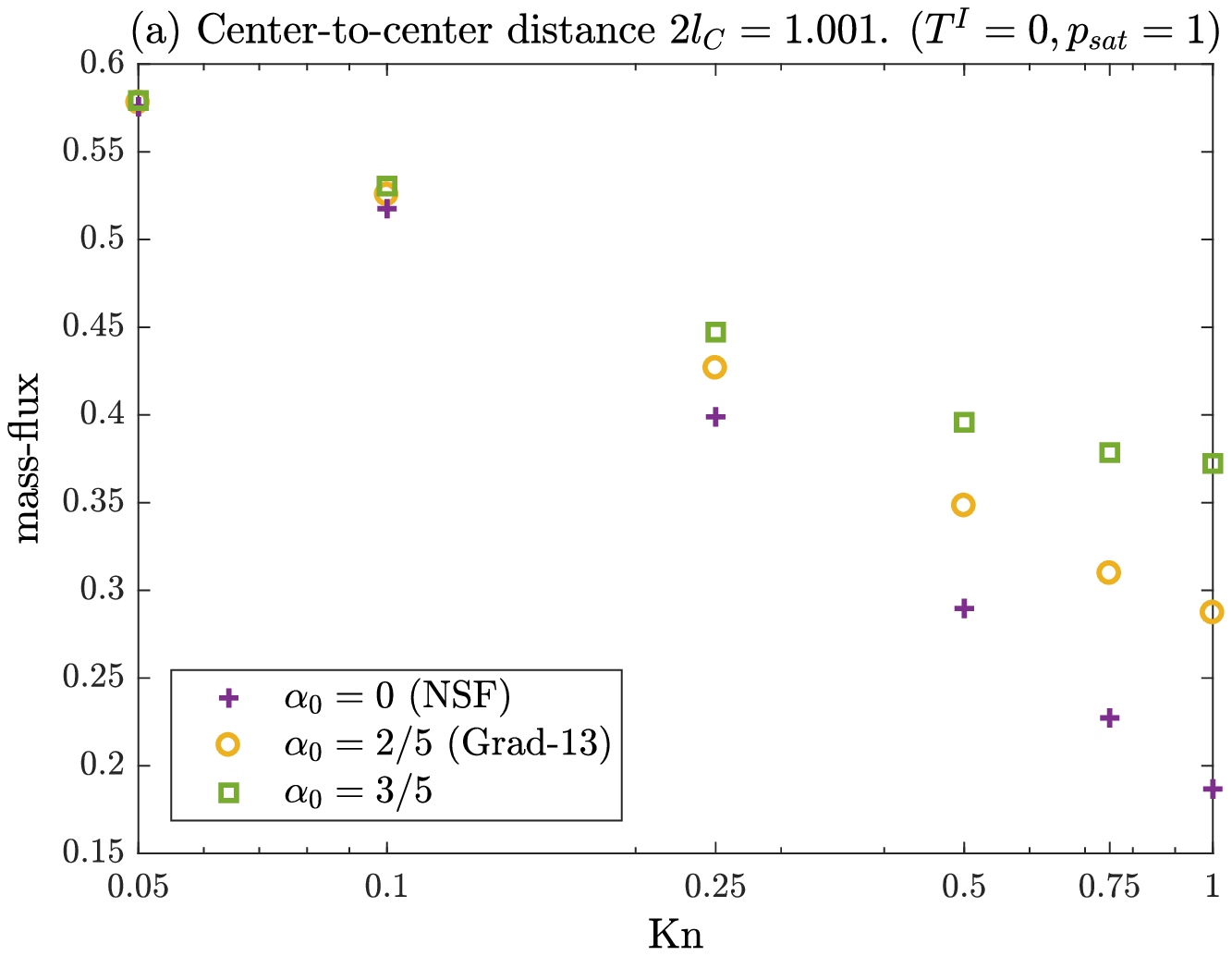}
\hfill
\includegraphics[height=50mm]{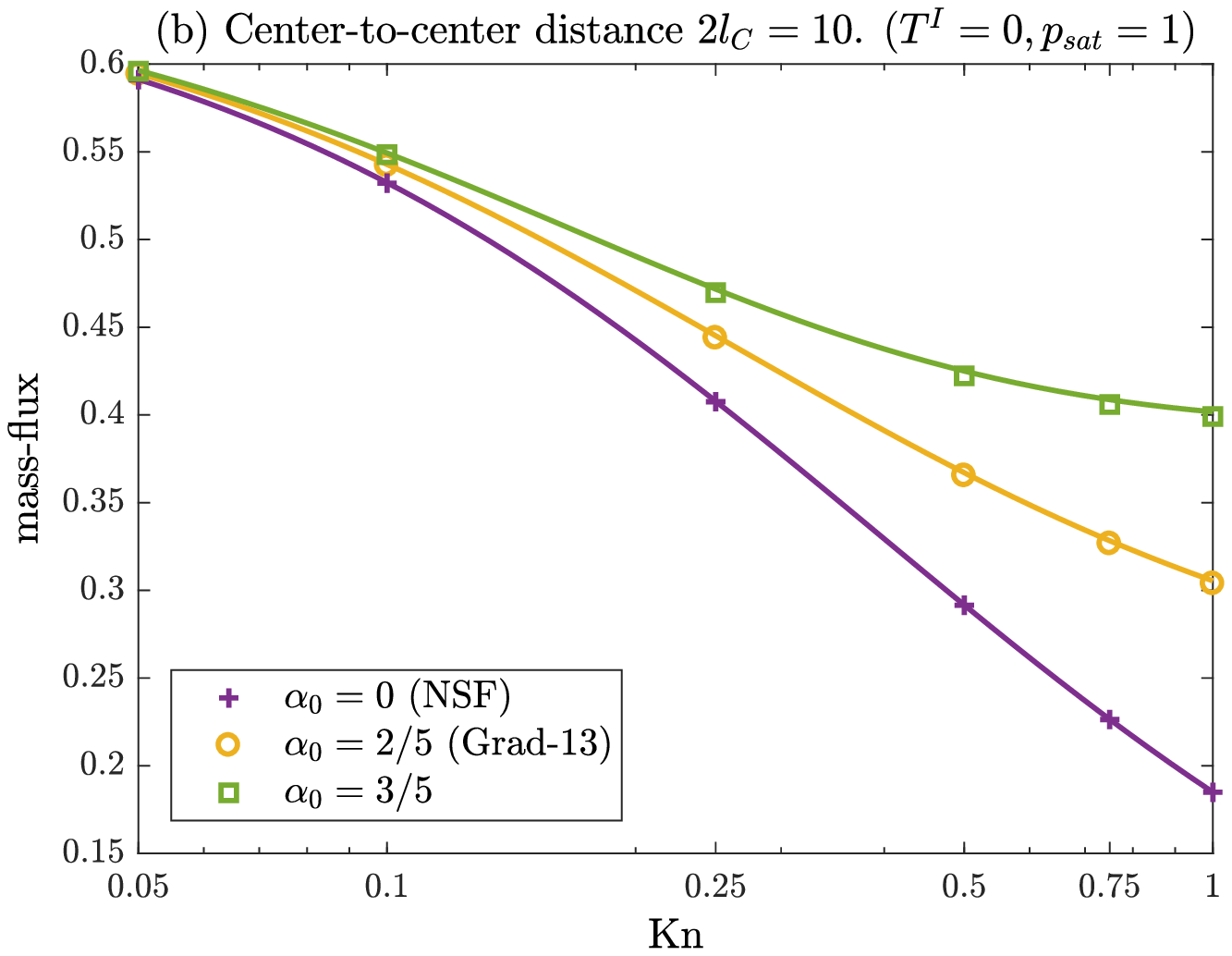}
\\
\includegraphics[height=50mm]{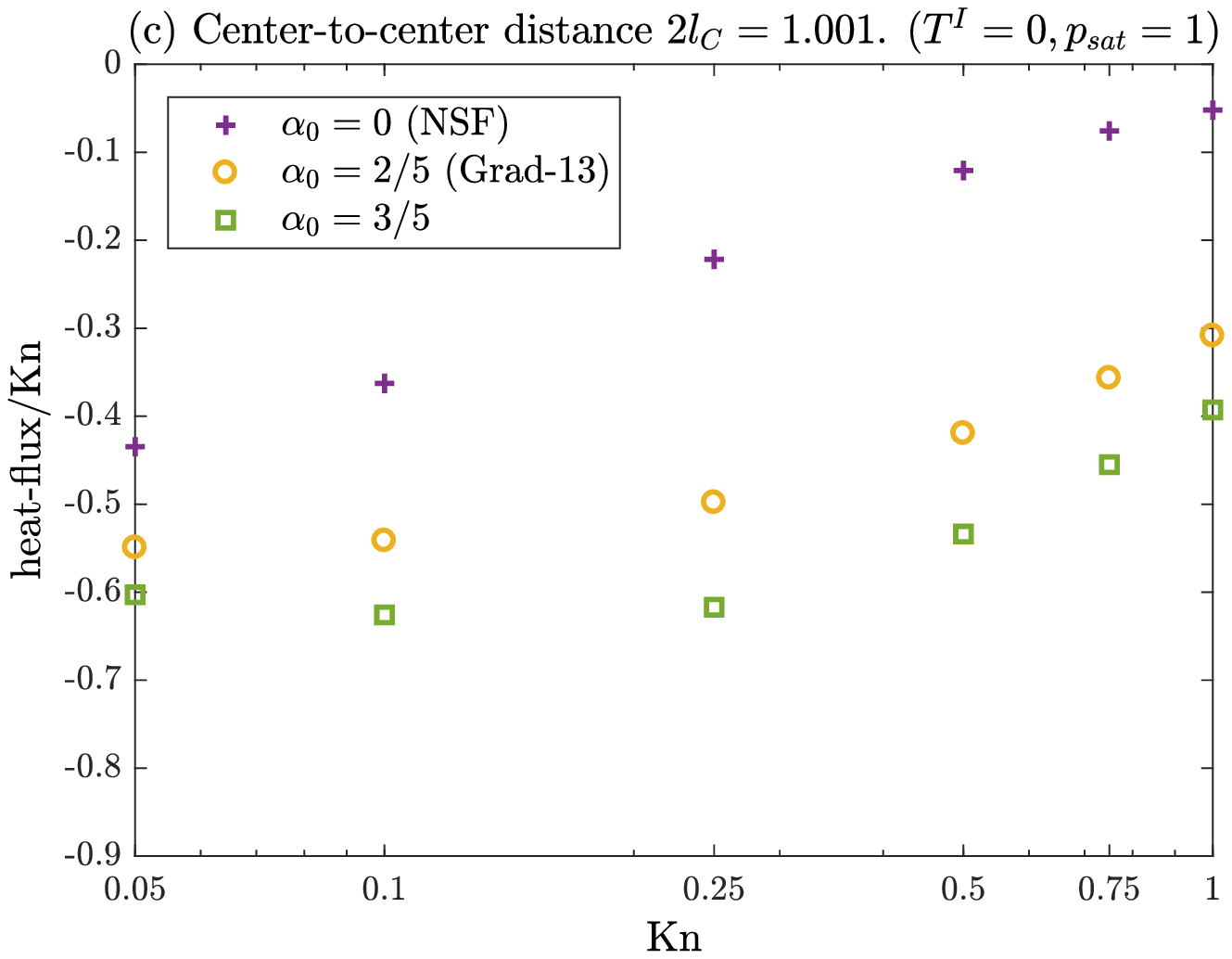}
\hfill
\includegraphics[height=50mm]{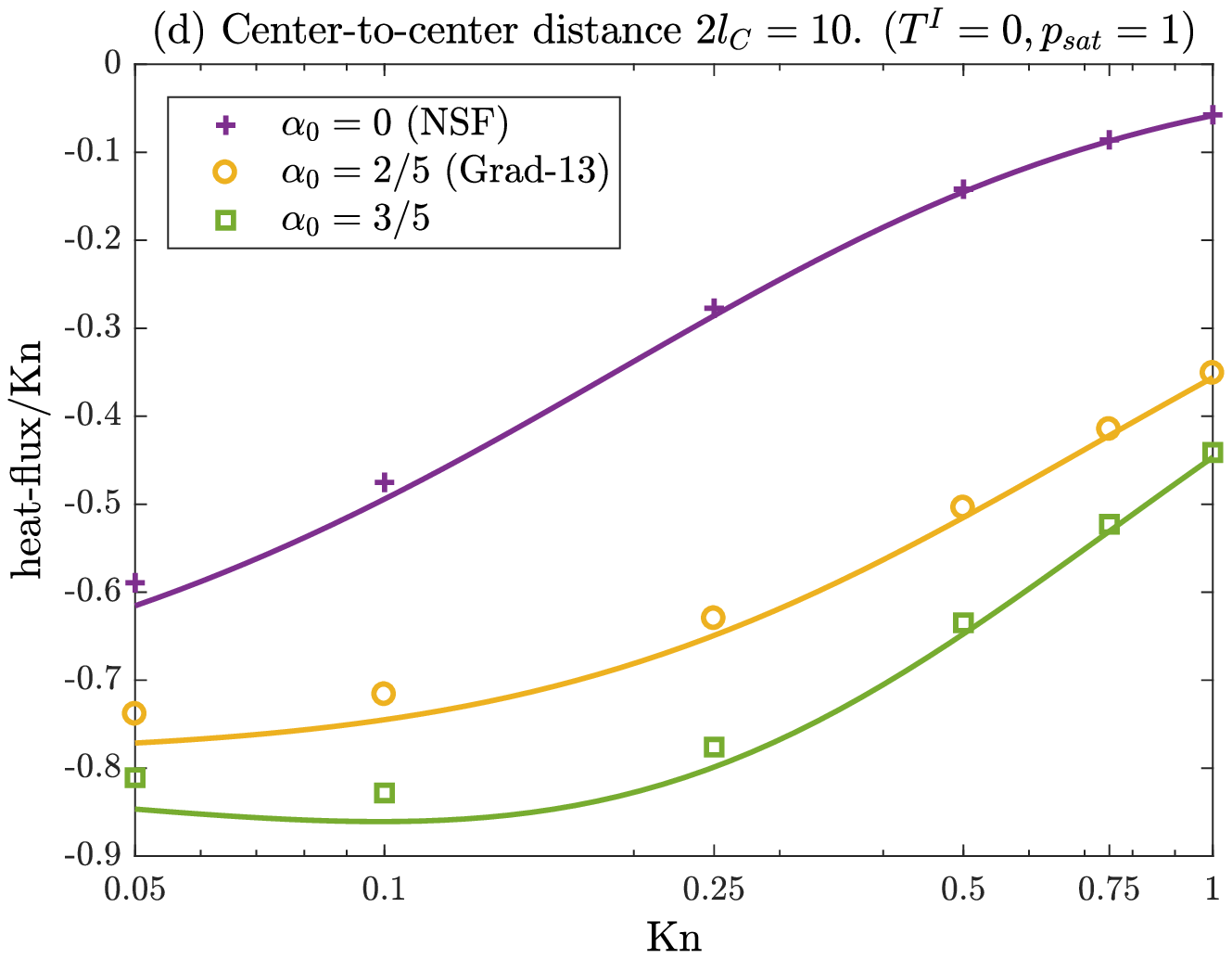}
\caption{\label{fig:twodropletspressure}Two interacting droplets for the pressure driven case ($T^{I}=0, p_{sat}=1$): Profiles of the (a) mass-flux with center-to-center distance $2l_C=0.001$, (b) mass-flux with $2l_C=10$ (c) heat-flux with  $2l_C=0.001$ and (d) heat-flux with $2l_C=10$ as functions of the Knudsen number for NSF, Grad-13 and CCR with $\alpha_0=3/5$.  Symbols denote MFS solutions and curves represent analytic solution from a single droplet.}
\end{figure}

\underline{Pressure driven case}: In Figure \ref{fig:twodropletspressure}, we show the  (a) mass-flux with center-to-center distance $2l_C=0.001$, (b) mass-flux with $2l_C=10$ (c) heat-flux with  $2l_C=0.001$ and (d) heat-flux with $2l_C=10$ as functions of the Knudsen number for the pressure driven case.  Again, in order to gain insights into the proximity effects, the numerical solution obtained via MFS (symbols) and the  analytic results from a single droplet case (continuous lines) are compared in Figure \ref{fig:twodropletspressure}. The this case, the mass-flux is slightly reduced ($\lesssim 3\%$ for $2l_C=0.001$ and $\lesssim 0.06\%$ for $2l_C=10$) between the range ($0.05\leq \mathrm{Kn}\leq 1$). On the other hand, the proximity effects on the heat flux are observed to be  significant. For $2l_C=0.001$, the magnitude of the heat-flux is reduced about $29.3\%$ for NSF ($28.7\%$ for Grad-13 and CCR) at $\mathrm{Kn}=0.05$ and about $10\%$ for NSF ($\leq 15\%$ for Grad-13 and CCR) at $\mathrm{Kn}=1$. For $2l_C=0.001$, all models give less than $4\%$ reduction in the heat flux, for all range of the Knudsen number considered. 

\begin{figure}
\centering
\includegraphics[scale=0.55]{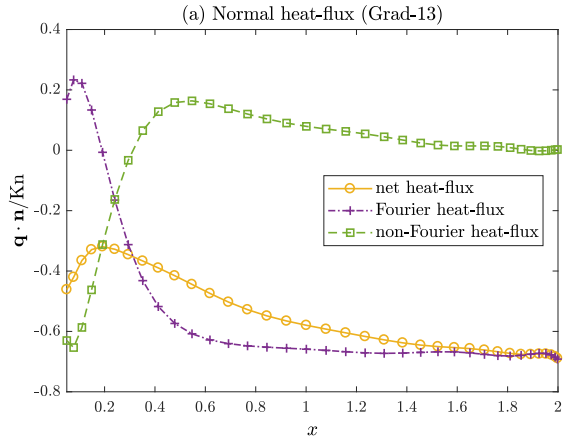}
\caption{\label{fig:FourierNonFourierGrad}Two interacting droplets for the pressure driven case ($T^{I}=0, p_{sat}=1$): Fourier and non-Fourier contributions to the normal heat-flux ($\mathbf{q}\cdot\mathbf{n}/\mathrm{Kn}$) given by the Grad-13 theory plotted along the top surface of the droplet in $x$-$z$ plane.}
\end{figure}

Another interesting point to be noted from Figures \ref{fig:twodropletspressure}(c), (d) is that the heat-flux predicted by the NSF equations is significantly lower in magnitude compared to the CCR and Grad-13 models, especially in the transition regime ($\mathrm{Kn} \geq 0.1$). Within this regime, the heat-flux is not only driven by the temperature gradient (i.e., Fourier's law) but also due to the pressure gradient; this second-order effect can easily be understood from the constitutive relations (\ref{CCR relations vector form}) and the momentum balance equation (\ref{conservation laws vector form}), which give
\begin{equation}
\mathbf{q}=-\frac{c_{p}\mathrm{Kn}}{\Pr }\left( \nabla T-\alpha _{0}\nabla p%
\right).  \label{non-Fourier heat flux}
\end{equation}%
Here, the second term on the right-hand-side is the non-Fourier contribution to the heat flux due to pressure gradient  \citep{RGS2018}. In Figure \ref{fig:FourierNonFourierGrad}, we show two different contributions to the scaled normal heat-flux ($\mathbf{q}\cdot\mathbf{n}/\mathrm{Kn}$) along the surface of the droplet (in  $x$-$z$ plane) for the pressure-driven case with $\mathrm{Kn}=0.1$. As predicted from the Grad-13 theory, the heat-flux owing to pressure gradient (non-Fourier) and heat-flux due to temperature gradient (Fourier) have opposing effects, and consequently the net heat-flux is reduced in the second-order theories.


%

\begin{figure}
\centering
\includegraphics[scale=0.45]{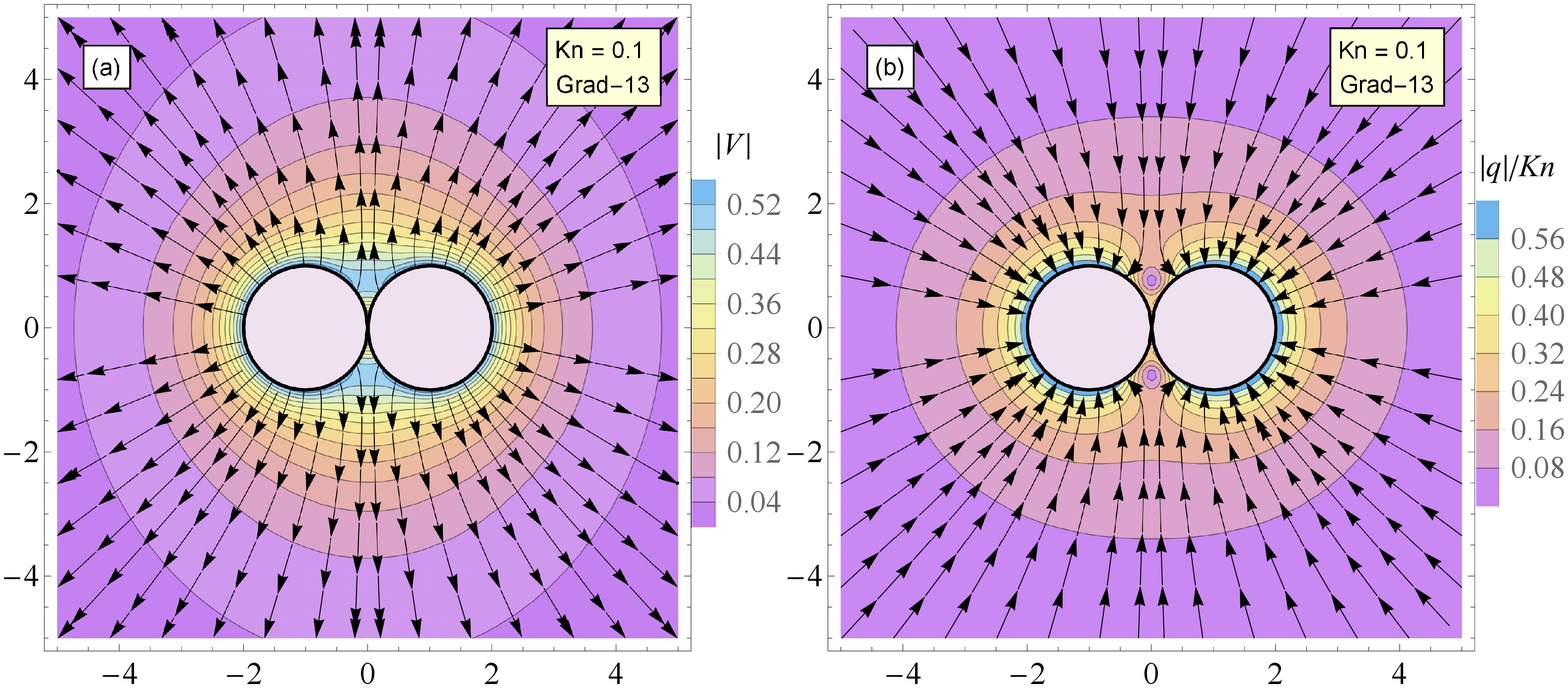}
\caption{\label{fig:G13evporatingdrops1p001} Two droplets  placed next to each other ($2l_c=1.001$) for the pressure driven case ($T^{I}=0, p_{sat}=1$): (a) stream lines and speed contours (b) heat-flux lines and magnitude, obtained from the Grad-13 MFS solution. Vertical and horizontal axis represents $x$- and $y$-directions. ($N_c = N_s = 650$ and $\gamma = R_s/R_c = 0.1$).}
\end{figure}

 In Figure \ref{fig:G13evporatingdrops1p001}, we again show the streamlines and the velocity magnitude contours (a), and the heat-flux lines and the magnitude contours (b) for the pressure driven case. In this case, evaporation occurs due to a negative pressure gradient and the heat flows inwards. The mass-flux and the heat-flux at the confined region between droplets are
reduced due to a high vapour pressure. 

\begin{figure}
\centering
\includegraphics[scale=0.5]{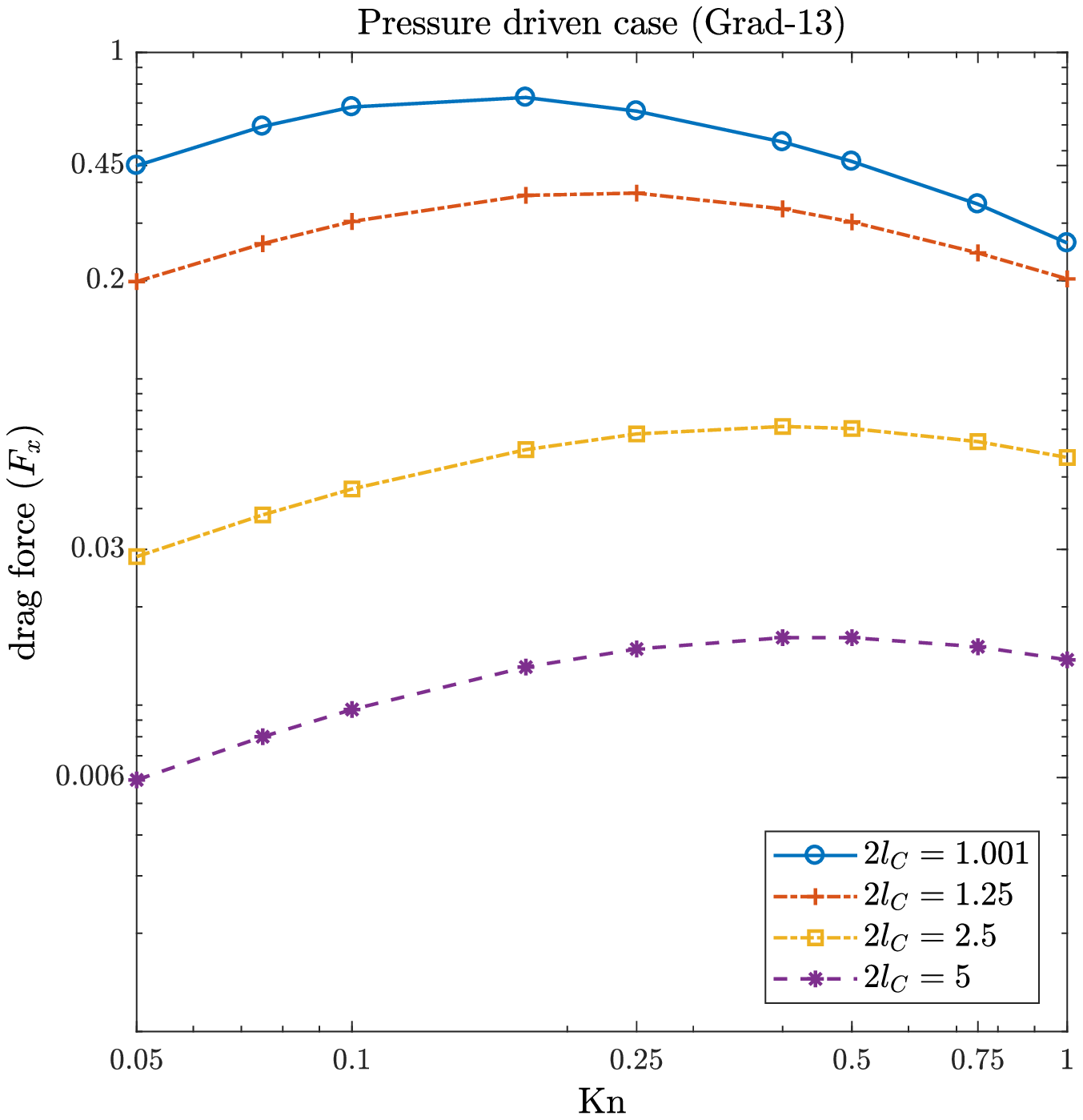}
\caption{\label{fig:forceFxcase2Grad13}
The drag force acting on the droplets vs Knudsen numbers for various values of the separation distance as predicted by the Grad-13 theory for the pressure driven case.}
\end{figure}

The high-pressure region created between two droplets pushes them away from each other. The net drag force (in the $x$ direction) acting on each droplet is plotted in Figure \ref{fig:forceFxcase2Grad13}  as a function $\mathrm{Kn}$. The results  for various values of center-to-center distance shown in Figure \ref{fig:forceFxcase2Grad13}  are computed from the Grad-13 model; the drag force predicted by the NSF model is within $10\%$ and not shown. 
The force decreases exponentially as droplets become further apart, i.e., as $l_C$ increases. Interestingly, the drag force attains a minimum value at a certain Knudsen number, which depends on how far the droplets are from each other. As $\mathrm{Kn}\rightarrow \infty $, the drag force vanishes.

%
%

\subsection{Evaporation of a deformed droplet}
\label{sec: evaporation of a deformed droplet}
In this section, we shall study evaporation/condensation  from a deformed droplet with fixed shape and sized, in order to gain some initial insight into evaporation occurring during oscillatory and/or deformed droplet events---leaving the step to unsteady flow as a focus for future work.  The effects of heat conduction and the flow inside the droplet are not considered. The parametric equation for the droplet surface in the second harmonic mode is given by (e.g., see \cite{Sprittles2012})
\begin{equation}
\vec{X}=\eta _{0}\left[ 1+\frac{\eta }{2}\left( 3\cos ^{2}\phi -1\right) %
\right] \left\{ \sin \phi \cos \theta ,\sin \phi \sin \theta ,\cos \phi
\right\} 
\end{equation}%
where $\phi \in \lbrack 0,\pi ]$ and $\theta \in \lbrack 0,2\pi ]$ and $\eta 
$ is the shape factor. The volume of the droplet is%
\begin{equation}
V=\eta _{0}\frac{4\pi }{3}\left(\frac{35+21\eta ^{2}+2\eta ^{3}}{35}\right)^{1/3}\text{.}
\end{equation}%
Therefore, in order to have the volume $V$ equal to a spherical droplet of
radius $1$, the normalizing factor $\eta _{0}$ is given by 
\begin{equation}
\eta _{0}=\left(\frac{35}{35+21\eta ^{2}+2\eta ^{3}}\right)^{1/3}\text{.}
\end{equation}%

\begin{figure}
\centering
\includegraphics[scale=0.4]{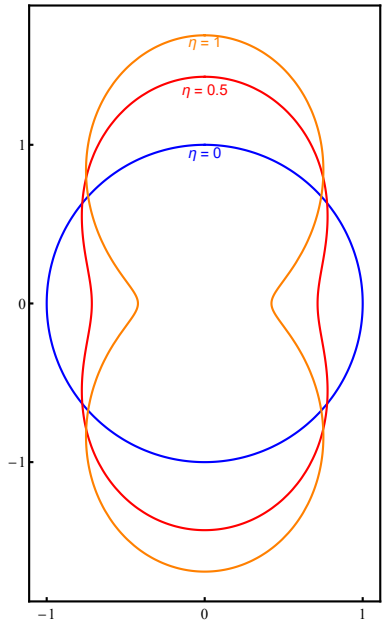}
\caption{\label{fig:sphericalharmonics}
Droplet shapes given by second spherical harmonics with shape parameter $\eta = 0, 0.5, 1$ and the normalizing factor $\eta_0 = 35^{1/3}/(35+21 \eta^2+2\eta^3)^{1/3}$ ensuring that the droplet has  fixed  volume.}
\end{figure}

The droplet shapes with different values of  shape parameter $\eta=0, 0.5$ and $1$ are depicted in Figure \ref{fig:sphericalharmonics}; the corresponding (non-dimensional) surface area of the droplets are $4\pi, 4\pi\times1.078$ and $4\pi\times1.192$, respectively.  

We shall again consider two cases: (i) flow due to a unit pressure difference while the temperature
of the liquid is equal to the far-field temperature (i.e., $p_{sat}-p_{\infty }=1$ and $%
T^{l}-T_{\infty }=0$) and (ii) flow due to a unit temperature difference while the
saturation pressure in the liquid is same as the far-field pressure (i.e., $%
p_{sat}-p_{\infty }=0$ and $T^{l}-T_{\infty }=1$).

\begin{figure}
\centering
\includegraphics[scale=0.5]{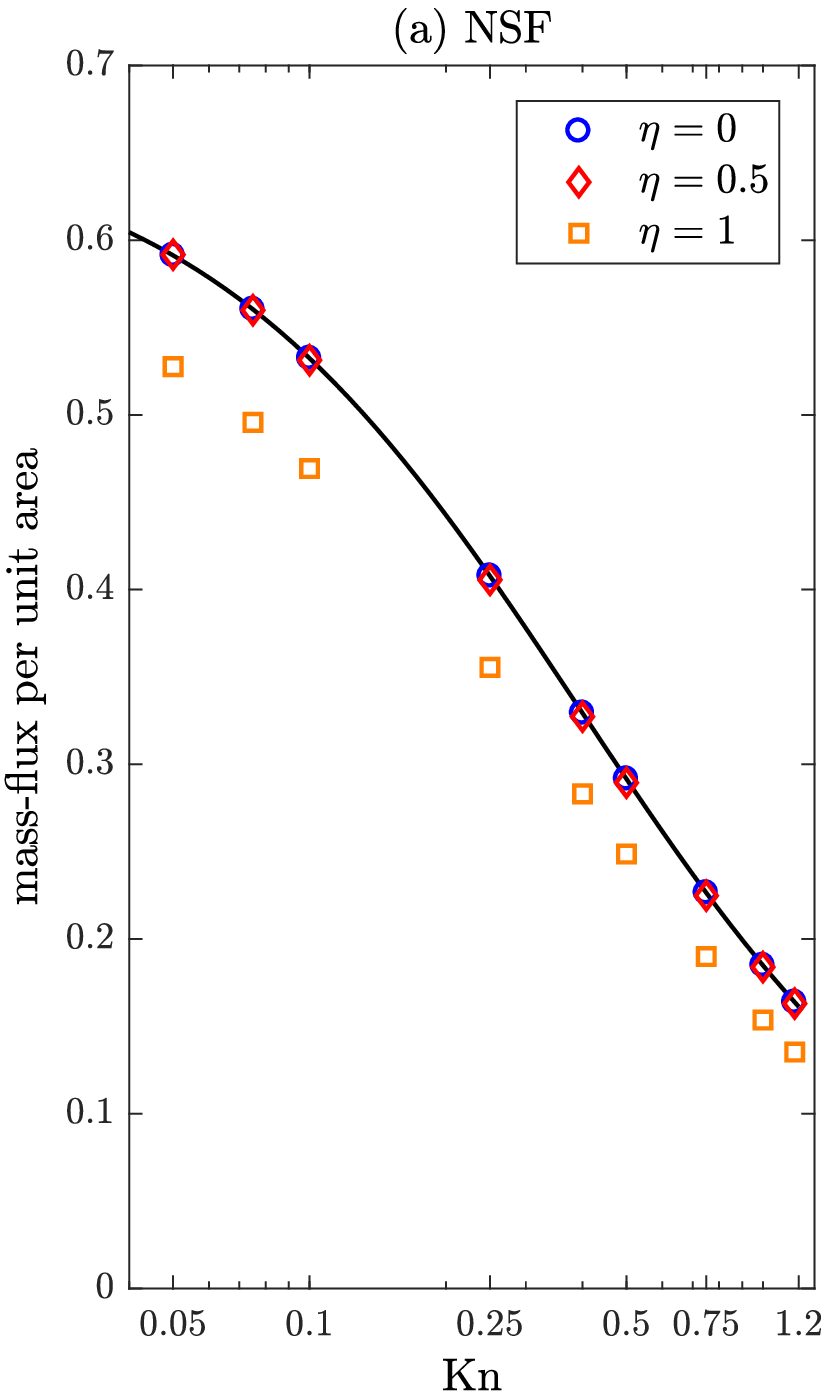}\hfill
\includegraphics[scale=0.5]{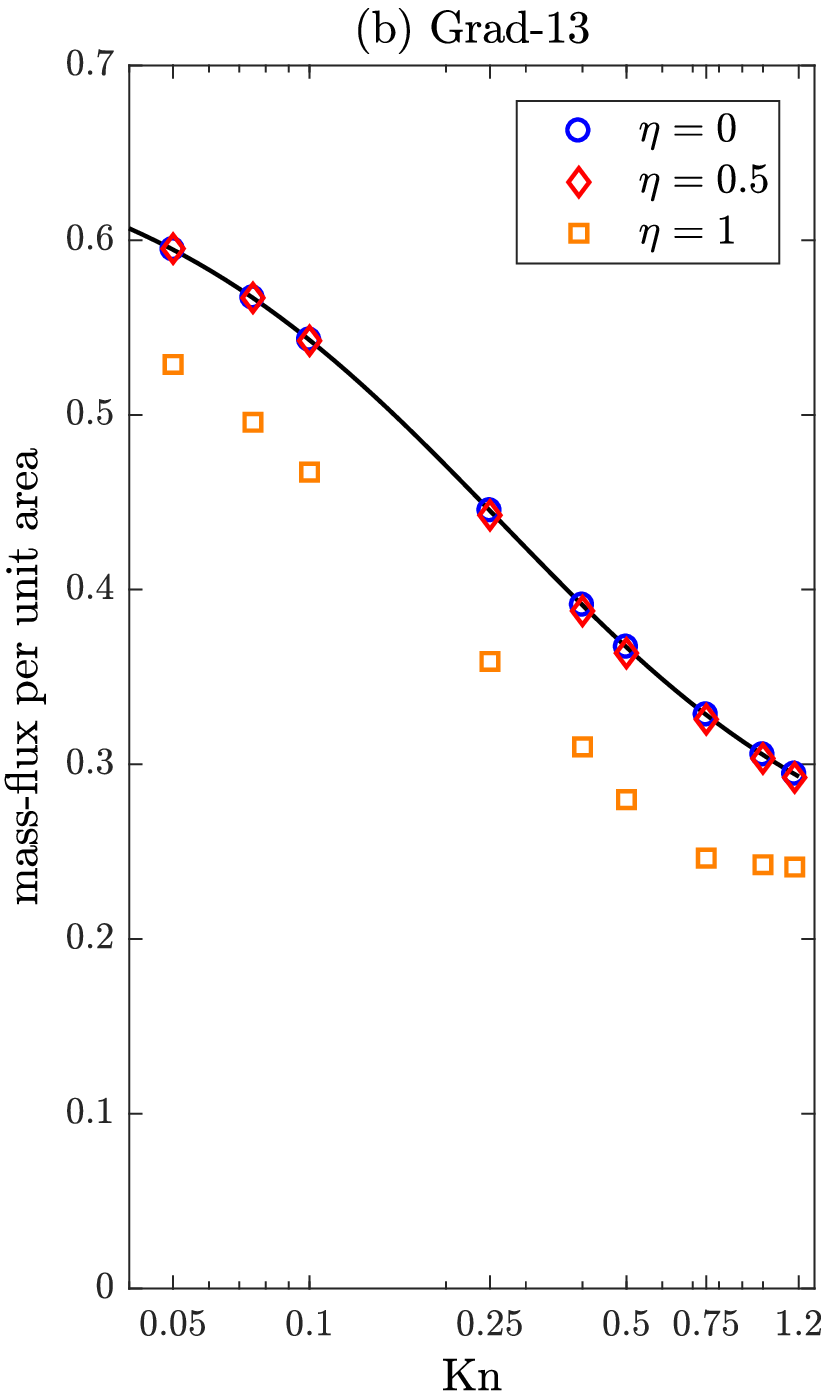}\hfill
\includegraphics[scale=0.5]{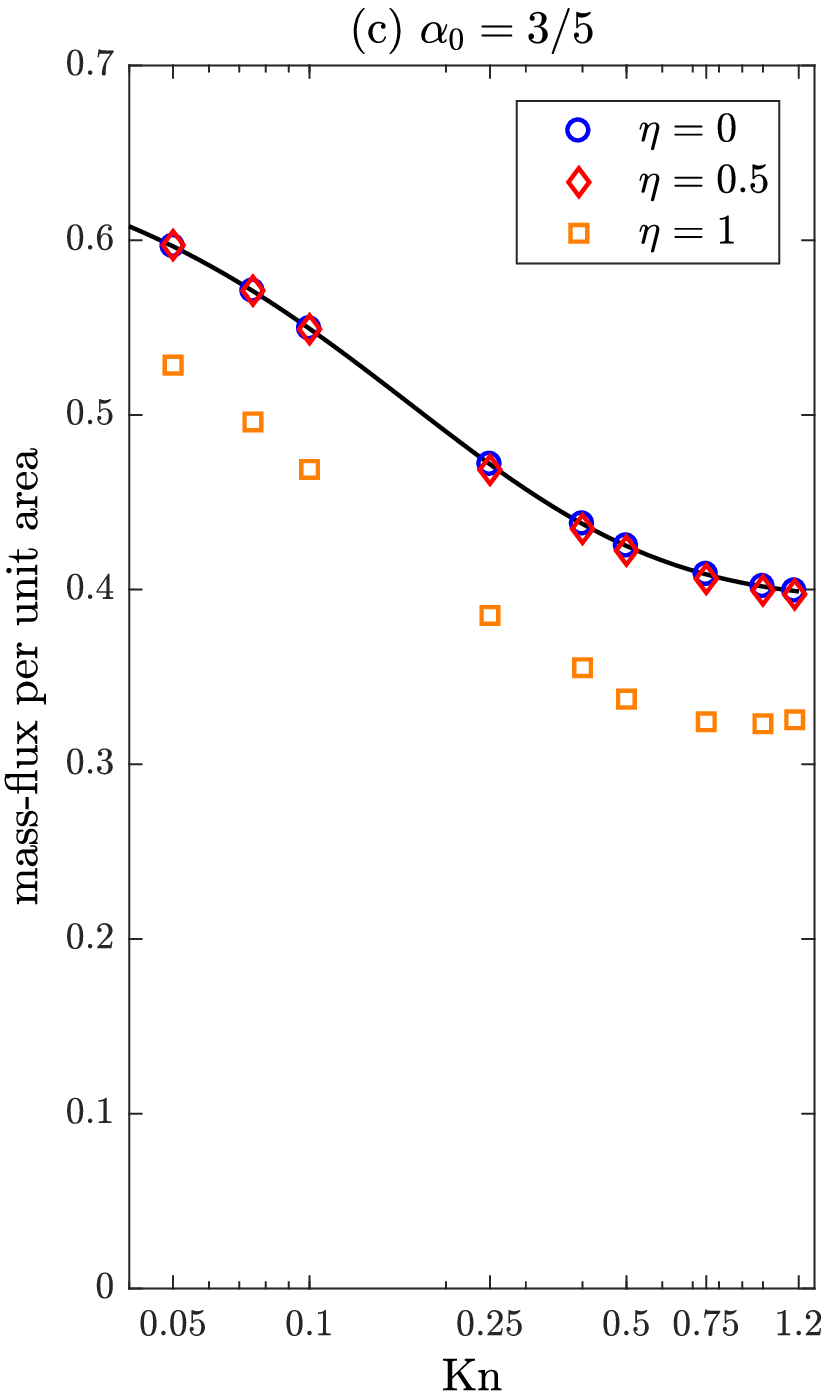}
\caption{\label{fig:massfluxcase2}
Pressure-driven case ($p_{sat}-p_{\infty }=1$, $T^{l}-T_{\infty }=0$): mass flux per unit area vs Knudsen number with $\eta = 0, 0.5, 1$. (a) NSF, (b) Grad-13, and (c) $\alpha_0=3/5$.} The symbols denote the MSF solutions and the solid lines represent analytic results for a spherical droplet. 
\end{figure}

\underline{Pressure-driven case}: In Figures \ref{fig:massfluxcase2}(a), (b) and (c), we show the mass-flux per unit area as a function of $\mathrm{Kn}$ computed using NSF, Grad-13 and $\alpha_0=3/5$ models, respectively for the shape parameter $\eta = 0, 0.5, 1$. Here, the numerical results from MFS are denoted by symbols and the analytic results for the spherical droplet ($\eta = 0$) are represented by the solid line.
For small deformity ( $\eta = 0, 0.5$) mass-flux is nearly same, however when the droplet is more deformed ($\eta = 1$)  the mass-flux is reduced. The percentage reduction in the mass-flux  (compared to a spherical droplet) is about $11\%$ for $\mathrm{Kn}=0.05$ and about $21\%$ for $\mathrm{Kn}=1$.
Similar to the case of two droplets placed next to each other, considered in the previous section, the reduction in the mass-flux is due to formation of a high pressure in the high curvature (for $\eta = 1$) cusp  region near $z=0$, which is not present at smaller $\eta$.

\begin{figure}
\centering
\includegraphics[scale=0.5]{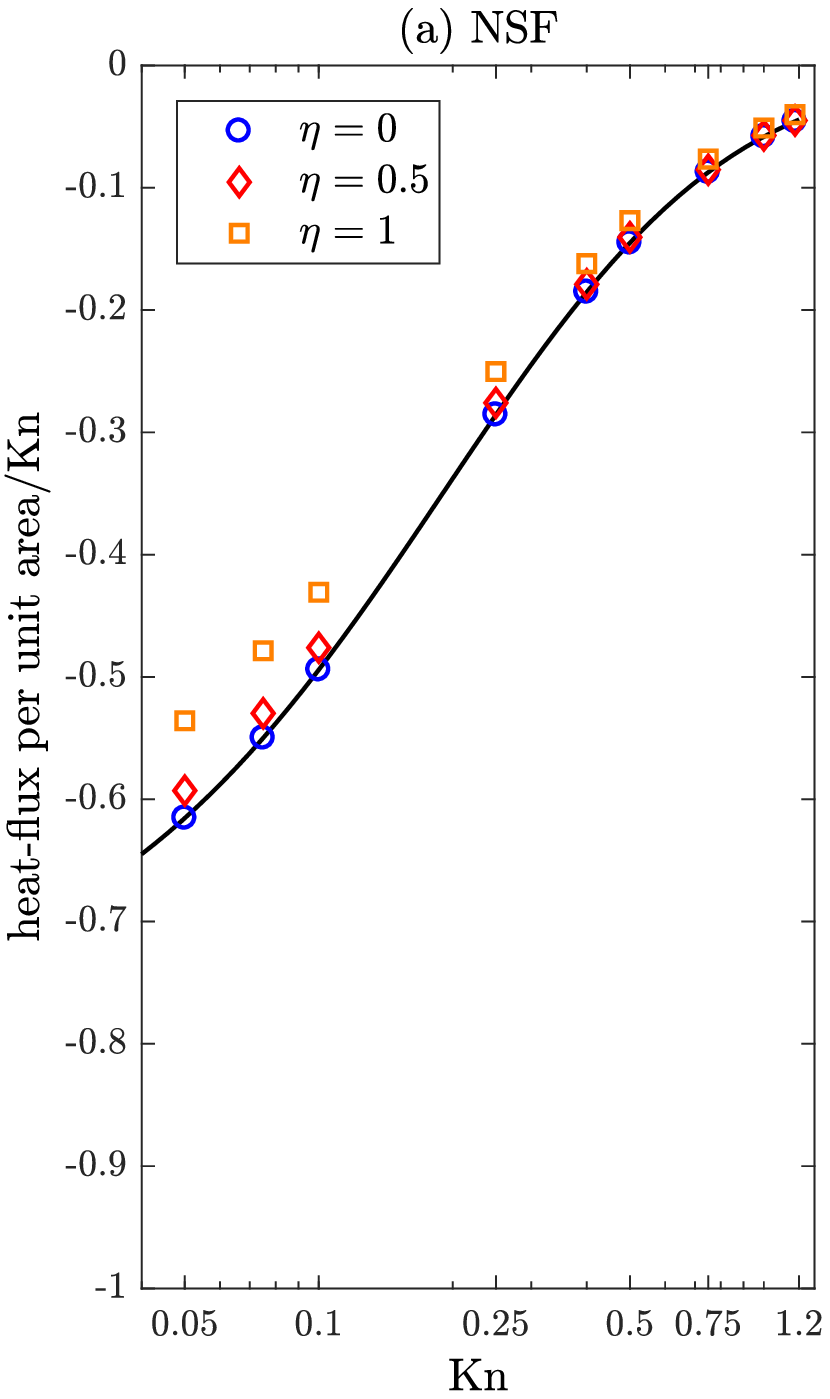}\hfill
\includegraphics[scale=0.5]{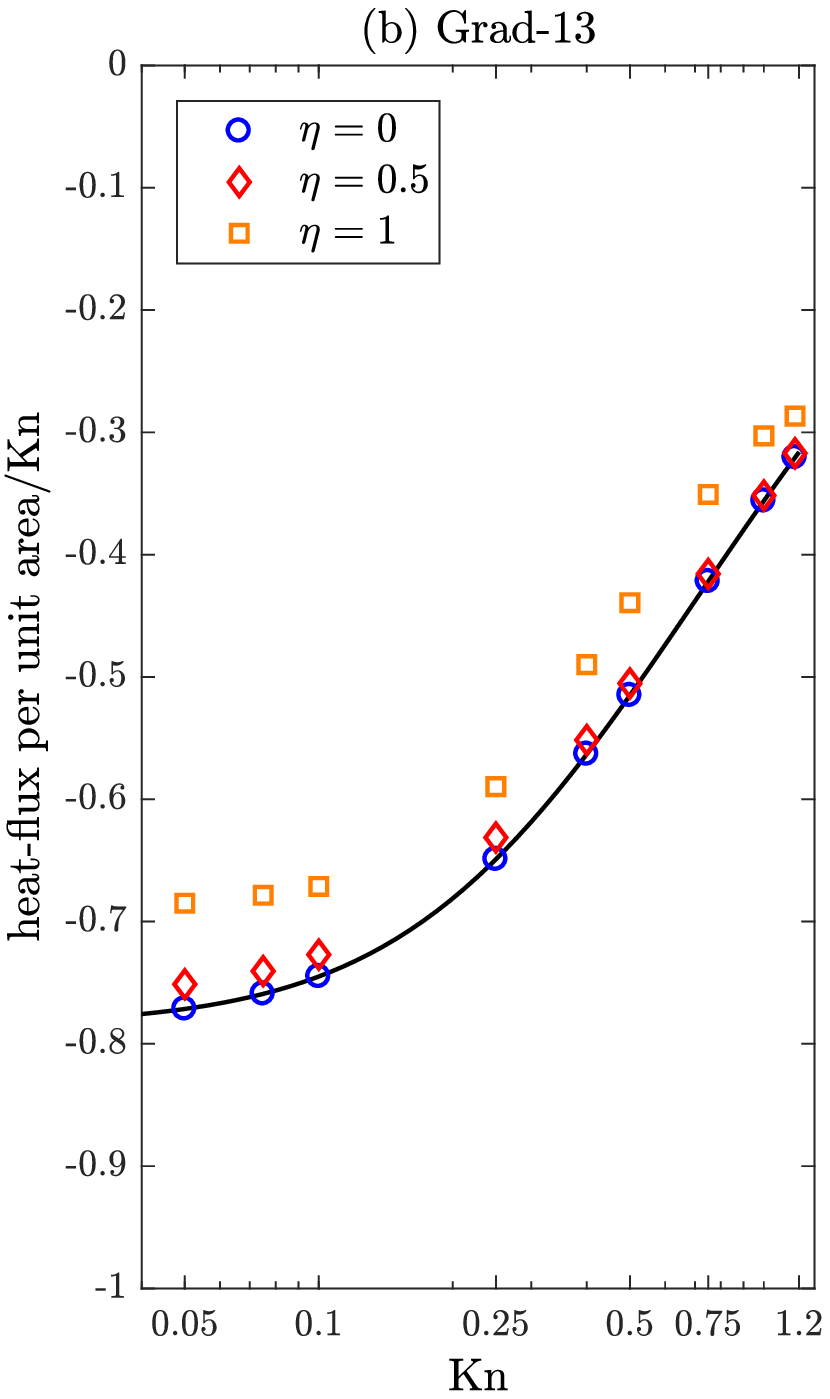}\hfill
\includegraphics[scale=0.5]{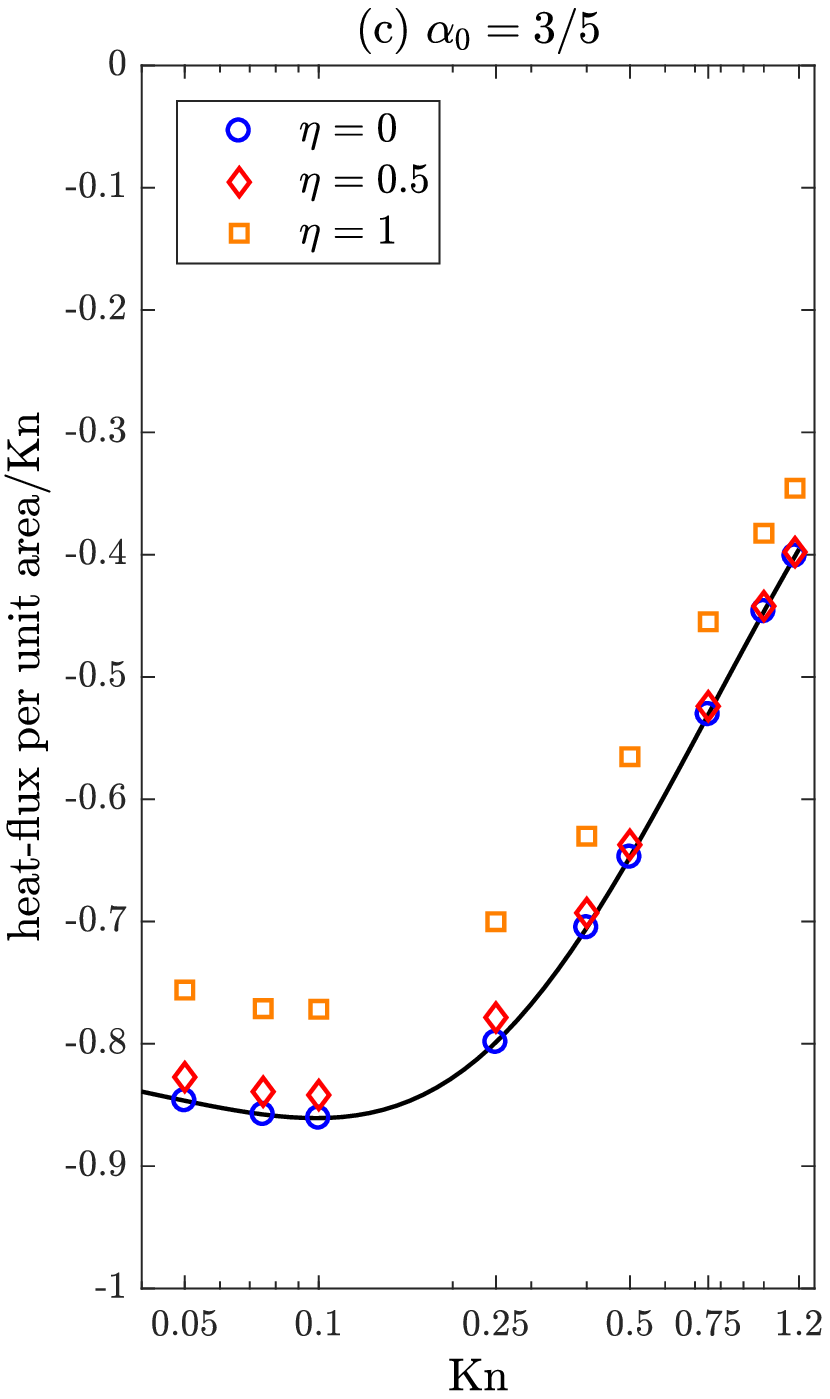}
\caption{\label{fig:heatfluxcase2}
Pressure driven case ($p_{sat}-p_{\infty }=1$, $T^{l}-T_{\infty }=0$): heat flux per unit area/$\mathrm{Kn}$ vs Knudsen number with $\eta = 0, 0.5, 1$. (a) NSF, (b) Grad-13, and (c) $\alpha_0=3/5$.} The symbols denote the MSF solutions and the solid lines represent analytic results for a spherical droplet. 
\end{figure}

Similarly, in Figures \ref{fig:heatfluxcase2}(a), (b) and (c) we plot the heat-flux per unit area as a function of $\mathrm{Kn}$ for the same shape parameters. Note that the heat-flux due to pressure difference is a first order quantity in terms of $\mathrm{Kn}$  (that is it vanishes as $\mathrm{Kn}\rightarrow 0$), hence in these figures we have scaled the heat-flux with inverse $\mathrm{Kn}$. Again, the magnitude heat flux magnitude  reduces as $\eta \rightarrow 1$. The reduction in the heat-flux is about 13\% 
Furthermore, the Grad-13 and CCR theories predict a larger heat-flux compared to the NSF results due to non-Fourier heat-flux contribution (\ref{non-Fourier heat flux}).

\begin{figure}
\centering
\includegraphics[scale=0.5]{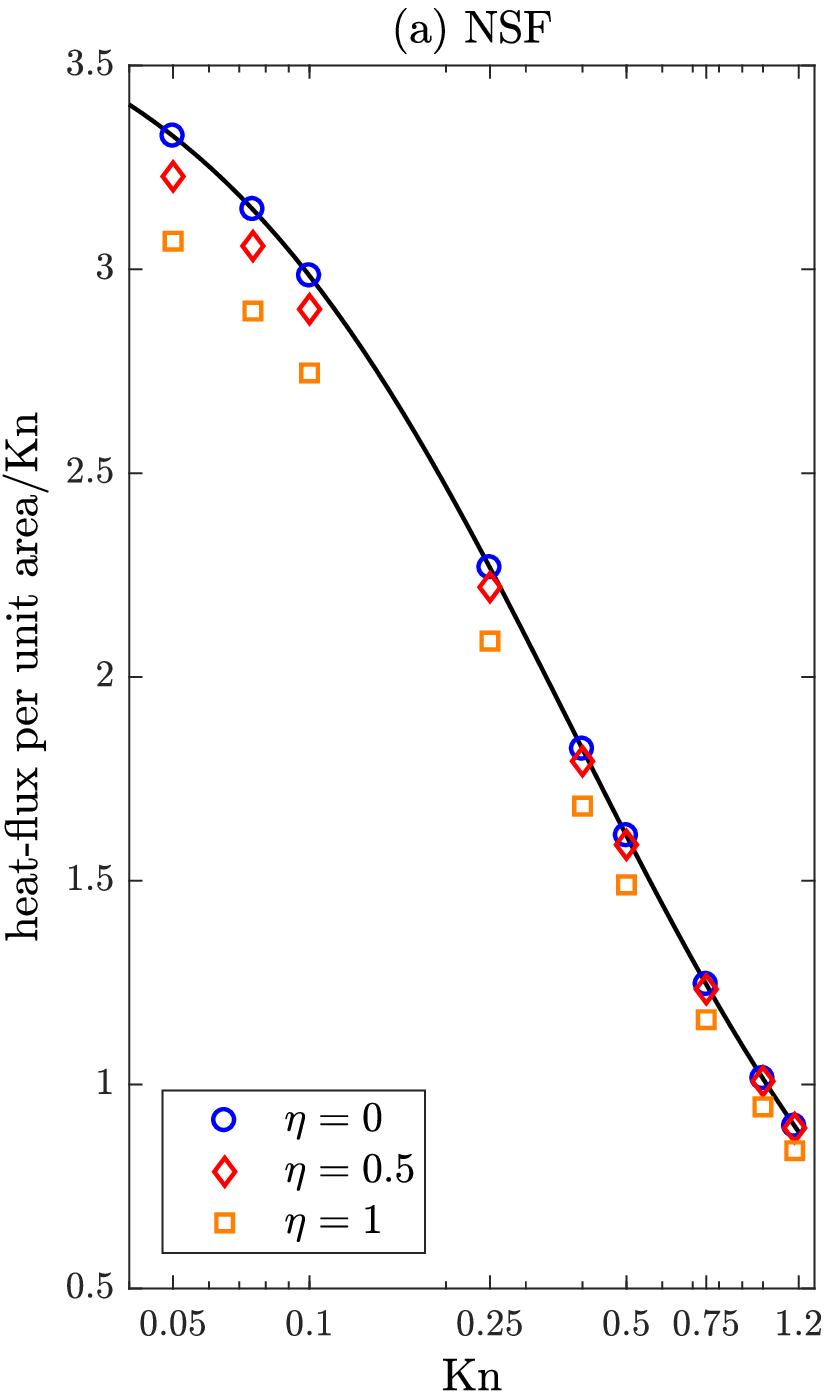}\hfill
\includegraphics[scale=0.5]{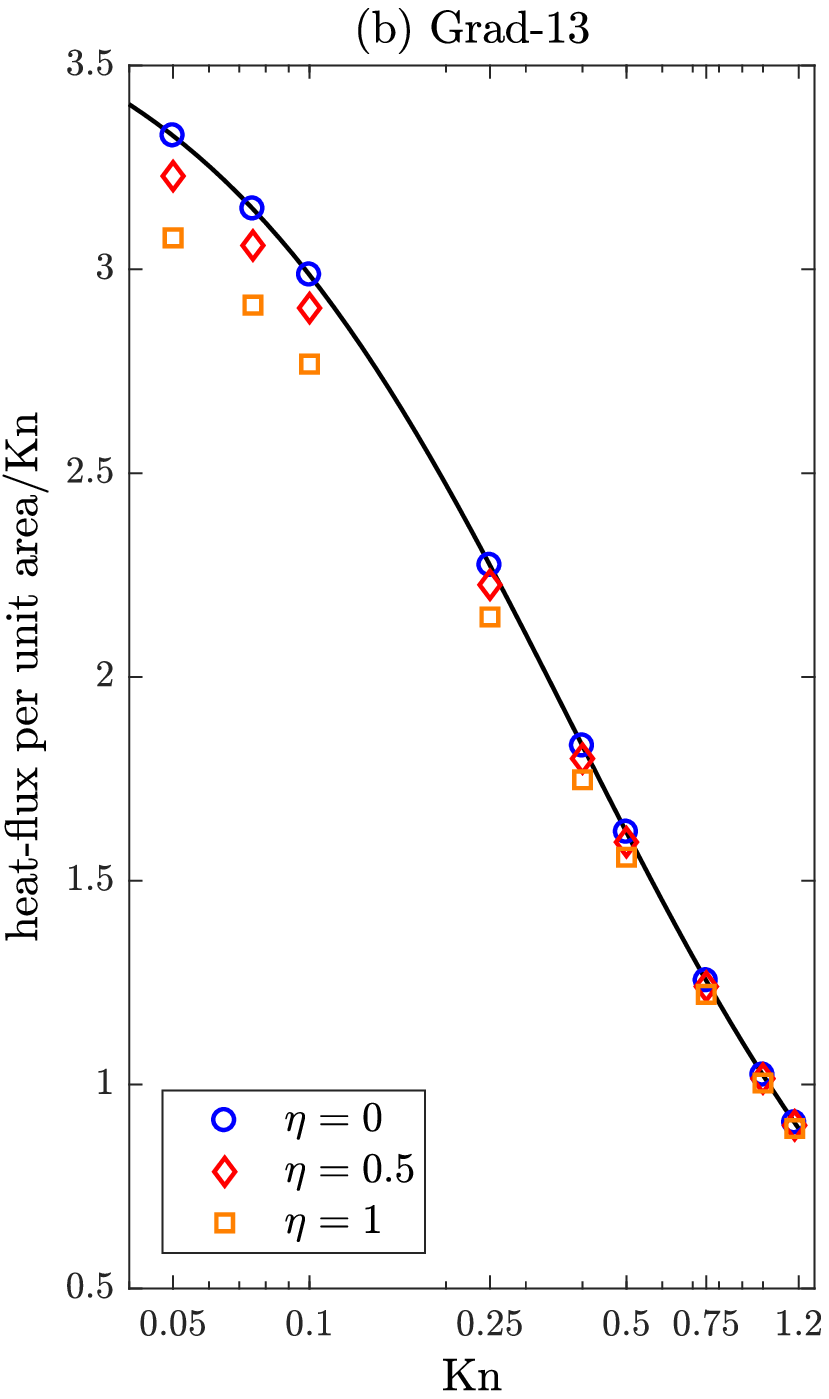}\hfill
\includegraphics[scale=0.5]{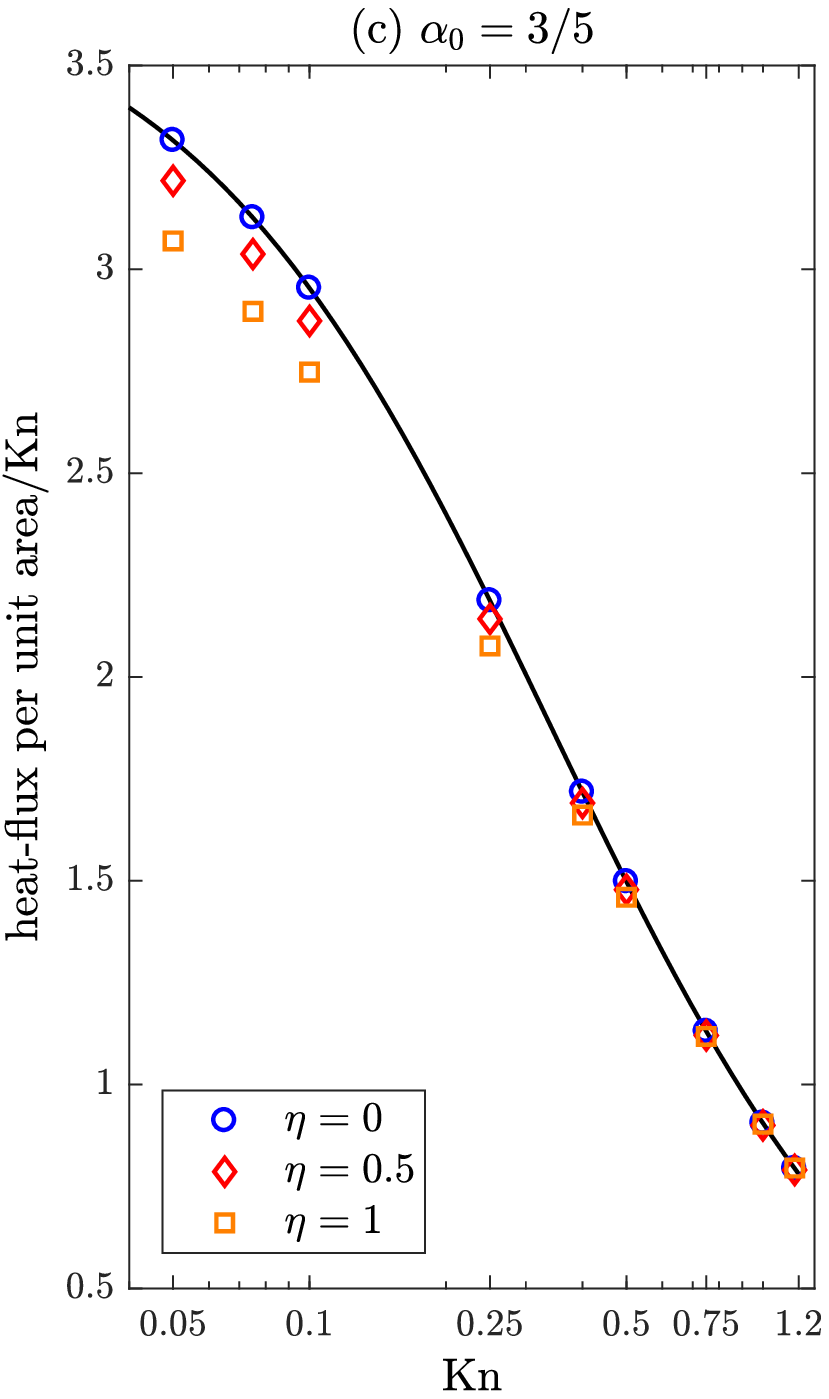}
\caption{\label{fig:heatfluxcase1}
Temperature driven case ($p_{sat}-p_{\infty }=0$, $T^{l}-T_{\infty }=1$): heat flux per unit area/$\mathrm{Kn}$ vs Knudsen number with $\eta = 0, 0.5, 1$. (left) NSF, (middle) Grad-13, and (right) $\alpha_0=3/5$.} The symbols denote the MSF solutions and the solid lines represent analytic results for a spherical droplet. 
\end{figure}

\underline{Temperature driven case}: In Figure \ref{fig:heatfluxcase1}, we present the heat-flux per unit area for the temperature driven case for different shape factors (see legend in the figures) for NSF (in sub Figure (a)), Grad-13 (sub Figure (b)) and CCR with $\alpha_0=3/5$ (sub Figure (c)). Note that, due to the Onsager reciprocity relations \citep{CHERNYAK1989}, the mass-flux for the temperature driven case is same as the heat-flux  for the pressure driven case (Figure \ref{fig:heatfluxcase2}) hence it is not shown. 
Evidently, form Figures \ref{fig:heatfluxcase1}, the magnitude of the heat-flux reduces with the deformation, i.e., as $\eta \rightarrow 1$. For NSF (Figure \ref{fig:heatfluxcase1}(a)); the reduction is about 3\% for $\eta=0.5$ and about 12\% for $\eta=1$.

\underline{Curvature dependence of the saturation pressure}: Throughout this article, we have taken $T^{I}$ and $p_{sat}$ as independent parameters. However, in
general, the Clausius-Clapeyron-Kelvin formula
provides a relation between the saturation pressure and the temperature at the
interface (linearized and dimensionless) \citep{MCELROY1979, BondStruchtrup2004}:%
\begin{equation}
p_{sat}\left( T^{I}\right) =p_{sat}^{p}\left( T^{I}\right) \underline{\exp \left(
2\gamma \kappa \right)} \text{.}  \label{saturationpressure}
\end{equation}%
Here, $\gamma$ is the dimensionless surface tension ($\gamma =\hat{\gamma}/%
\hat{\rho}_{l}\hat{\ell}\hat{R}\hat{T}_{0}$), $\kappa $ is the
dimensionless curvature ($\kappa =\hat{\kappa}\hat{\ell}$) and $%
p_{sat}^{p}\left( T^{I}\right) =H_{0}T^{I}$ is the saturation pressure
for a planar surface with $H_{0}$ ($=\hat{H}_{0}/\hat{R}\hat{T}_{0}$) being
the dimensionless heat of evaporation.

Thus far, the modelling assumption has been that $H_{0}$ (and $\gamma$) can be independently chosen to give a desired value of the saturation pressure. Notably, for the noble gases (vapour/liquid) the  Kelvin correction term (underlined) in equation (\ref{saturationpressure}) are often small, unless the radius of curvature $\hat{\kappa}$ is in nano-meters. For example, for Argon, if we choose $\hat{T}_0=103\,\mathrm{K}$, $\hat{p}_0=p^p_{sat} \left(\hat{T}_0\right) = 0.4\,\mathrm{MPa}$, $\hat{\rho}_l = 1294.6\,\mathrm{kg/m^3}$, $\hat{\gamma} = 8.75\, \mathrm{mN/m}$ [a case considered in \citep{RanaPRL2019}], the Kelvin correction term gives $\exp\left(0.6305/a\right)$, where $a$ (in $\mathrm{nm}$) is the inverse of local curvature $\hat{\kappa}$. Therefore, for a drop of size greater than $10\, \mathrm{nm}$, the effect of the  Kelvin correction term will be less then $6.1\%$.

In order to compare our results with existing literature, in \S\ref{sec:Inverted temperature profile} we shall use the dimensionless heat of evaporation $H_{0}$  as an independent parameter.


\section{Inverted temperature profile and Knudsen layer}
\label{sec:Inverted temperature profile}
A notorious deficiency of NSF and CCR theories is their inability to capture the Knudsen layer---a kinetic boundary layer extending within a few mean free path into the domain \citep{Sone2002, StruchtrupTorrilhon2008, TorrilhonARFM}. We have seen in previous sections that the CCR model is able to predict global flow features, such as net mass flux,  heat flux and drag; however, it may fail to describe finer flow features especially those in which Knudsen layer and other rarefaction effects are dominant. In this section, we shall consider such a case. 
\begin{figure}
\centering
\includegraphics[height=45mm]{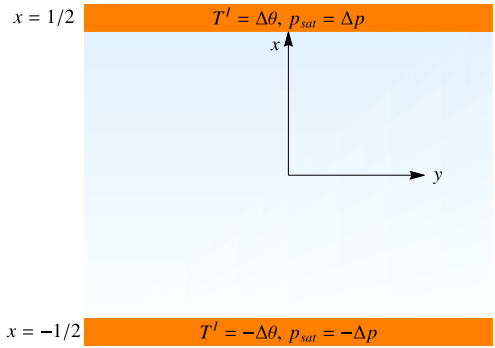}
\caption{\label{fig:scamatic1DHT}
Schematic for heat and mass transfer between two liquid layers.}
\end{figure}

Let us consider heat and mass
transfer between two liquid layers located at locations $x=\pm 1/2$, see Figure~\ref{fig:scamatic1DHT}.  We
prescribe the dimensionless temperatures and saturation pressures of the liquid layers as $T^{I}\left( \pm 1/2\right) =\pm \Delta \theta$ and $p_{sat}\left( \pm 1/2\right) =\pm \Delta p$.
For a planar surface, $\hat{\kappa}=0$, and thus $\Delta p = \Delta
\theta H_{0}$.

After some calculation, the analytic solution of the linear problems \eqref{conservation laws vector form} and \eqref{CCR relations vector form}
assumes the form 
\begin{subequations}
\label{1dcase}
\begin{eqnarray}
v_{x} &=&c_{1}\text{, }q_{x}=c_{2}\text{, }T=-\frac{\Pr c_{2}}{c_{p} \mathrm{Kn}}x\text{,}  \label{1dcase1} \\
p &=&\sigma _{xx}=0\text{,}  \label{1dcase2}
\end{eqnarray}%
\end{subequations}
where $c_{1}$ and $c_{2}$ are integration constants, which need to
be evaluated using boundary conditions. Solving \eqref{BCs evaporation} for $c_{1}$ and $c_{2}$ gives

\begin{subequations}
\begin{eqnarray}
c_1 &=&  \sqrt{\frac{2}{\pi}}\left(\frac{(0.306763 +1.62499  \mathrm{Kn})H_0-0.741358 \mathrm{Kn}}{0.370583+1.77245 \mathrm{Kn}}\right) \Delta\theta\text{,} \label{1dcasec1} \\
 c_2  &=& \frac{5 \mathrm{Kn}}{2 \mathrm{Pr}}\left(\frac{0.741165 -0.157738 H_0}{0.370583+1.77245 \mathrm{Kn}}\right)\Delta\theta \text{.} \label{1dcasec2} 
\end{eqnarray}
\end{subequations}
Note that, for this problem, the CCR reduces to
NSF constitutive relations. 

Interestingly, as can be seen from inspection of  (\ref{1dcasec2}), the conductive heat flux $q_{x}$ switches sign for $H_{0} \geq 4.7$, which leads
to an inverted temperature profile in the vapour---this is qualitatively consistent with kinetic theory \citep{Pao1971} and MD simulations \citep{Frezzotti2003}. Notably, the NSF equations with classical boundary conditions,
i.e., with no-temperature jump boundary conditions along with the
Hertz-Knudsen-Schrage can not describe the inverted temperature  phenomenon. 

\begin{figure}
\centering
\includegraphics[height=50mm]{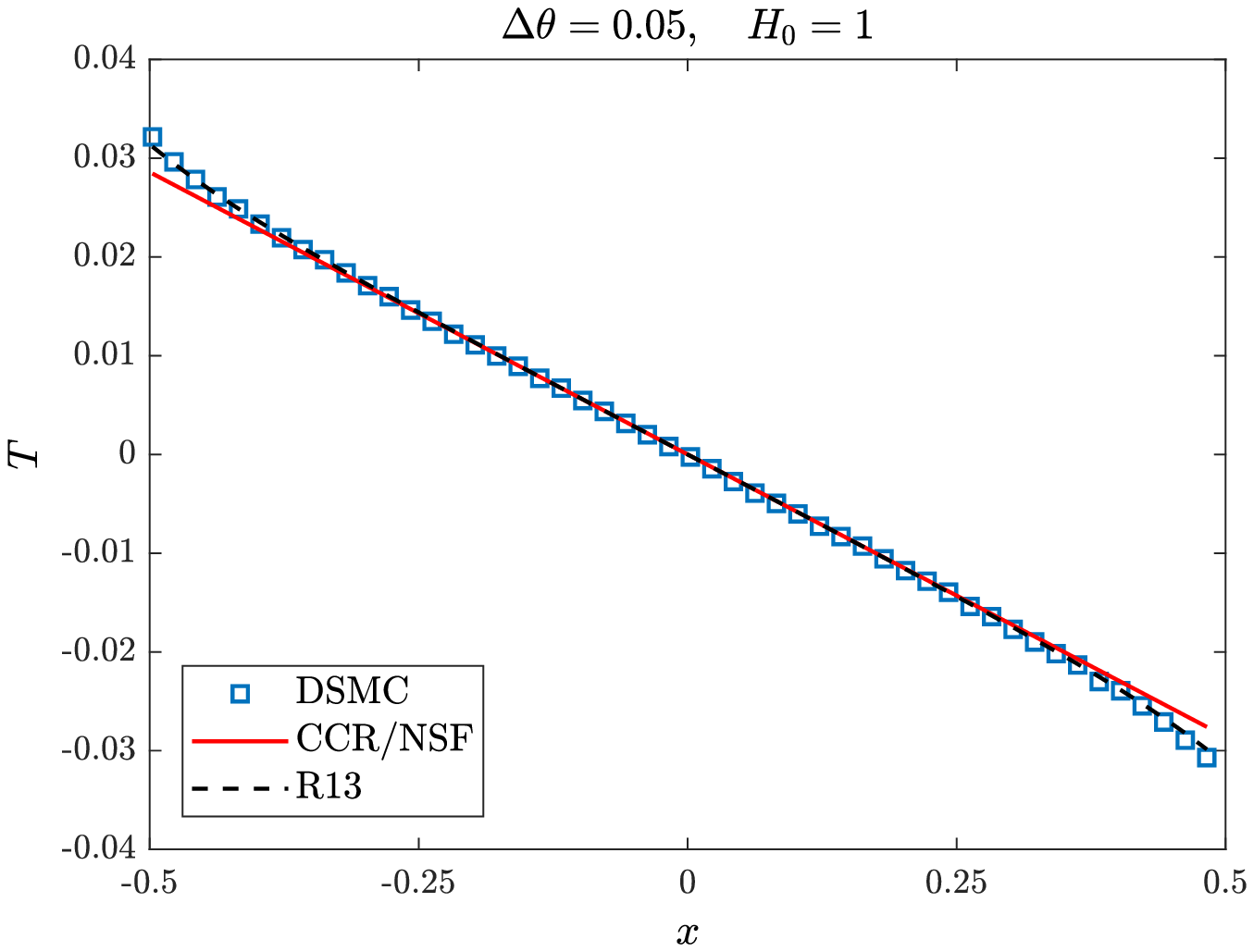}\hfill
\includegraphics[height=50mm]{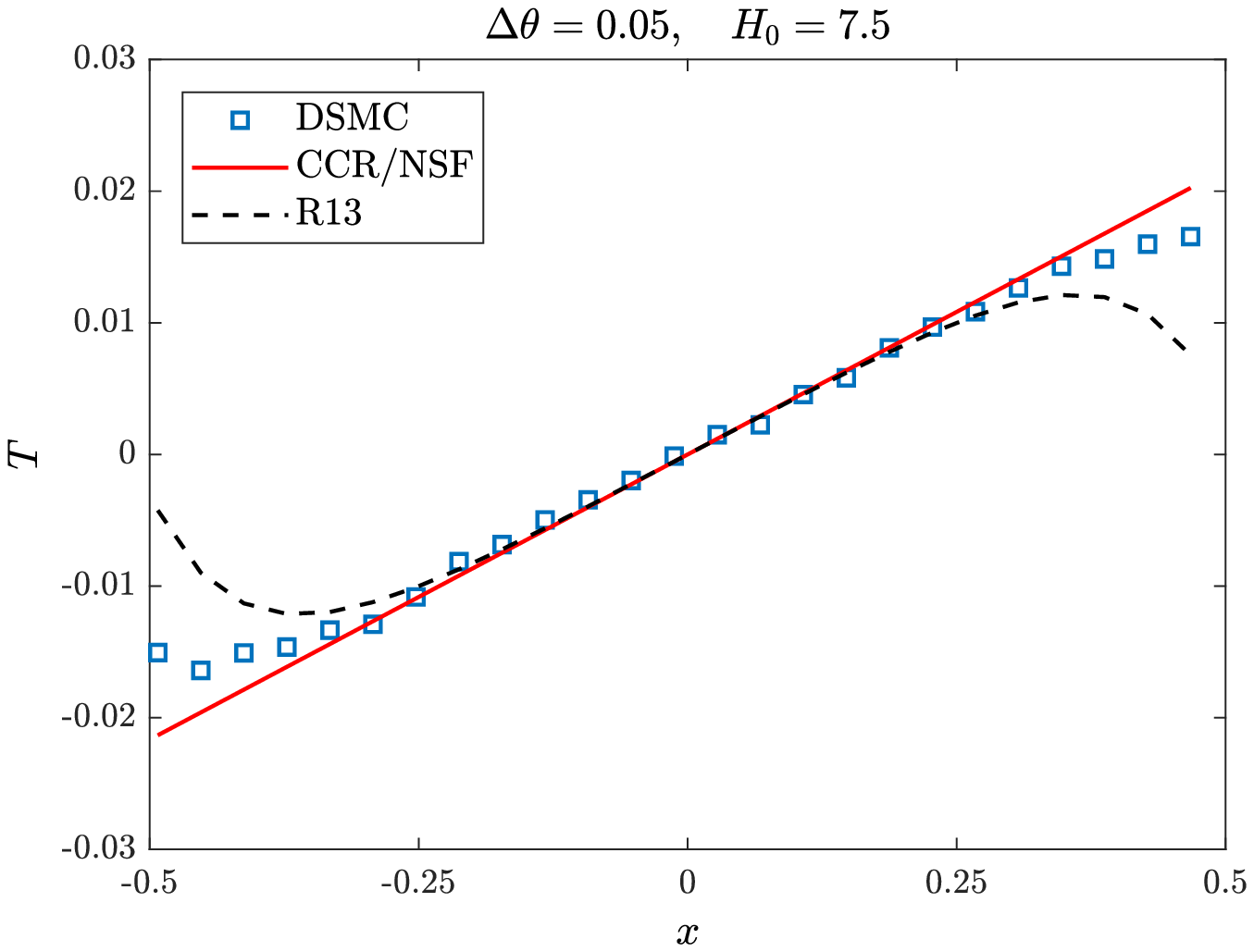}
\caption{\label{fig:invertedprofiles}
Temperature profiles as functions of location $x$ for $\left(
\Delta \theta =0.05\text{, }H_{0}=1\right) $ (left) and $\left( \Delta \theta =0.05
\text{, }H_{0}=7.5\right)$ (right) for $\mathrm{Kn}=0.078$. The figure on right demonstrates the inverted temperature profile.}
\end{figure}

Figures~\ref{fig:invertedprofiles}(i) and (ii) show the temperature curves as functions of $x$ for $
\Delta \theta =0.05\text{, }H_{0} =1 \left(< 4.7\right) $ and $\Delta \theta =0.05
\text{, }H_{0}=7.5 \left( > 4.7\right)$, respectively, for $\mathrm{Kn}=0.078$. These flow
parameters are chosen so that our results can be compared with those
obtained from the DSMC and R13 theories given in \citep{BRTS2018}. Comparing both cases, when $H_0$ is smaller than the critical heat of evaporation (case 1) the temperature profile
shows a large jump at both boundaries with higher temperature on the lower boundary. In this case, the mass and heat fluxes are both positive (i.e., from lower boundary to upper) and as expected, the temperature is higher at the lower boundary. On the other hand, in case 2, $H_0$ is larger then the critical heat of evaporation and the temperature profile is inverted. For this case the heat flows from the hot side (where condensation occurs) towards the cold side (where evaporation occurs). The results from the DSMC are denoted by symbols. 

Evidently, the CCR theory qualitatively explains the intriguing inverted temperature phenomenon. However, the
CCR model does not provide
 detail of the temperature profiles near to the boundaries; the region which is dominated by the Knudsen layer. The R13 theory provides improvement in predicting the temperature profile at this value of the Knudsen number because, as is well known, it can capture some basic features of the Knudsen layer.    

%
%

\section{Conclusion and future directions}
\label{sec: Conclusion and future directions}

The present article studies liquid vapour phase-transition processes in which $\mathrm{Kn}\lesssim 1$, i.e., down to sub-micron scales. In particular, we derive a set of fundamental solutions for the CCR system which allows
for three-dimensional multiphase micro flows at remarkably low computational cost---the fundamental solutions to the linearised NSF and Grad-13 equations are obtained as a special case. A set of thermodynamically consistent liquid-vapour interface conditions for the linearised CCR system are derived within the framework of irreversible thermodynamics.

Different macroscopic models were applied to solve the motion of a  sphere is a gas and compared the results with theoretical and experimental results from the literature. We observed that the results for the drag force obtained from the Grad-13 and CCR model are both in good agreement with the experimental results; a slightly better match at larger Knudsen number is obtained by taking $\alpha_0=3/5$.

The motion of two spherical particles was investigated and compared to
the classical Stokes solutions and experimental results. The drag force reduces as the Knudsen number increases, mainly due to velocity slip at the surface. The effect of proximity is investigated on the drag force, which decreases as particles gets closer.  For a doublet, contrary to the experimental observations, the Stokes and NSF equations predict a non-zero drag while the Grad-13 and CCR with coupling coefficient $\alpha_0=3/5$ provide excellent match with the experimental data. Proximity effects on mass and heat transfer coefficients of two interacting droplets were investigated over a range of Knudsen numbers. Two far-field conditions were examined: (i) Pressure driven where $p_{sat}-p_{\infty }=1$, $T^{l}-T_{\infty }=0$, and (ii) Temperature driven where $p_{sat}-p_{\infty }=0$, $T^{l}-T_{\infty }=1$.  However, due linearity of the governing equations and the boundary conditions involved, these two cases can be combined to give results for a general case. For the pressure-driven case a slight reduction ($\lesssim 3\%$) in the mass-flux was observed when the droplets are placed next to each other. On the other hand, the heat-flux was significantly reduced ($\sim 30\%$). For the temperature driven case, the situation is reversed, where a significant reduction in the mass-flux was observed due to shielding effects.
We also considered the case of a single deformed droplet and studied the effects of deformity on the mass and heat transfer characteristics over a range of the Knudsen number.

Motivated by these findings, future work
could be to extend this work to study phase transition
process in polyatomic fluids and binary mixtures.  

Extension of the current work  to unsteady flows and coupling to liquid dynamics \citep{RanaPRL2019, Mykyta2020} can be considered in future---the liquid phase can be modelled as an incompressible fluid  (Stokeset/Oseenlet and heatlet) and the vapor phase modelled via CCR model. However, it will require solving the moving boundary problem efficiently within the MFS framework \citep{jiang2014}. Another line of inquiry would be to implement these fundamental solutions via boundary integral methods which offer more flexibility and robustness as compared to the MFS. A prominent deficiency of NSF and CCR model is their inability to capture the Knudsen layer---a kinetic boundary layer extending  a few mean free paths into the domain. Future work can also include finding fundamental solutions to
the linearised R26-equations in order to incorporate Knudsen layers. 
The cornerstone of this work is the linearity of the differential operators involved, which allow us to obtain Green's functions and formulate the MFS. As such, the method developed here can not be directly applied to non-linear problems. In another context, MFS has been applied to some non-linear problems using appropriate fixed point iterative schemes \citep{FCM2009}. It would be interesting to attempt a similar approach for solution of the nonlinear CCR model and/or nonlinear boundary conditions.

\section{Acknowledgments}
This work has been partially  supported via Scheme for Promotion of Academic and Research Collaboration (SPARC) grant (IDN\_SKI 444) funded by the Ministry of Human Resources Development (MHRD), India and EPSRC
Grants No. EP/N016602/1, EP/P031684/1,
EP/S029966/1. The authors are grateful to Dr. Juan Padrino from the University of Warwick for many fruitful discussions on Green's functions.

\section{Declaration of Interests}
The authors report no conflict of interest.
\appendix

\section{Derivation of the Green's functions for the Sourcelet}
\label{sec:appendixA0}
The Fourier transformation of the energy conservation law (\ref{conservation laws perturebed source-sink}$_3$) and the heat
flux constitutive relation (\ref{CCR relations vector form}$_2$) gives%
\begin{equation}
\mathbf{k}\cdot \mathbf{\tilde{q}}=0\text{, and \quad }\mathbf{\tilde{q}}=i\frac{%
c_{p}Kn}{\Pr }\mathbf{k}\tilde{\Theta}\text{,}
\label{FT temperature problem}
\end{equation}%
where we have defined $\Theta =T-\alpha _{0}p$ and its Fourier transformation 
$\mathcal{F} \left[ \Theta \left( \mathbf{r}\right) \right] =\tilde{\Theta}%
\left( \mathbf{k}\right) $. Clearly, the inverse Fourier transformation of (%
\ref{FT temperature problem}) leads to 
\begin{equation}
\mathbf{q=0}\text{, and \quad}\Theta =0\text{.}  \label{IFT temperature problem}
\end{equation}%
Similarly, taking a Fourier transformation of (\ref{conservation laws perturebed source-sink}$_{1,2}$), one obtains 
\begin{equation}
\mathbf{k}\cdot \mathbf{\tilde{v}}=ih\text{, and \quad }\tilde{p}\mathbf{k}+\tilde{\Pi}%
\cdot \mathbf{k}=\mathbf{0}\text{,}
\label{FT velocity problem}
\end{equation}
where from (\ref{CCR relations vector form}$_1$) 

\begin{equation}
\tilde{\bm{\Pi }}=i\mathrm{Kn}\left( \mathbf{\tilde{v}k}+\mathbf{k\tilde{v}}-\frac{2}{3}%
\left( \mathbf{k}\cdot \mathbf{\tilde{v}}\right) \mathbf{I}\right) \text{.}
\label{FT CCR velocity}
\end{equation}%
Equations (\ref{FT velocity problem} and \ref{FT CCR velocity}) give%
\begin{equation}
\tilde{v}=ih\frac{\mathbf{k}}{|k|^{2}},\text{\quad }\tilde{\bm{\Pi}}=-2\mathrm{Kn}h\left( \frac{%
\mathbf{kk}}{|k|^{2}}-\frac{1}{3}\mathbf{I}\right) \text{, \quad }\tilde{p}=\frac{4%
}{3}\mathrm{Kn}h\text{.}  \label{FT v sigma and p}
\end{equation}%
Finally, taking the inverse Fourier transformation of (\ref{FT v sigma and p}%
) yields%
\begin{eqnarray*}
\mathbf{v} &=&ih\mathcal{F}^{-1}\left[ \frac{\mathbf{k}}{|k|^{2}}\right] =%
\frac{h}{4\pi }\frac{\mathbf{r}}{|r|^{3}}, \quad p = 0\text{, and} \\
\mathbf{\bm{\Pi} } &=&-2\mathrm{Kn}h\mathcal{F}^{-1}\left[ \frac{\mathbf{kk}}{|k|^{2}}-%
\frac{1}{3}\mathbf{I}\right] =\frac{3\mathrm{Kn}h}{2\pi }\frac{1}{|r|^{5}}\left[ \mathbf{rr}-\frac{1}{3}|r|^{2}%
\mathbf{I}\right]\text{.}
\end{eqnarray*}%
Here, we have used the following inverse Fourier
transformation identities:%
\begin{eqnarray}
\mathcal{F}^{-1}\left[ \frac{\mathbf{k}}{|k|^{2}}\right]  =-\frac{i}{4\pi }%
\frac{\mathbf{r}}{|r|^{3}}\text{, and }\mathcal{F}^{-1}\left[ \frac{\mathbf{kk}}{|k|^{2}}\right]  =-\frac{3}{4\pi 
}\left[ \frac{\mathbf{rr}}{|r|^{5}}-\frac{1}{3|r|^{3}}\mathbf{I}\right] 
\text{.}
\end{eqnarray}

\section{Phenomenological coefficients in the boundary conditions}
\label{sec:appendixPBC}
The numerical coefficients appearing in the phenomenological boundary
conditions are based on \citet{Sone2002}, where these were obtained by an asymptotic
expansion of the linearized Boltzmann equation for $\mathrm{Kn\rightarrow 0}$%
\textrm{.} The coefficients $\eta _{ij}$ appearing in equations (\ref{PBC coeffcients}) can be
directly compared to \citet{Sone2002}, as%
\begin{eqnarray}
\eta _{11} &=&-\frac{4\sqrt{\pi }d_{1}}{4C_{4}^{\ast }d_{1}+5\text{$\gamma $}%
_{2}d_{4}^{\ast }}\text{, \quad}\eta _{12}=\frac{5\sqrt{\pi }\text{$\gamma $}%
_{2}d_{4}^{\ast }}{4C_{4}^{\ast }d_{1}+5\text{$\gamma $}_{2}d_{4}^{\ast 2}}
\\
\eta _{12} &=&-\frac{25\sqrt{\pi }\text{$\gamma $}_{2}^{2}d_{4}^{\ast 2}}{%
16C_{4}^{\ast }d_{1}^{2}+20\text{$\gamma $}_{2}d_{1}d_{4}^{\ast 2}}\text{,
and \quad}\tau _{0}=\frac{5\sqrt{\pi }\text{$\gamma $}_{2}}{8d_{1}}
\end{eqnarray}%
Here, the values of $\gamma $$_{1}$, $\gamma $$_{2}$, $d_{1}$, $d_{4}^{\ast }
$ and $C_{4}^{\ast }$ for a hard-sphere gas (HS) and for the BGK model (BGK)
are given in Table \ref{Tab:Table1}:

\begin{table}
\begin{center}
\begin{tabular}{l l l l l l}
& $\gamma $$_{1}$ & $\gamma $$_{2}$ & $d_{1}$ & $d_{4}^{\ast }$ & $%
C_{4}^{\ast }$ \\ 
\hline
HS & $1.27004$ & $1.9223$ & $2.4001$ & $-0.4557$ &  $-2.1412$ \\ 
\addlinespace[0.4cm]
BGK & $1$ & $1$ & $1.3027$ & $-0.4467$ & $-2.1320$ \\
\end{tabular}
\end{center}
\caption{\label{Tab:Table1}The values of 
$\gamma_{1}$, $\gamma_{2}$, $d_{1}$, $d_{4}^{\ast }
$ and $C_{4}^{\ast }$ for a hard-sphere gas (HS) and for the BGK model
(BGK). Data taken from \citet{Sone2002}.}
\end{table}

\section{Analytic solutions obtained from fundamental solutions}
\label{sec:appendixA}
The analytic solutions (\ref{excat solution CCR}) for an evaporating spherical droplet can also
be obtained from the fundamental solutions (\ref{velocity solution all}--\ref{stress solution all}). Due to the spherical
symmetry let us place one singular mass source of strength $h$ and a heat
source of strength $g$ at the origin. Let $\mathbf{r}=r\left\{ \cos \phi
\sin \theta ,\sin \phi \sin \theta ,\cos \theta \right\} $ be an arbitrary
point outside the droplet ($r\geq 1$). From (3.7--3.8) we get 

\begin{equation}
\mathbf{v}\left( \mathbf{r}\right) \cdot \mathbf{n}=\frac{h}{4\pi r^{2}}%
\text{, }p\left( \mathbf{r}\right) =0\text{, }\mathbf{n}\cdot \sigma \left( 
\mathbf{r}\right) \cdot \mathbf{n=}\frac{\mathrm{Kn}\left( h+\text{$\alpha
_{0}$}g\mathrm{Kn}\right) }{\pi r^{3}}  \label{Single source 1}
\end{equation}%
and%
\begin{equation}
T\left( \mathbf{r}\right) =\frac{\text{$\Pr $}}{\text{$c_{p}$}}\frac{g}{4\pi
r}\text{, }\mathbf{q}\left( \mathbf{r}\right) \cdot \mathbf{n}=\frac{\mathrm{%
Kn}g}{4\pi r^{2}}  \label{Single source 2}
\end{equation}%
Comparing (\ref{Single source 1}--\ref{Single source 2}) with (\ref{excat solution CCR}), one finds 
$c_{1}=h/4\pi $, $c_{2}=\mathrm{Kn}g/4\pi $.

\bibliography{refer}
\bibliographystyle{jfm_doi}

\end{document}